\documentclass[traditabstract]{aa}
\usepackage{graphicx}
\usepackage{txfonts}
\usepackage{longtable}
\usepackage{natbib}
\usepackage{float,color}
\usepackage[figuresright]{rotating}
\bibpunct{(}{)}{;}{a}{}{,}

\newcommand{\dnu}{\Delta \nu}
\newcommand{\num}{\nu_{\rm max}}

\newcommand{\teff}{$T_{\!\mbox{\scriptsize\em \rm eff}}$}

\newcommand{\feh}{[\rm{Fe}/\rm{H}]}

\newcommand{\Msun}{$M_\odot$}

\newcommand{\wh}{W_{H_\alpha}}

\newcommand{\Teff}{T_{\rm eff}}
\newcommand{\logg}{\log{g}}

\newcommand{\ha}{H$_\alpha$}

\begin{document}

\title{The Gaia-ESO survey: Hydrogen lines in red giants directly trace stellar mass}
\author{Maria Bergemann\inst{\ref{inst1}} \and Aldo Serenelli\inst{\ref{inst2}} \and Ralph Sch{\"o}nrich\inst{\ref{inst3}} \and Greg Ruchti\inst{\ref{inst4}} \and Andreas Korn\inst{\ref{inst5}} \and Saskia Hekker\inst{\ref{inst6},\ref{inst13}} \and Mikhail Kovalev\inst{\ref{inst1}} \and Lyudmila Mashonkina\inst{\ref{inst18}} \and Gerry Gilmore\inst{\ref{inst7}} \and Sofia Randich\inst{\ref{inst8}} \and Martin Asplund\inst{\ref{inst9}} \and  Hans-Walter Rix\inst{\ref{inst1}} \and Andrew R. Casey\inst{\ref{inst7}} \and 
Paula Jofre\inst{\ref{inst7}} \and Elena Pancino\inst{\ref{inst10},\ref{inst15}} \and Alejandra Recio-Blanco\inst{\ref{inst11}} \and Patrick de Laverny\inst{\ref{inst11}} \and Rodolfo Smiljanic\inst{\ref{inst12}} \and Grazina Tautvaisiene\inst{\ref{inst14}} \and
Amelia Bayo\inst{\ref{inst16}}  \and Jim Lewis\inst{\ref{inst7}} \and Sergey Koposov\inst{\ref{inst7}} \and Anna Hourihane\inst{\ref{inst7}} \and Clare Worley\inst{\ref{inst7}} \and 
Lorenzo Morbidelli\inst{\ref{inst8}} \and Elena Franciosini\inst{\ref{inst8}} \and Germano Sacco\inst{\ref{inst8}} \and Laura Magrini\inst{\ref{inst8}} \and Francesco Damiani\inst{\ref{inst17}} and Joachim M. Bestenlehner\inst{\ref{inst1}}
}
\institute{
Max-Planck Institute for Astronomy, 69117, Heidelberg, Germany \email{bergemann@mpia-hd.mpg.de}
\label{inst1}
\and
Instituto de Ciencias del Espacio (ICE-CSIC/IEEC) Campus UAB, Carrer Can Magrans S/N, Bellaterra, 08193, Spain
\label{inst2}
\and
Rudolf-Peierls Centre for Theoretical Physics, University of Oxford, 1 Keble Road, OX1 3NP, Oxford, United Kingdom 
\label{inst3}
\and
Lund Observatory, Box 43, SE-221 00 Lund, Sweden 
\label{inst4}
\and
Department of Physics and Astronomy, Division of Astronomy and Space Physics, Angstrom laboratory, Uppsala University, Box 516, 75120 Uppsala, Sweden 
\label{inst5}
\and
Max Planck Institute for Solar System Research, Justus-von-Liebig-Weg 3, 37077, G�ttingen Germany 
\label{inst6}
\and
Institute of Astronomy, University of Cambridge, Madingley Road, CB3 0HA, Cambridge, UK 
\label{inst7}
\and
INAF - Osservatorio Astrofisico di Arcetri, Largo E. Fermi 5, 50125, Florence, Italy 
\label{inst8}
\and
Research School of Astronomy \& Astrophysics, Mount Stromlo Observatory, The Australian National University, ACT 2611, Australia 
\label{inst9}
\and
INAF - Osservatorio Astronomico di Bologna, via Ranzani 1, 40127, Bologna, Italy 
\label{inst10}
\and
Laboratoire Lagrange, Universite de Nice Sophia Antipolis, CNRS, Observatoire de la Cote d'Azur, BP 4229,F-06304 Nice cedex 4, France 
\label{inst11}
\and
Department for Astrophysics, Nicolaus Copernicus Astronomical Center, ul. Rabianska 8, 87-100 Torun, Poland 
\label{inst12}
\and
Stellar Astrophysics Centre, Department of Physics and Astronomy, Aarhus University, Ny Munckegade 120, DK-8000 Aarhus, Denmark 
\label{inst13}
\and
Institute of Theoretical Physics and Astronomy, Vilnius University, Go\v{s}tauto 12, LT-01108 Vilnius, Lithuania 
\label{inst14}
\and
ASI Science Data Center, Via del Politecnico SNC, 00133 Roma, Italy 
\label{inst15}
\and
Instituto de Fisica y Astronomia, Fac. de Ciencias, U de Valparaiso, Gran Bretana 1111, Playa Ancha, Chile
\label{inst16}
\and
INAF - Osservatorio Astronomico di Palermo, Piazza del Parlamento, 1, 90134 Palermo, Italy
\label{inst17}
\and
Institute of Astronomy, Russian Academy of Sciences, Pyatnitskaya st. 48, RU-119017 Moscow, Russia
\label{inst18}
}
\date{Received date / Accepted date}
\abstract{Red giant stars are perhaps the most important type of stars for Galactic and extra-galactic archaeology: they are luminous, occur in all stellar populations, and their surface temperatures allow precise abundance determinations for many different chemical elements. Yet, the full star formation and enrichment history of a galaxy can be traced directly only if two key observables can be determined for large stellar samples - age and chemical composition. While spectroscopy is a powerful method to analyse the detailed abundances of stars, stellar ages are the "missing link in the chain", since they are not a direct observable. However, spectroscopy should be able to estimate stellar masses, which for red giants directly infer ages provided their chemical composition is known. 

Here we establish a new empirical relation between the shape of the hydrogen line in the observed spectra of red giants and stellar mass determined from asteroseismology. The relation allows to determine stellar masses and ages with the accuracy of 10-15\%. The method can be used with confidence for stars in the following range of stellar parameters: $4000 < \Teff < 5000$ K, $0.5 < \log g < 3.5$, $-2.0 <$ [Fe$/$H] $< 0.3$, and luminosities $\log L/L_{\rm Sun} < 2.5$. Our analysis provides observational evidence that the \ha~spectral characteristics of red giant stars are tightly correlated with their mass and therefore their age. We also show that the method samples well all stellar populations with ages above $1$ Gyr. Targeting bright giants, the method allows to obtain simultaneous age and chemical abundance information far deeper than would be possible with asteroseismology, extending the possible survey volume to remote regions of the Milky Way and even to neighbouring galaxies like Andromeda or the Magellanic Clouds already with present instrumentation, like VLT and Keck facilities.}

\titlerunning{Stellar Masses and Balmer lines}

\authorrunning{Bergemann et al.}

\maketitle
%
%
%
%
\section{Introduction}

One of the key problems in stellar and galactic astrophysics is to determine the age of a star. The star formation history of a population or galaxy can be best traced if we know how to connect the chemo-dynamical data of stars to their formation time.

Age determinations for the Galactic field stars have traditionally been limited to stars on the upper main sequence and on the subgiant branch. The most convenient and widely-used approach relies on fitting stellar isochrones to the classical observables (e.g. \teff, $\logg$, and $\feh$) (Pont \& Eyer 2004, Jorgensen \& Lindegren 2005). To a lesser extent, empirically calibrated methods are used (Soderblom 2010, 2015). They rely on various observational findings, such as emission lines that are related to the chromospheric activity, the depletion of Li with stellar age as the convection zone of a star thickens, and surface rotation, which slows down with increasing age. For the cool main-sequence stars, more elegant and accurate adaptations of the rotation method have been proposed (Garcia et al. 2014, Meibom et al. 2015), but they also rely on ultra-precise light-curves.

For red giant stars, the stars of most prominent interest in the context of Galactic and extra-galactic archaeology, classical isochrone methods are not useful: even if the metal content is known, isochrones of different ages are very close and would require \teff~and $\logg$ determinations with better than $0.5$\% accuracy\footnote{In addition to the systematic uncertainty of the theoretical \teff~scale.}. Only if luminosities of the stars are accurately known, e.g. because distances are available and reddening is not a problem, can evolutionary models be used to determine stellar mass \citep{2015ApJ...812...96G}. However, also these determinations suffer from systematic errors in input stellar parameters  and uncertainties of the stellar models. One may relate the process of ageing to in situ stellar nucleosynthesis, e.g. the ratio of carbon over nitrogen abundance increases as the star experiences multiple dredge-up episodes and the products of stellar nucleosynthesis are mixed to the surface (Masseron \& Gilmore 2014, Martig et al. 2016, Ness et al. submitted).

Recently, asteroseismology has emerged as a promising tool to determine masses of stars that display solar-like oscillations, i.e. stochastic oscillations excited by turbulent motions in the stellar convective envelope. Asteroseismic observations provide the surface gravity and mean density of a star, and, in combination with \teff\ and $\feh$, yield precise masses for stars of all types (Chaplin et al. 2014, Pinsonneault et al. 2014), from the main sequence (MS) through the hydrogen-shell-burning (RGB) to the core-helium burning (Red Clump, hereafter RC) phase. Asteroseismology is a very promising method, however it has two caveats. First, the accuracy of masses and ages derived depends on asteroseismic scaling relations. These relations may suffer from systematic biases of a few percent\footnote{A systematic error in $\num$ propagates to the 3rd power in mass and the error in $\dnu$ to the fourth power.} (e.g. Belkacem et al. 2011, Miglio et al. 2012, Coelho et al. 2015, White et al. 2015). Secondly, the method is mainly useful for sub-Galactic studies in the limited volume of the Milky Way, since the photometric data quality quickly deteriorates with magnitude and hence distance. Thus the technique cannot provide any mass or age estimates for larger distances, in- and outside the Galaxy. 

The  determination of the mass and composition of an evolved red giant star is, unlike for a main sequence star, an excellent proxy for its age. The time a star lives as a red giant is a small fraction of its total main sequence lifetime. The latter is determined through stellar evolutionary models by the initial mass and composition of the star. Then, the age of a red giant star is to a very good approximation only a function of those two quantities. If masses, or proxies for masses, can be estimated from observations, age-dating of giants is more straightforward than for main-sequence stars (Soderblom 2010).

Here we report on a newly observed relationship between the shape of the optical hydrogen line (\ha) and the mass of red giant stars. More massive stars appear to have fainter, i.e. weaker, \ha~lines. We make use of high-resolution and high signal-to-noise spectra from different observational programs, including the Gaia-ESO large spectroscopic survey (Gilmore et al. 2012, Randich et al. 2013) and asteroseismic data from the CoRoT and Kepler space missions. Our approach is purely empirical. We cannot yet identify the physical mechanism underlying the relation, because there are no red giant model atmospheres built from first principles that could be used to reproduce the observed \ha~line, to guide our understanding of the relationship with the stellar mass. The relationship, which we established on Galactic field stars, also gives accurate results for the stars in open and globular clusters, enabling direct spectroscopic stellar mass measurements without isochrone fitting.

The paper is structured as follows. Our observations and the data analysis are presented in Sec. \ref{sec:obs}, along with a brief summary of the state-of-the-art in modelling Balmer lines in the spectra of cool stars. The new method to derive masses and their implied ages is described in Section \ref{sec:massage}. In Sec. \ref{sec:discussion} we outline some applications of the method in the context of Galactic and extra-galactic astrophysics.
%
%
\section{Observations and analysis}{\label{sec:obs}}

\subsection{Spectroscopic observations}
Our main dataset is the high-resolution\footnote{The high-resolution Gaia-ESO spectra are obtained with UVES spectrograph at the VLT, $R \sim 47\,000$. The data are publicly available through the ESO archive.} stellar spectra from the Gaia-ESO 2nd and 3rd data releases (iDR2iDR3). The majority of the spectra have signal-to-noise (S/N) ratios between $50$ and $220$. In the top tier we have the stars with high-quality asteroseismic data from CoRoT, but we, too, include red giants observed in open and globular clusters (Table \ref{table5}). The stars in clusters do not belong to our core dataset, though, since their masses\footnote{We assume the same mass for all stars in a cluster. The stars are nearly coeval (to about 10\%), thus their initial masses differ by no more than 3\%. As most of the stars are not so high on the RGB to be significantly affected by mass loss, we may assume that their present day mass is a good proxy for their  initial mass.} and ages are determined by another method, the cluster colour-magnitude diagram (CMD) fitting. The data reduction of the UVES spectra has been described in \citet{2014A&A...565A.113S} in detail. The radial velocity correction was performed masking the Balmer lines and regions affected by telluric contamination. The spectra were normalised by dividing them with a function, which describes the stellar continuum emission convolved with the FLAMES-UVES instrumental response as described in \citet{2014A&A...565A.113S}. In total, the Gaia-ESO sub-sample contains $21$ CoRoT stars in the Galactic field and $73$ stars in $7$ star clusters. The observational details are listed in Tables \ref{table1} and \ref{table3} in Appendix.

In addition, we include $47$ stars from \citet{2012A&A...543A.160T} that were observed by \rm{Kepler} and included in the first APOKASC Catalog (Pinsonneault et al. 2014). The evolutionary status is known for some of them. The stellar spectra were taken on different facilities with resolving powers ranging from $67\,000$ (NOT) to $80\,000$ (NARVAL). The spectra are available from \citet{2012yCat..35430160T}. The S$/$N ratios are between $80-100$ to $200$. Our sample, too, includes two very metal-poor stars, KIC 4671239 (Hennes) and KIC 7693833 (Rogue). These objects are the targets of Silva Aguirre et al. (2016, in preparation). The spectrum of KIC 4671239 is now publicly available from the NOT database, and KIC 7693833 was observed by \citet{2012A&A...543A.160T}. The data reduction of these spectra is described in \citet{2012A&A...543A.160T}.

\begin{figure}[!ht]
\includegraphics[width=0.44\textwidth, angle=0]{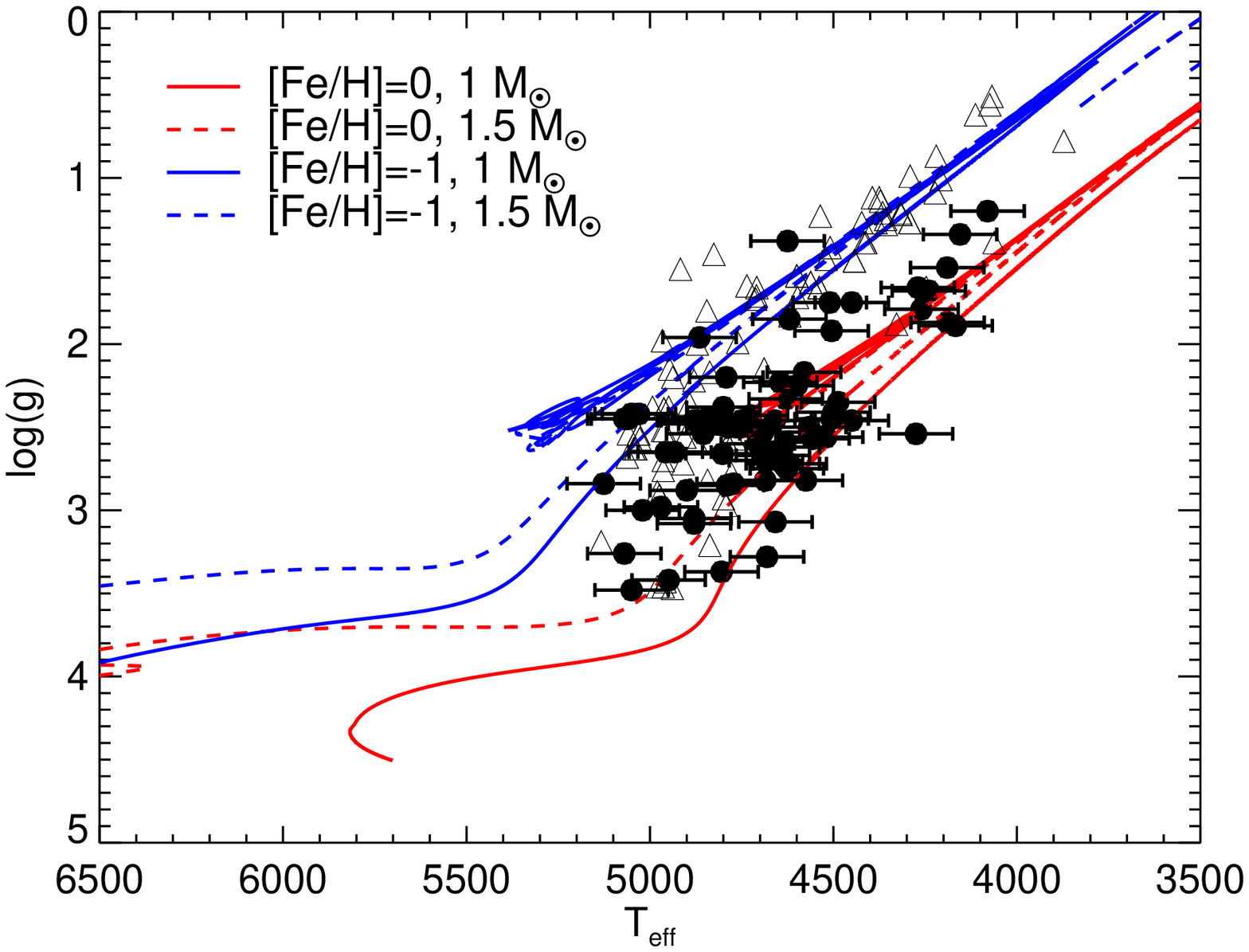}
\includegraphics[width=0.44\textwidth, angle=0]{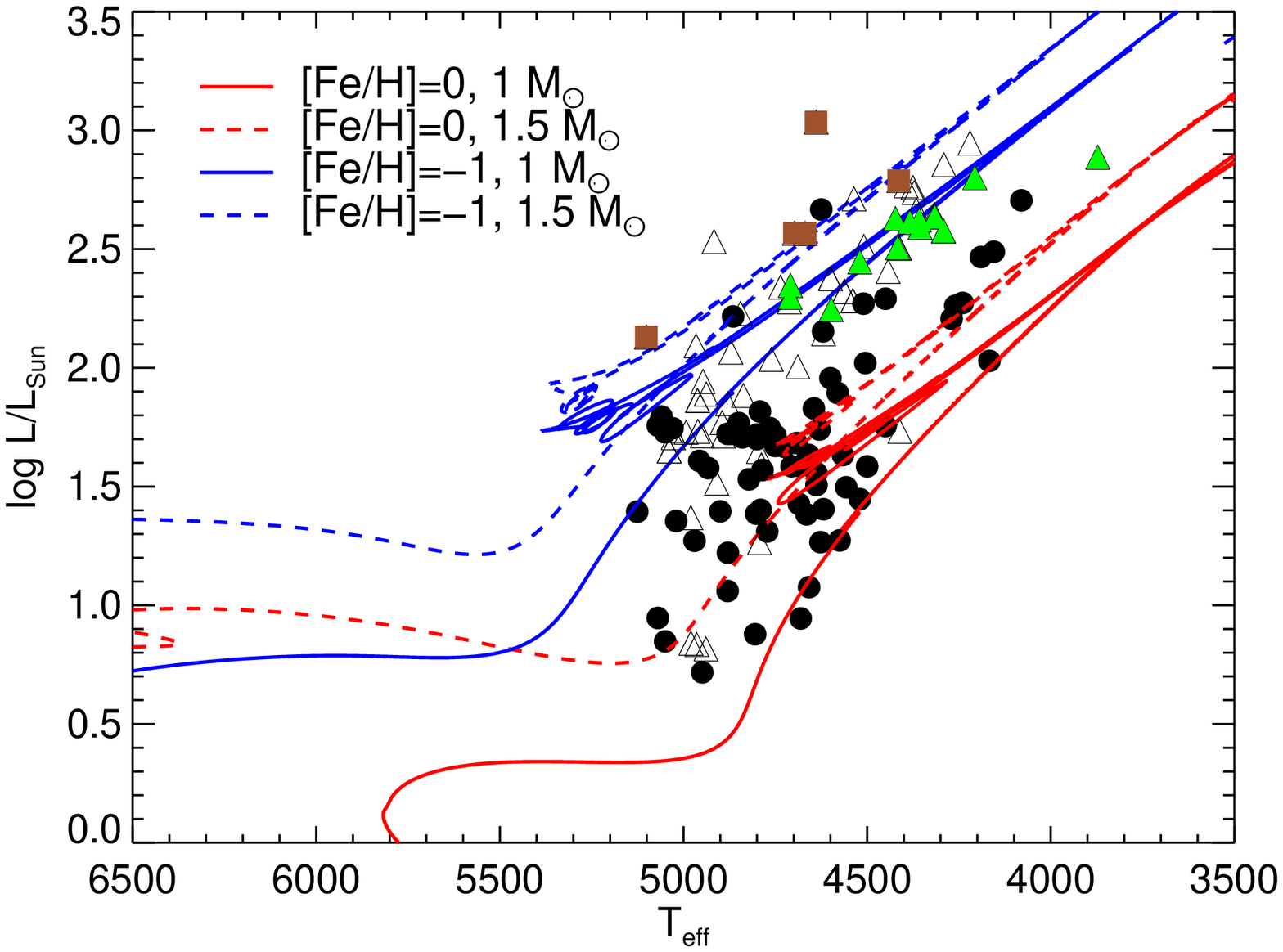}
\caption{The Hertzsprung-Russell diagram for the observed stellar sample: black filled points - the stars with asteroseismic masses, triangles - the stars in stellar clusters. The stars in the globular clusters NGC 2808 and NGC 4372 are shown with green triangles and brown squares, respectively. The Garstec evolutionary tracks \citep{2013MNRAS.429.3645S} for different masses and metallicities are also shown.}
\label{hrd}
\end{figure}

The full sample contains $69$ Galactic field stars with asteroseismic masses and $73$ stars in the clusters. Figure~\ref{hrd} shows the location of the stars on the Hertzsprung-Russell diagram. These are red giants or red clump stars with \teff~ in the range $4000 - 5000$ K, and $\log$ g from $0.5$ to $3.6$. The stars span a wide range of metallicity from $-2.64$ to $0.5$.

For consistency, we renormalise all available spectra using the same procedure as in \citet{2014MNRAS.443..698S} that fits selected continuum windows with a low-order polynomial.

%
%
%
\begin{table}
\caption{Stellar parameters for star clusters. The references to metallicities and ages are also given.}
\label{table5}
\tabcolsep1.1mm
\begin{center}
\begin{tabular}{rrlll}
\noalign{\smallskip}\hline\noalign{\smallskip} 
\noalign{\smallskip}\hline\noalign{\smallskip} 
          Cluster & [Fe/H] &  Age & Mass & Reference \\
                      &  dex  &   Gyr & M$\odot$ &  \\
\noalign{\smallskip}\hline\noalign{\smallskip}
M 67          & ~~0.06  & 4.30           & 1.32 & Salaris et al. (2004)  \\
NGC 2243 & $-$0.48  & 4.66           & 1.20 & Salaris et al. (2004)  \\
NGC 5927 & $-$0.50  & 10.75 $\pm$ 0.38 & 0.94  & Vandenberg et al. (2013) \\
NGC 2808 & $-$1.18  & 11 $\pm$ 0.38    & 0.84 & Vandenberg et al. (2013) \\
NGC 1851 & $-$1.18  & 11 $\pm$ 0.25    & 0.84 & Vandenberg et al. (2013)  \\
NGC 6752 & $-$1.54  & 12.50 $\pm$ 0.25 & 0.79 & Vandenberg et al. (2013)  \\
NGC 4372 & $-$2.17  & 11.2 - 12.5   & 0.81 & De Angeli et al. (2005) \\
\noalign{\smallskip}\hline\noalign{\smallskip}
\end{tabular}
\end{center}
\end{table}

%
\subsection{Model atmospheres and spectroscopic analysis}{\label{sec:specmod}}
For the Gaia-ESO dataset, we use the recommended \teff~ and \feh\ available from the iDR3 stellar parameter analysis run (Table \ref{table2}). The stellar parameters values were determined as described in Smiljanic et al. (2014) and in Bergemann et al. (2014), using several different methods of spectroscopic analysis. The radiation transport is solved in LTE with MARCS model atmospheres (Gustafsson et al. 2008). The typical uncertainty in the parameter values is 100~K in \teff~and $0.1$~dex in $\feh$. For one of the NGC 4372 stars, 12253419-7235252, we adopt the following uncertainties: 
$\delta$ \teff$~= 100$ K, ~$\delta \log g = 0.5$ dex, $\delta \feh = 0.2$ dex, because no data is available in the survey database. The uncertainties are conservative and reflect the deviation of individual spectroscopic estimates. These uncertainties have a very small impact on the mass estimates: the 100 K error in \teff~propagates as a $5\%$ error in mass, and the error of $0.1$ dex in $\feh$ propagates as a 1.5\% error in mass. The comparison with the independent photometric estimates gives us confidence in the the accuracy of stellar parameters. The spectroscopic \teff~ estimates agree to better than $70$ K with the temperatures determined using the method of Infra-Red Fluxes \citet{2014MNRAS.439.2060C}.

For the Thygesen et al. (2012) stellar sample, we use the values of \teff~ and $\feh$ that were determined by Bruntt et al. (2012). The stellar parameters are provided in Table \ref{table6}. One of the very metal-poor (VMP) stars, Rogue, has metallicity $\feh = -2.23$ and was analysed by Thygesen et al. (2012). For the other, Hennes, the mean of two values given in Silva Aguirre et al. (2016, in preparation) and used by Guggenberger et al. (in preparation) is $\feh = -2.64 \pm 0.22$.
\subsection{Asteroseismic data and analysis}\label{sec:astero}
Stellar masses are determined using the global seismic quantities, available from the CoRoT  and Kepler space mission observations. For the Gaia-ESO stellar sample, the asteroseismic data taken from \citet{2010A&A...517A..22M}. The large frequency separation, $\dnu$, is approximately the average frequency separation of radial oscillation modes of consecutive  order. The second seismic quantity is $\num$, the frequency at which the oscillations power spectrum exhibits its maximum power. $\dnu$ and $\num$ scale with global stellar parameters as 
\begin{equation}
\frac{\dnu}{\dnu_\odot} \simeq \sqrt{\frac{M/{M_\odot}}{(R/{R_\odot})^3}}, \ \ \ 
\frac{\num}{\num{_\odot}} \simeq  \frac{M/M_\odot}{(R/{R_\odot})^2\sqrt{\Teff/\Teff{_\odot}}}.
\end{equation}

For each star, $\dnu$ and $\num$ together with \teff\ and $\feh$ values (Section \ref{sec:specmod}) are used as inputs to the BeSPP code (Serenelli et al. 2013), a Bayesian grid-based modelling algorithm for determination of fundamental stellar properties (mass, age, evolutionary stage). BeSPP uses a large grid of Garstec stellar evolutionary tracks ($6\times10^7$ stellar models, $0.6 \leq M/~M_{\odot} \leq 3.0$), from the pre-main sequence to the beginning of the thermal pulses on the asymptotic giant branch \citep{2008Ap&SS.316...99W}. The posterior probability of each model is constructed as the product of the likelihood of the observables given the model and a set of prior probabilities related to the initial mass function, star formation rate and age-metallicity relation \citep[see][ for details]{2013MNRAS.429.3645S}. The evolutionary state of stars, when known as it is the case for some stars in Thygesen's sample, is also added as a binary prior. The full probability distribution function is then marginalised to obtain the posterior probability distributions for stellar mass and age. For the VMP stars, Rogue and Hennes, we use the best available asteroseismic masses (Silva Aguirre et al. 2016, in preparation), where a new version of the scaling relations by Guggenberger et al. (in preparation) was employed. We use their results, because the validity of standard scaling relations at low metallicity is currently under debate (Epstein et al. 2014).
%
%
%
\subsection{State-of-the-art in modelling \ha~ spectra of cool stars}\label{sec:failures}
It is well-known that strong spectral lines, including the \ha, Mg II UV doublet,  and Ca II near-IR triplet cannot be fit in the spectra of cool stars with the classical models of stellar photospheres, i.e. the models based on the standard assumptions of 1D hydrostatic equilibrium and LTE \citep[][and references therein]{2008ASPC..397...54R, 2012A&A...540A..86R}.

\begin{figure}[!ht]
\includegraphics[width=0.6\columnwidth, angle=-90]{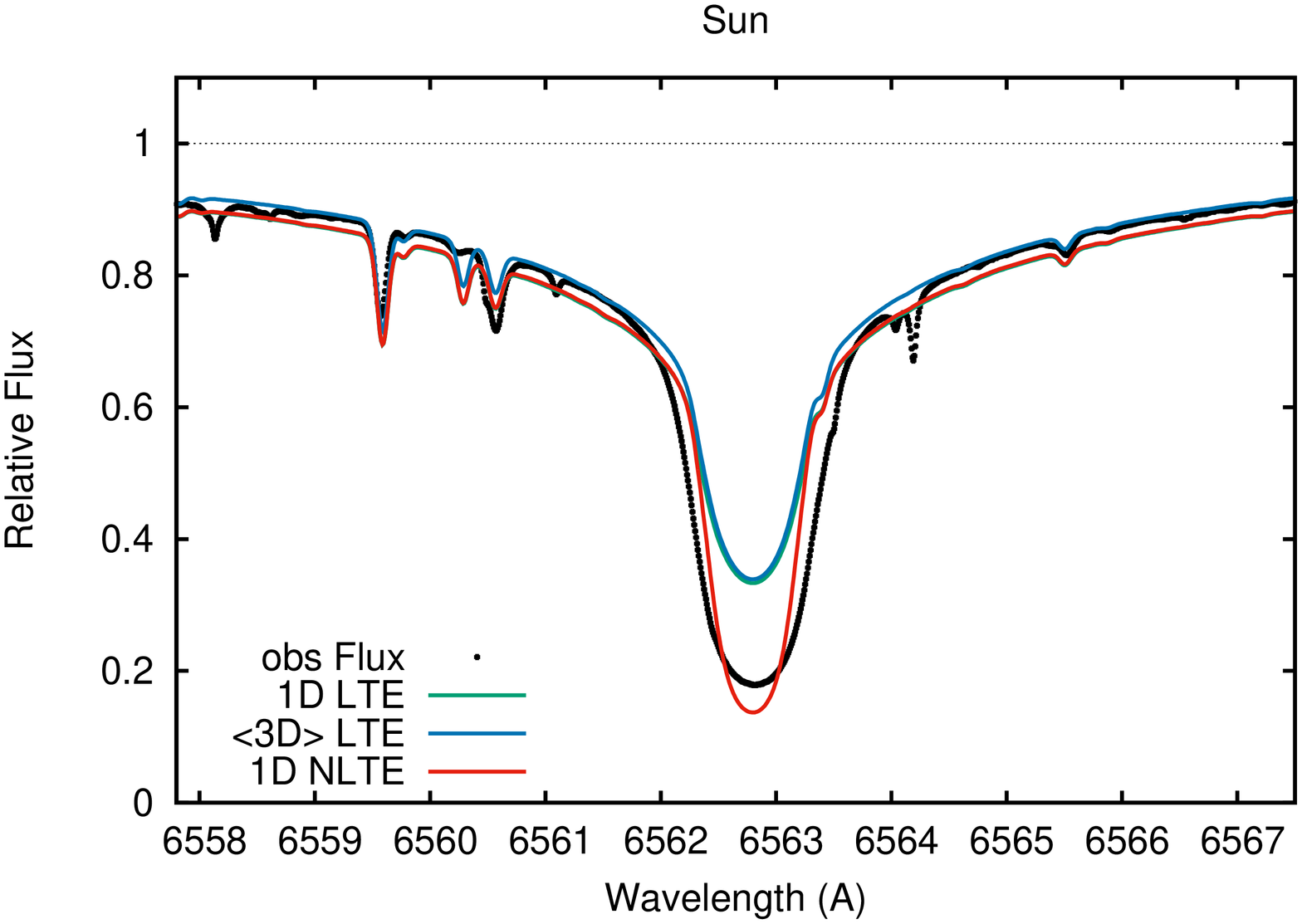}
\includegraphics[width=0.6\columnwidth, angle=-90]{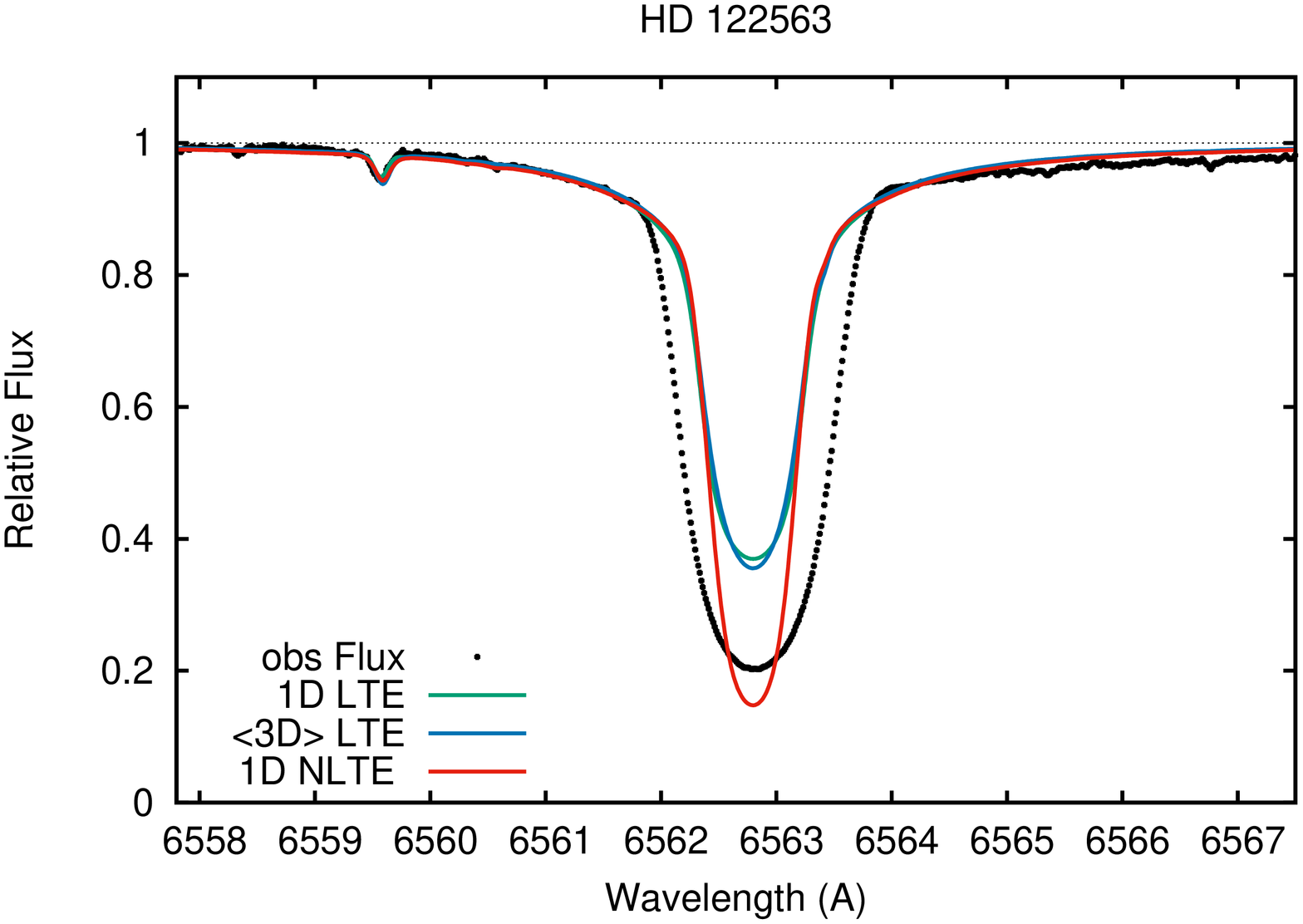}
\caption{Model line profiles (colour lines) computed with different physics (1D LTE, mean 3D LTE, 1D NLTE) in comparison with the observed spectrum of the Sun (dots, top) and the metal-poor red giant HD 122563 (dots, bottom). The models lie far away from the data, thus we resort to an empirical method (Sect. 2.5).}
\label{fig:profile_cal}
\end{figure}

To illustrate the problem, we show the observed and synthetic line profiles for the two reference stars, the Sun (KPNO FTS atlas) and HD 122563 (the UVES POP spectrum, Bagnulo et al. 2003) in Fig. \ref{fig:profile_cal}. The model spectra were computed in LTE using the broadening theory of Barklem et al. (2000) using the time- and spatially-averaged 3D hydrodynamical models from the \textsc{Stagger} grid (see Collet et al. 2011, Bergemann et al. 2012, Magic et al. 2013a,b) and the MARCS models (Gustafsson et al. 2008). In both cases, the models fail to describe the observed line shape: the observed profile is stronger than the best-fit model. The plot also shows non-local thermodynamic equilibrium (NLTE) line profiles computed using the H model atom from \citet{mashonkina2008}. While NLTE radiation transport does improve the fit in the \ha~core, the model line is still too narrow compared to the observed data.

\begin{figure*}[!ht]
\hbox{
\includegraphics[width=0.63\columnwidth, angle=0]{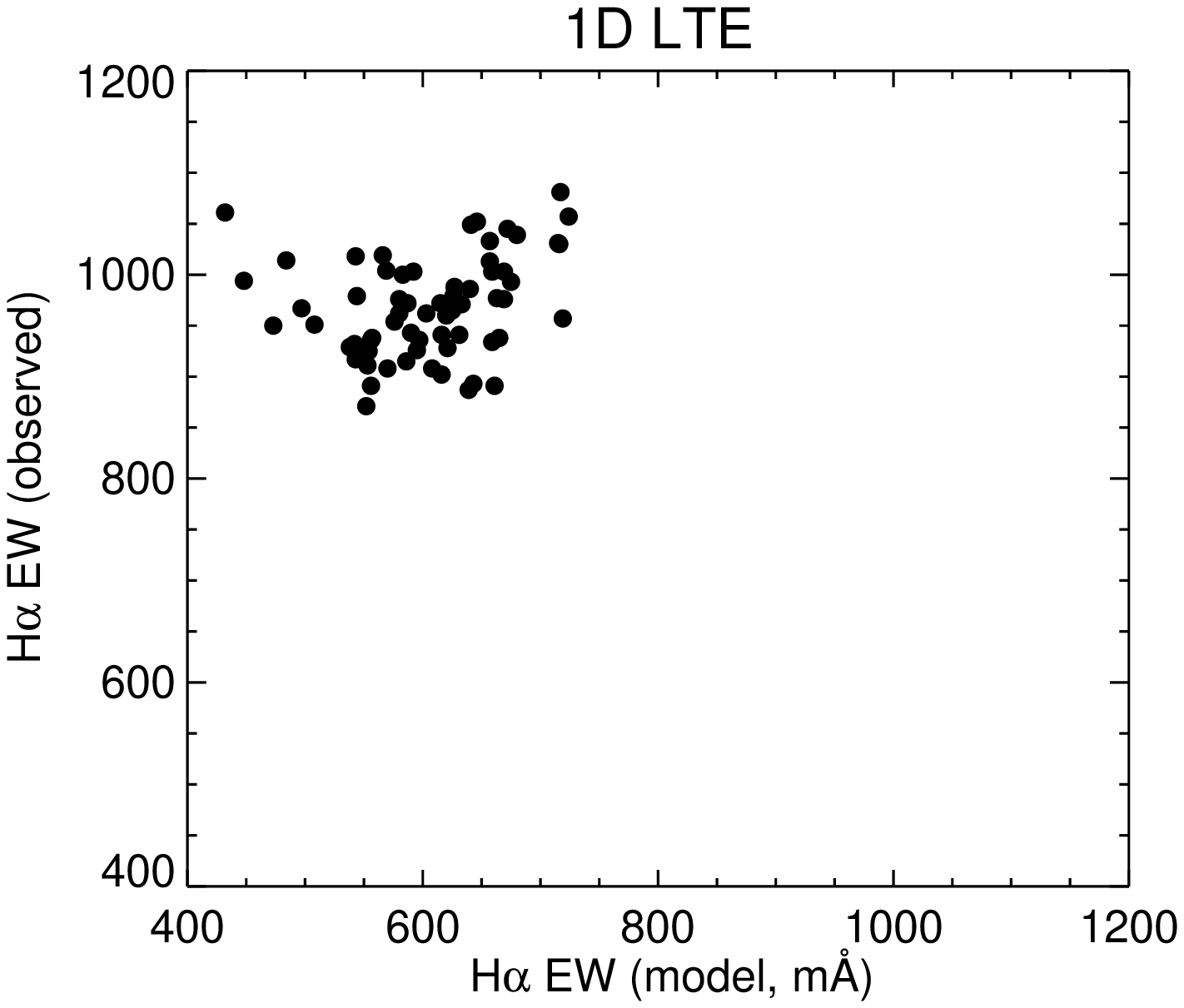}
\includegraphics[width=0.63\columnwidth, angle=0]{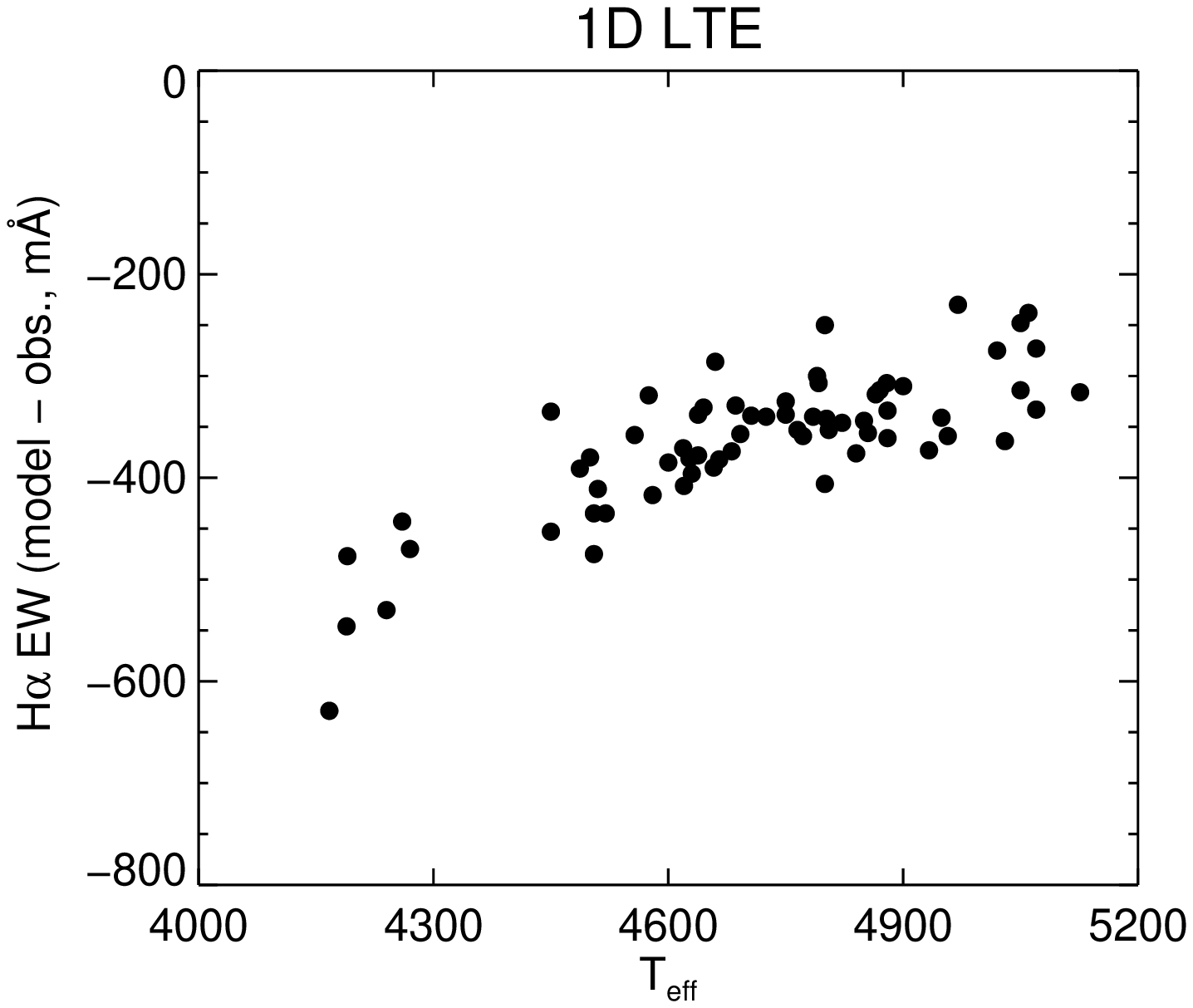}
\includegraphics[width=0.63\columnwidth, angle=0]{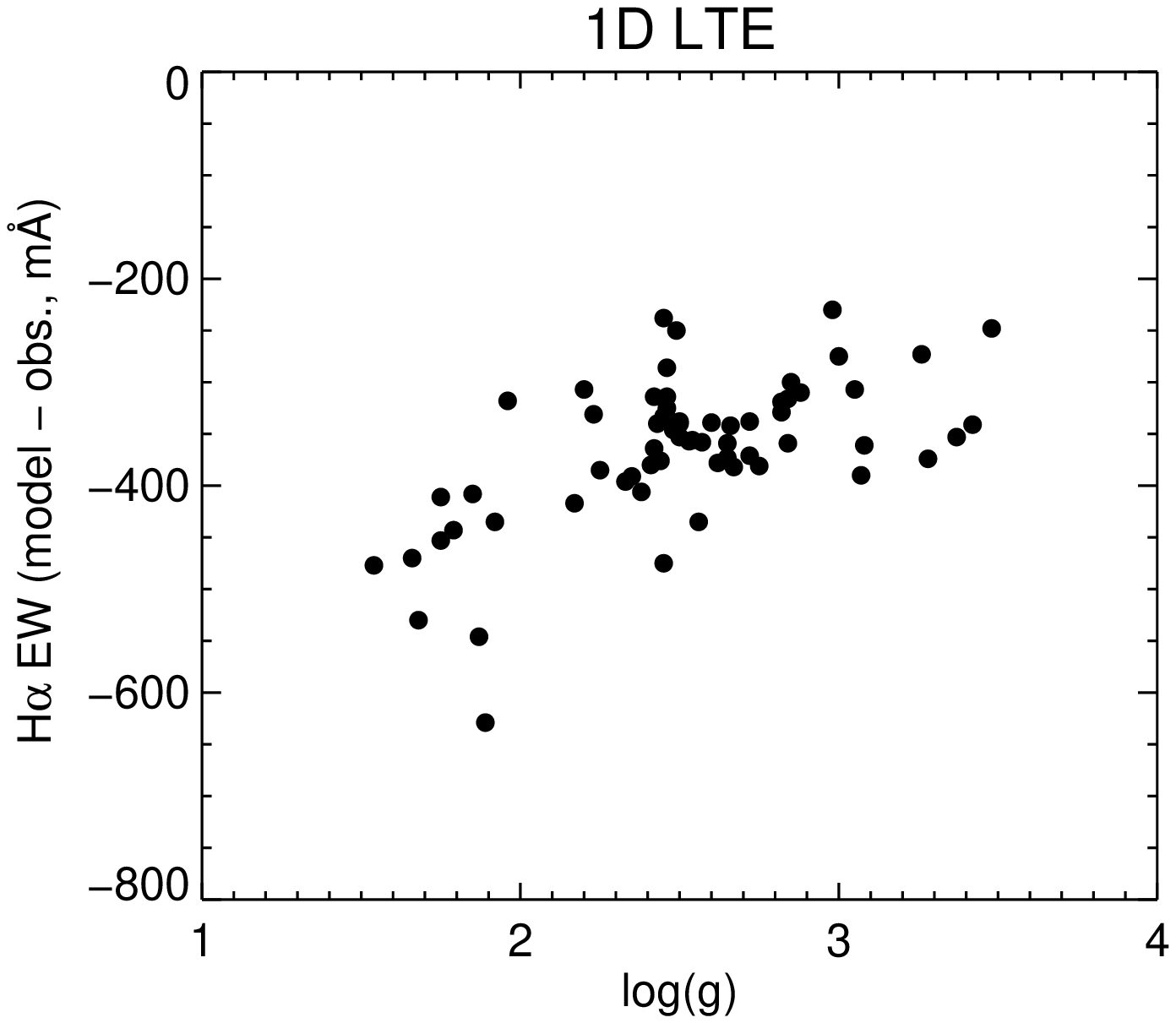}}
\hbox{
\includegraphics[width=0.63\columnwidth, angle=0]{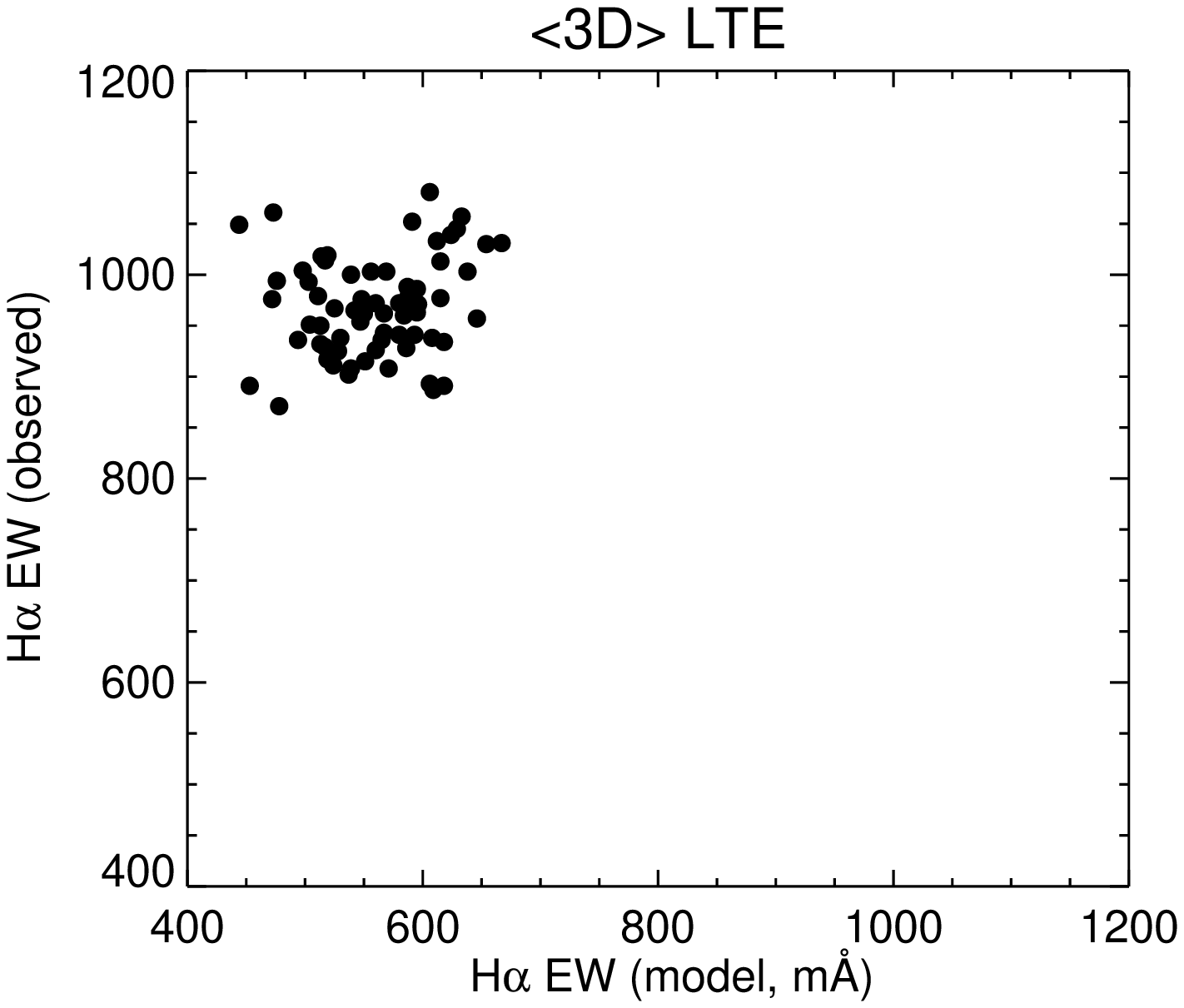}
\includegraphics[width=0.63\columnwidth, angle=0]{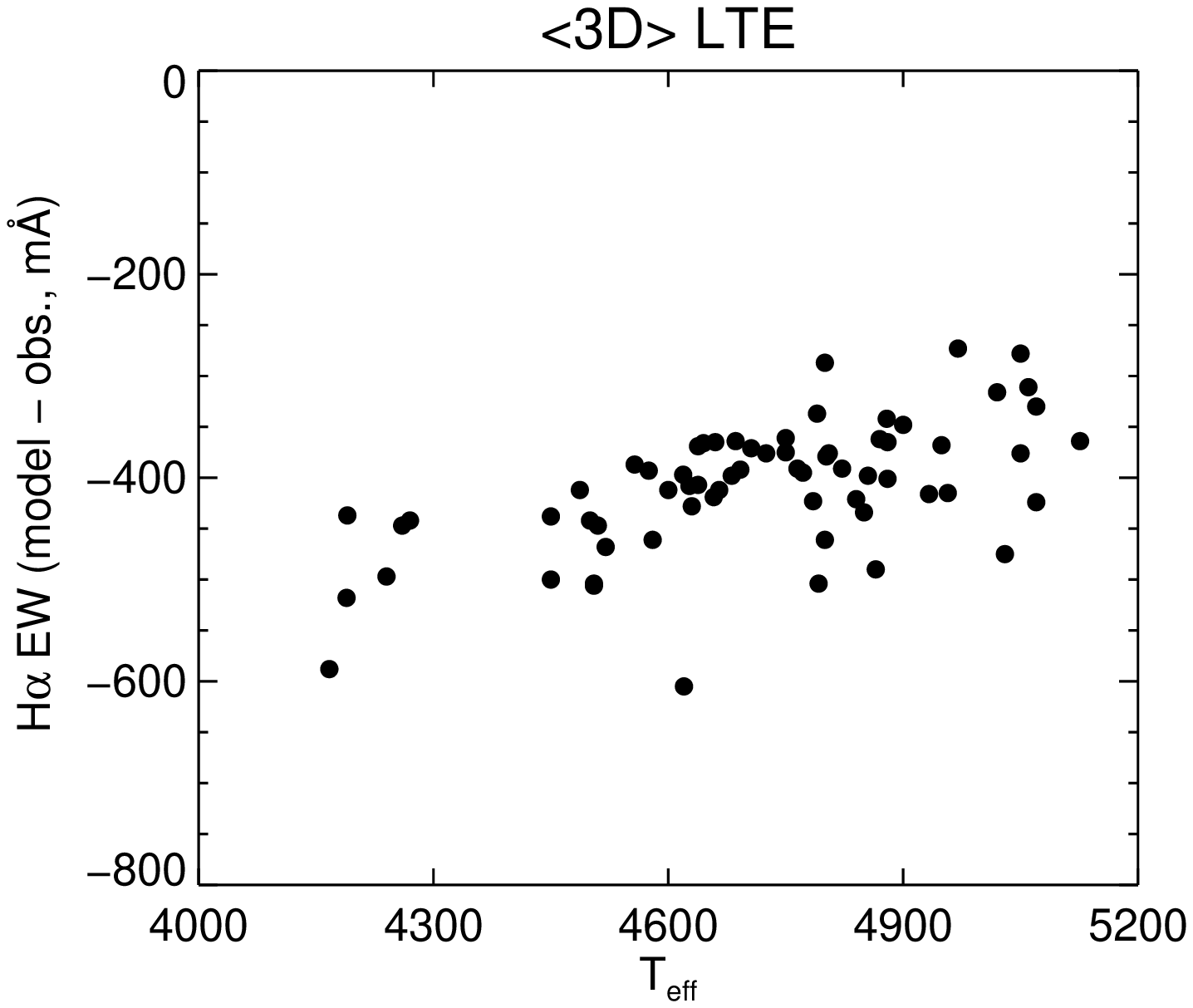}
\includegraphics[width=0.63\columnwidth, angle=0]{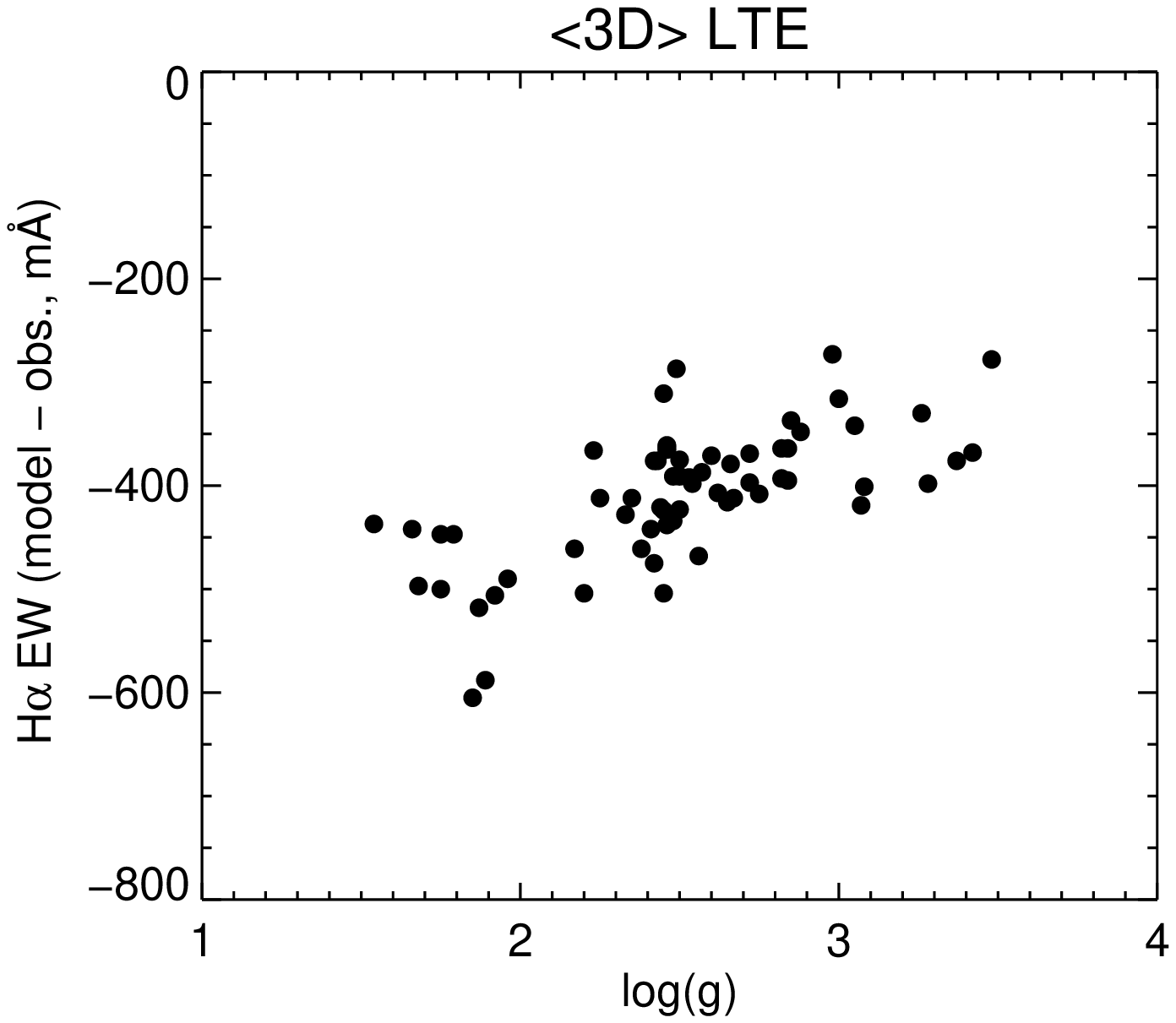}}
\caption{Comparison of the observed and synthetic equivalent widths of the \ha~ line for the reference asteroseismic stellar sample. See Sect 2.4 for more detail.}
\label{fig:ews}
\end{figure*}

Remarkably, every RGB and RC star in the sample suffers from a similar systematic offset. Figure \ref{fig:ews} compares the observed equivalent widths (EWs) of the \ha~line for our stellar sample with asteroseismic masses (Section 2.1) with the best-fit 1D and average 3D model atmospheres. Comparison with the observed EWs shows a much greater than expected spread of the line strengths: the differences amount to almost $\sim 700$ m\AA~ for the coolest stars, i.e. more than a $50$\% error. The error shows a striking correlation with the surface stellar parameters, the effective temperature and surface gravity. Taking into account the mean 3D structures (Fig.~\ref{fig:ews}, bottom panel) does not improve the agreement with the \ha~observations suggesting that some other physical process is needed to explain the data.

While the systematic discrepancy has been known for decades, the methodological strategies to improve the models have been quite fragmented so far. Dupree et al. (1984) were among the first to suggest that adding chromospheres and mass flows to the 1D static model photospheres in the framework of "extended atmosphere models" may work for cool stars. Such semi-empirical models assume that the chromospheres are excited by magnetic and mechanical disturbances propagating as electromagnetic-hydrodynamical waves (Alfv\'en 1942, Babcock 1961, Hartmann \& McGregor 1980). The deposition of energy and momentum due to shock waves defines chromospheric energy balance, thus, at the very least, incident wave energy flux and the magnetic field are needed to compute the model. The ability of such models to fit the observed stellar spectra, however, critically depends on the availability of additional information. The observed \ha~line profiles in the stellar spectra are used to iteratively constrain the free parameters, including the temperature (hereafter, $T$)-depth slope, the location of the $T$ minimum, and the velocity field \citep{meszaros09,dupree16}. All these parameters can not yet be calculated from first principles: in particular, the choice of the profile of turbulent velocity with depth is essential to fit the \ha~width \citep{dupree16}. Moreover, one essential ingredient in the chromospheric models is NLTE radiation transport, since under the LTE assumption the line source function couples to the outwardly increasing temperature and causes an unphysical emission in the line cores \citep{przybilla04}.

\citet{2012A&A...540A..86R} explored the formation of the solar \ha~line using the 1D solar plane-parallel Kurucz model atmosphere and the semi-empirical solar chromospheric model by Fontenla et al. (2009). Their results (see their Fig. 8) suggest that NLTE radiation transport and chromospheric back-radiation are two necessary, albeit not sufficient, ingredients to describe the \ha~profile. Indeed, while the Kurucz \citep{1979ApJS...40....1K} model predicts too narrow \ha~line cores, in the model with an overlying (but \textit{empirically constrained}\footnote{Based on the SUMER and UVSP data.}) chromosphere the \ha~core forms much higher in the atmosphere and is more opaque. On the other hand, their best model also produces too little opacity in the \ha~core to fit the observed solar spectrum, similar to what is shown in Fig. \ref{fig:profile_cal} (top panel), although the NLTE line centre intensity (at 6562.8 \AA) is closer to the observational data.
A similar approach has been explored by \citet{przybilla04}, who showed that the irradiation of the inner photospheric layers by the chromosphere does little for the \ha~profile in the solar spectrum (their Fig.2, top panel). Remarkably, four different solar model atmospheres, with and without chromospheres, produce essentially \textit{identical} NLTE \ha~profiles, challenging the argument that chromospheres are essential in describing the \ha~line shapes. Thus, apart from the conceptual difficulty with computing chromospheres from first principles, it is still unclear what effect do chromospheres have on the stellar \ha~lines. Disconcerting as it may be, as such, the 1D static or dynamical models with parametrised chromospheres do not yet offer a suitable framework for the quantitative analysis of \ha~profiles in stellar spectra.

More advanced 3D radiation magneto-hydrodynamic (MHD) simulations of chromospheres have been developed only recently (Carlsson \& Stein 2002, Hansteen 2004, Gudiksen et al. 2010, Leenaarts 2010, Hansteen et al. 2015). The models are clearly more physically realistic than the 1D or 2D implementations, however, at present their applicability to other stars than the Sun remains questionable, since the simulations are, too, parametrised and require making use of spatially-resolved stellar observations, such as those provided by the MDI instrument for the Sun (Leenaarts et al. 2012)\footnote{In the MHD simulations of the solar chromosphere, the velocity fields that generate acoustic waves at the bottom of the MHD solar simulation are set to reproduce the velocity profiles deduced from the Ni line-core imaging, the type of observations presently unavailable for any other star.}. Also, there are still some open issues that pose a challenge to the simulations (Leenaarts 2010). First, the MHD models combined with the Hinode/SOT Ca II observations of the Sun predict that the power in acoustic waves\footnote{In this discussion, the relevant waves are those with frequencies above 20 mHz and 50 mHz, above the limit most of the acoustic energy produced in the convective zone is strongly damped in the photosphere (Carlsson \& Stein 2002).} is far too small to explain the chromospheric radiative energy losses (Carlsson et al. 2007). Second, fitting the \ha~core with the MHD simulations has traditionally been problematic, because of the missing \ha~opacity in the chromosphere. Leenaarts et al. (2012, 2015) showed that including non-equilibrium ionisation of hydrogen in the calculations of the equation-of-state in the 3D MHD models could help to resolve the long-standing disagreement with the \ha~line-core imaging observations of the Sun. The simulations show that the opacity in the outer \ha~core is correlated with the gas temperature, while the opacity in the inner core is mostly sensitive to the gas density. However it remains yet to be investigated whether the MHD models with prescribed radiative cooling rates (from the 1D chromospheric simulations that is only computationally tractable presently) describe the observed solar \ha~line profile shape, not only the monochromatic brightness contrast across small areas on the solar surface. Finally, the 3D MHD chromospheric simulations for stars other than the Sun are currently not available, therefore neither quantitative nor qualitative statements about the behaviour of \ha~lines in other stars, especially those with very different interior properties, such as RGB or RC, can be made.
%
%
\subsection{Empirical fits to the \ha~ profile}\label{sec:fits}
As discussed in the previous section, it is presently not possible to construct a grid of stellar atmosphere models that include all the sophisticated physical processes needed to describe the hydrogen line profiles in the observed spectra of RGB and RC stars. On the other hand, the obvious systematic misfit in the \ha~line (Figs. \ref{fig:profile_cal}, \ref{fig:ews}) is puzzling and it has prompted us to take a different, empirically motivated, approach to the problem. We ask whether there is a mathematical function that fits the \ha~line profiles in the spectra of all our sample stars. 

Since the red wing of the \ha~ line is blended by a metallic feature (the line of neutral cobalt), we focus on the unblended left side of the core of the line, as shown by arrows in Fig. \ref{fig:profile}. The line mask is the same for all stars, it covers the wavelength range $6562.0 - 6562.8$ \AA. By selecting this mask, we ensure that the Co I blend has no effect on the fit. The left limit at 6562.0 \AA\ is set to minimise the contribution of the Si I and Fe I blends. None of the stars in the sample show evidence of emission in the wings or significant core shifts that could introduce spurious artefacts in the fit. Such features are usually seen in the spectra of metal-poor, $\feh < -1$, red giants brighter than $\log L/ L_{\odot} \sim 2.5$ \citep{2004A&A...413..343C,meszaros09}, while all but two of our reference stars with asteroseismic masses are fainter than this luminosity threshold (Fig. \ref{hrd}). We also do not see signatures of emission or line core shift in the most metal-poor stars in our sample, e.g. KIC 7693833, KIC 8017159 (Fig. \ref{fig:profile}).

We have explored different functions and find that the observed blue wing of the \ha~line is well described by a cubic exponential function:
\begin{equation}
f (\lambda) = 1 - f_0 \cdot \exp( -\left(|\lambda - \lambda_0|/\wh \right)^3)
\end{equation}
where $\lambda_0$ is the central wavelength of the line, $6562.819$ \AA\ \citep{baker} and is fixed, $f_0$ the minimum flux in the line core, and $\wh$ - a free parameter which correlates with the width of the spectral line and, as it will be shown later, with the mass of a star. It is important to stress that this is a purely mathematical function that has no input from the stellar atmosphere models or instrumental effects, such as the spectrograph line spread function. For our spectra, the resolving power is so high that the latter effects do not matter. However, for low-resolution spectra, values of $\wh$ derived from different observational set-ups can not be compared directly. The free parameters to be fit to an observed profile are the line depth parameter, $f_0$, and the line width parameter, $\wh$; their measured values for the full stellar sample are given in Tables \ref{table2},\ref{table4},\ref{table6}. The uncertainty of the fitting is determined by shifting the continuum by $\pm 2 \sigma$ from the best-fit solution and repeating the function fit. The difference with the best-fit normalisation then gives the uncertainty in the $\wh$ estimate, which for all stars in the sample is below $0.01$ \AA. It is thus reasonable to adopt the same error of $\wh$ for all stars.

Figure \ref{fig:profile} compares the function fit from the Eq. 2 with the observed \ha~line profiles in several program stars. The theoretical line profiles computed in 1D LTE and NLTE are also shown.

\begin{figure}
\begin{center}
\includegraphics[width=0.46\textwidth, angle=0]{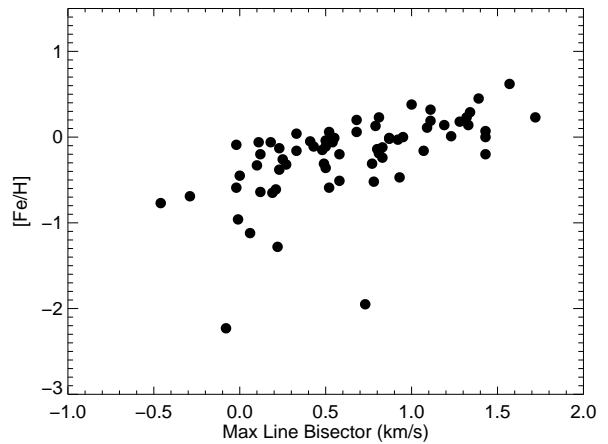}
\caption{Bisector velocity at the FWHM of the \ha~profile (km$/$s) as a function of metallicity $\feh$ for the observed stellar sample.}
\label{fig:bisec}
\end{center}
\end{figure}

\begin{figure*}
\begin{center}
\hbox{
\includegraphics[width=0.46\textwidth, angle=0]{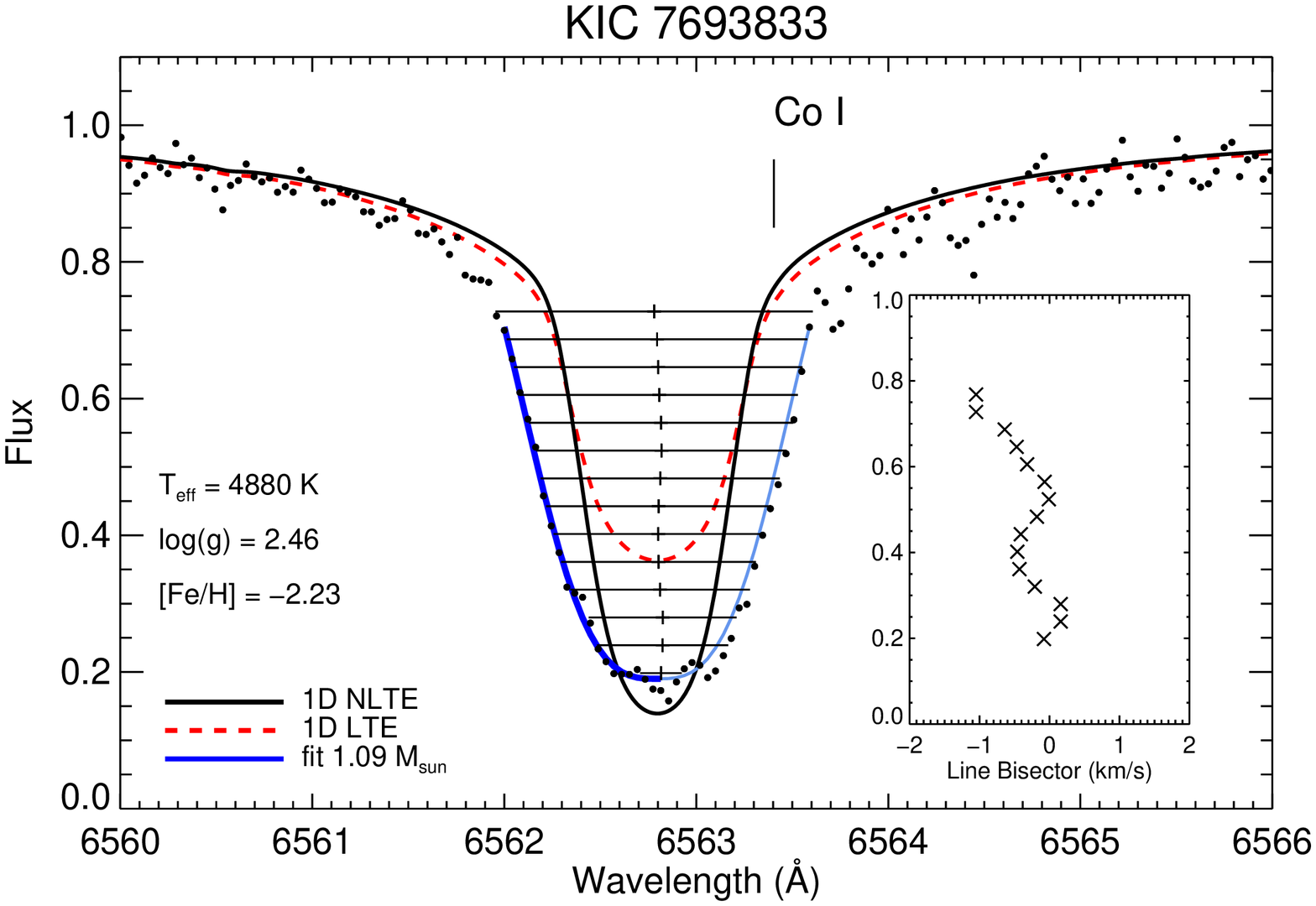} 
\includegraphics[width=0.46\textwidth, angle=0]{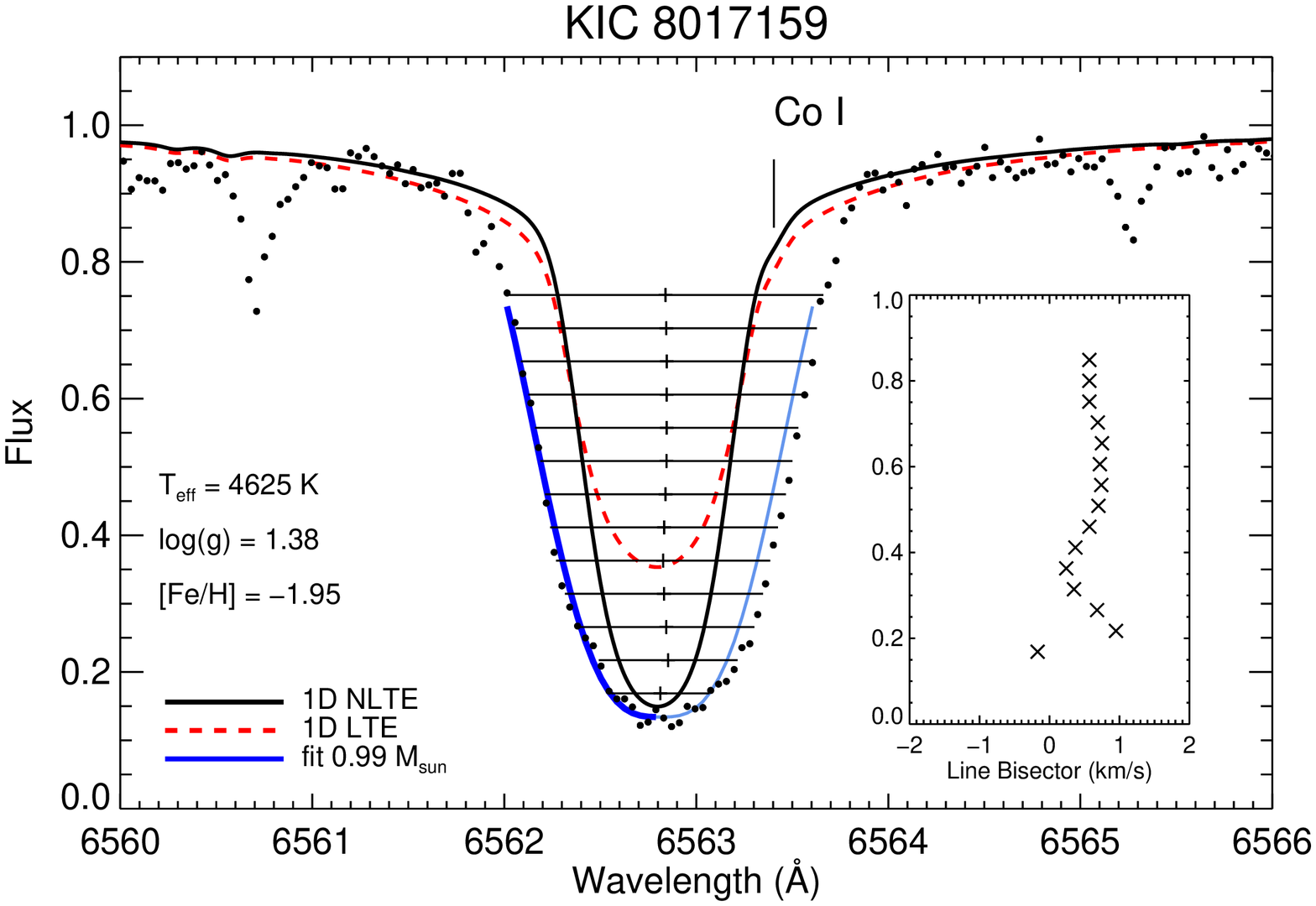} 
}
\hbox{
\includegraphics[width=0.46\textwidth, angle=0]{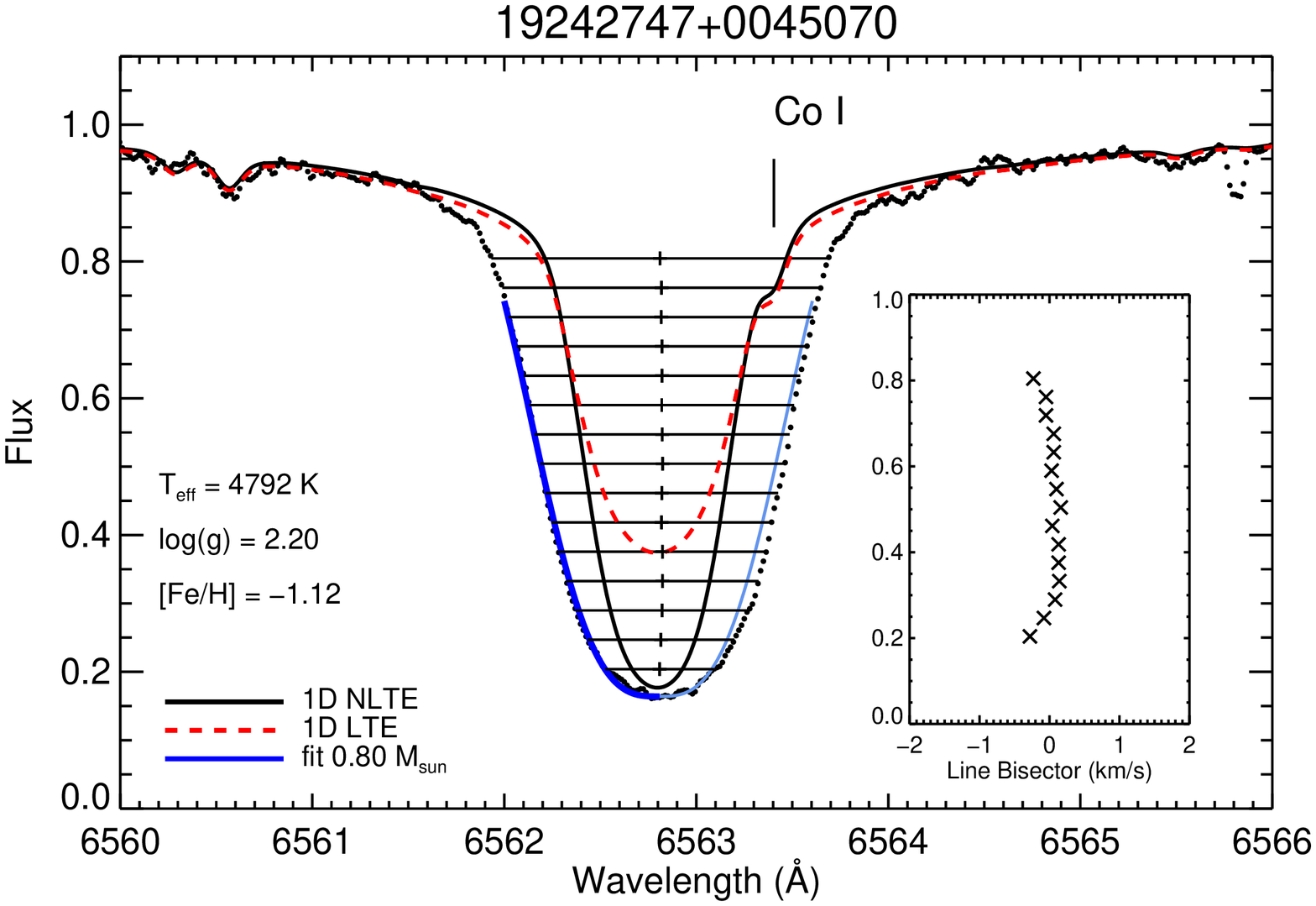}  
\includegraphics[width=0.46\textwidth, angle=0]{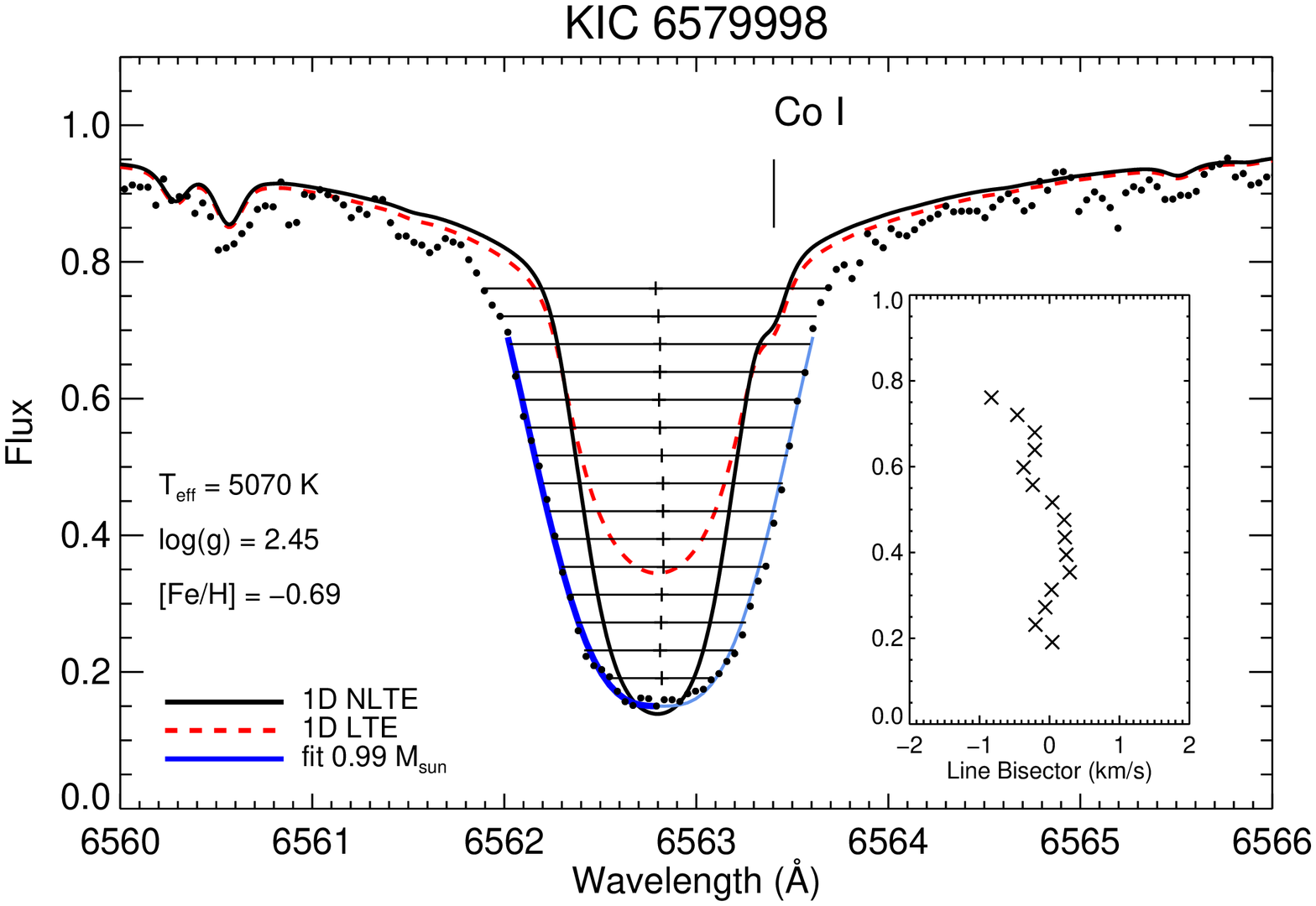} 
}
\hbox{
\includegraphics[width=0.46\textwidth, angle=0]{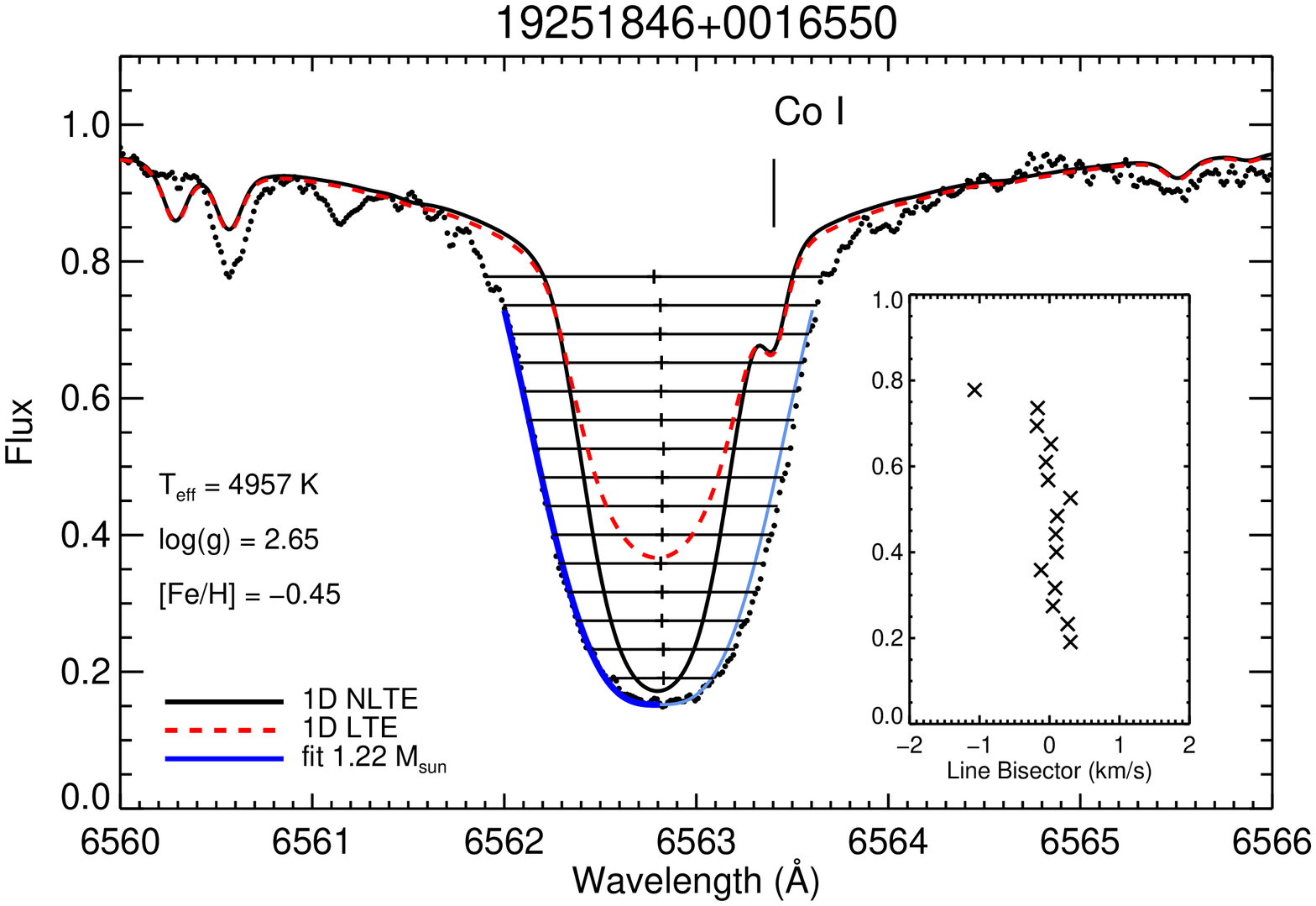}  
\includegraphics[width=0.46\textwidth, angle=0]{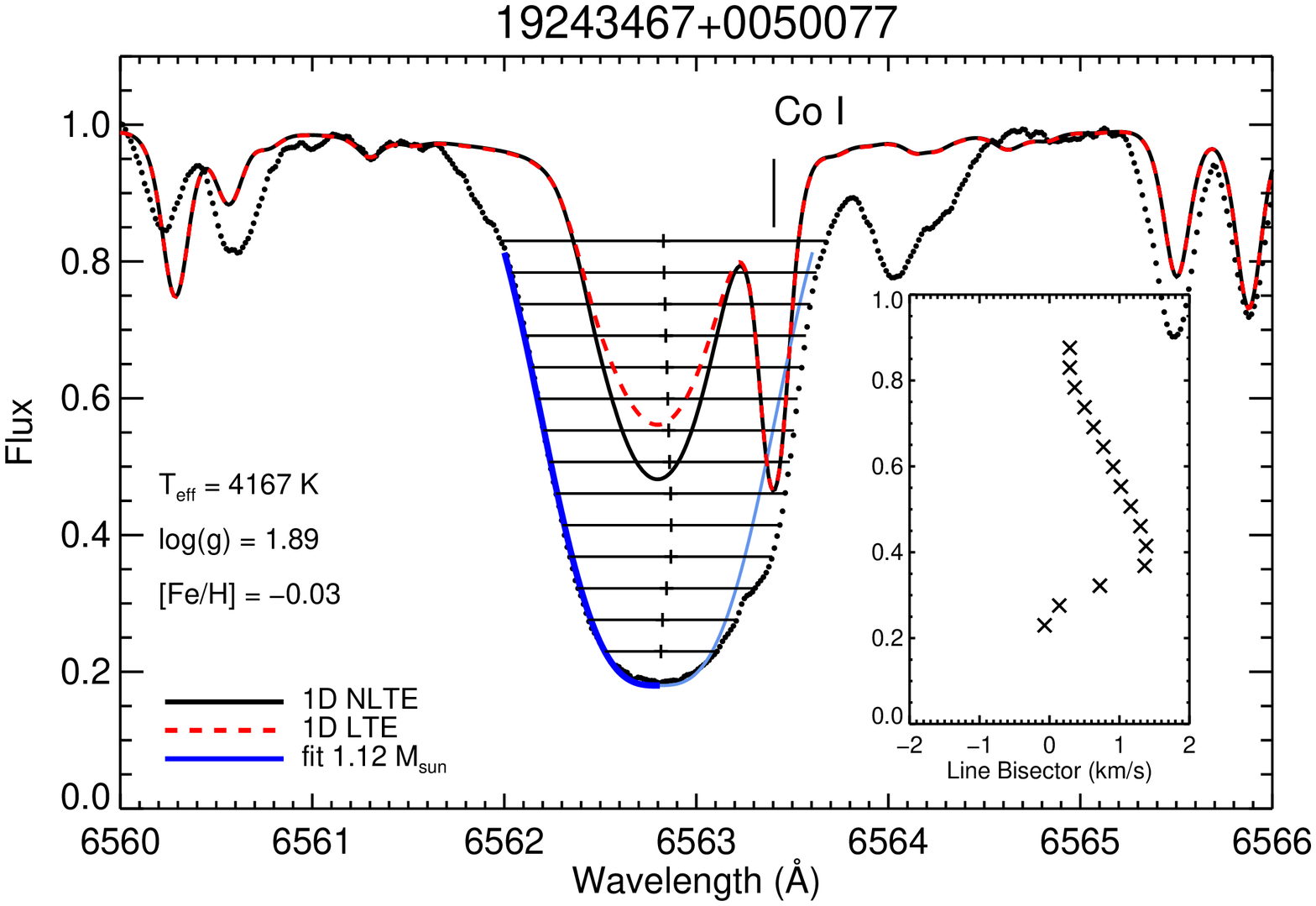}  
}
\hbox{
\includegraphics[width=0.46\textwidth, angle=0]{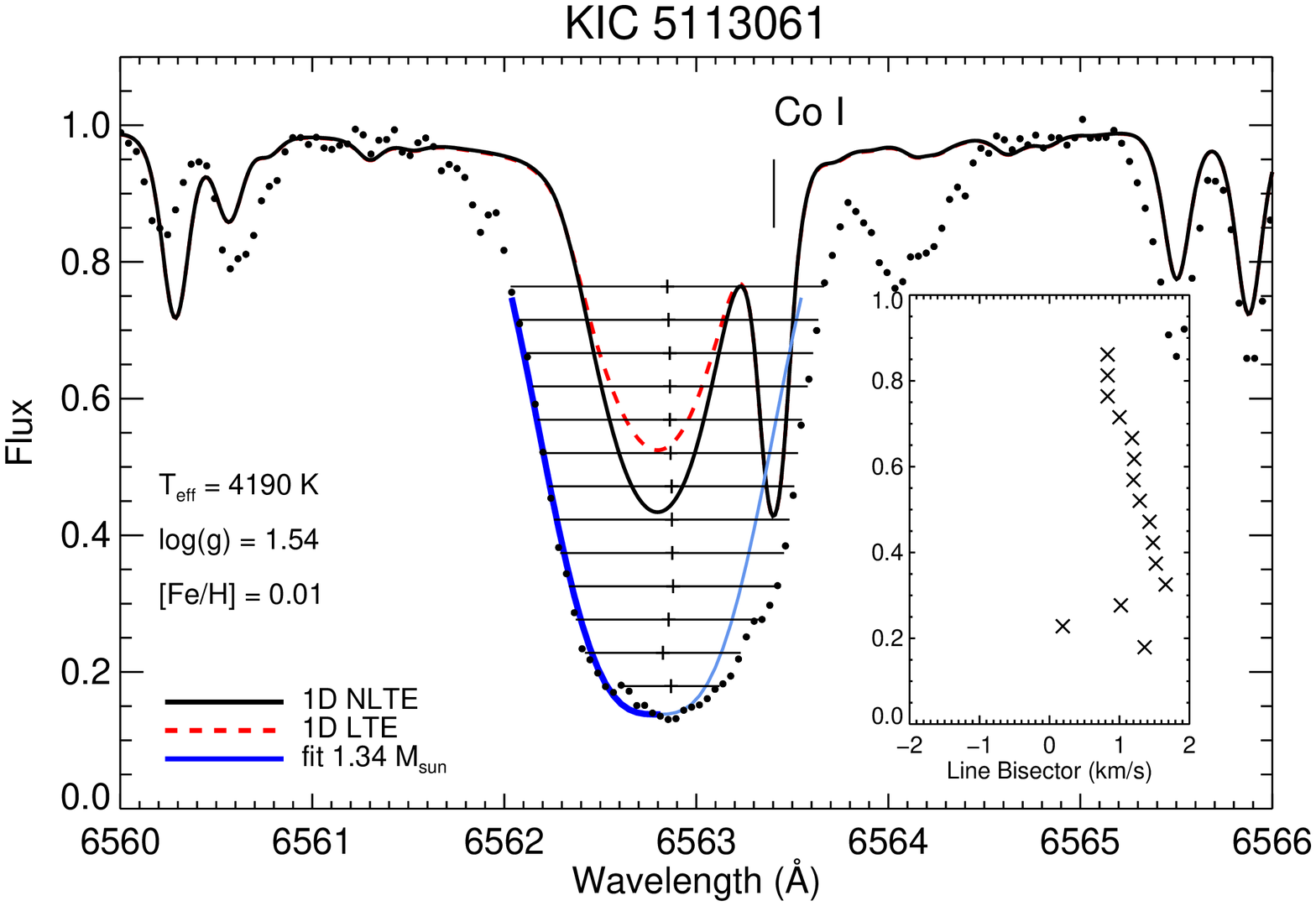} 
\includegraphics[width=0.46\textwidth, angle=0]{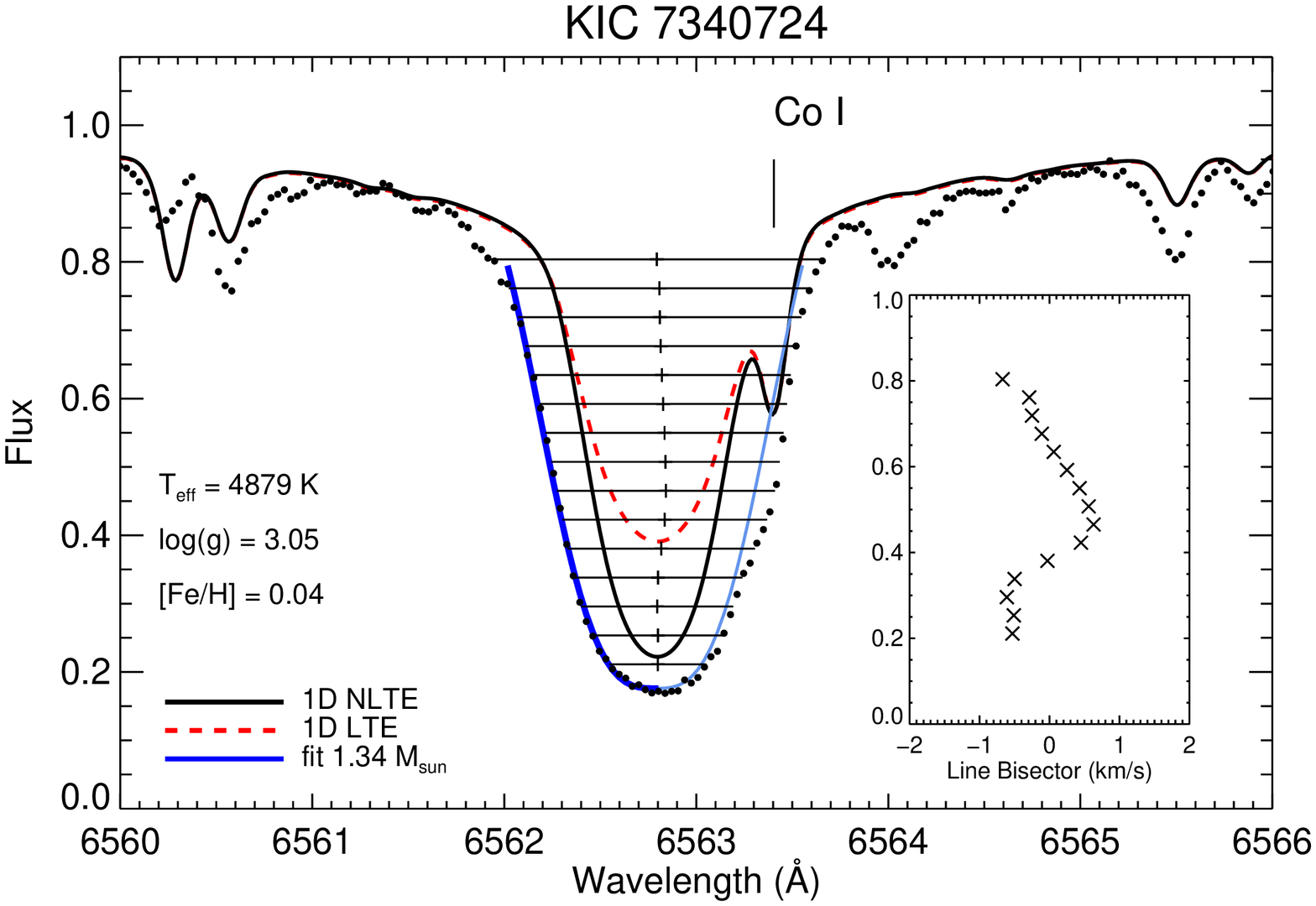} 
}
\caption{The observed \ha~ profile (black dots) in the selected program stars and the best-fit empirical model from the eq. 2 (bold blue line, $6562.0$ to $6562.8$ \AA). The red and black lines correspond to the best-fit theoretical models computed in 1D LTE respectively 1D NLTE. The part of the \ha~line used in the fit is the same for all stars in the sample. The thin blue line is shown to help guide the eye only - the red wing of \ha\ is \textit{not used} in the fit.  Line bisectors are indicated in the inset. The spectra are arranged in order of increasing metallicity.}
\label{fig:profile}
\end{center}
\end{figure*}

Line bisectors, included in the plot insets, help to understand how the observed \ha\ shape changes as a function of stellar parameters. The bisectors were computed by dividing the spectral line into $15$ segments. Then the bisector velocity $v_{\rm bis}$ was estimated at full width at half-maximum of the line profile, as the  wavelength shift relative to the central \ha~wavelength. The Co I blend at 6563.42 \AA, which is very strong in the solar metallicity giants ($\feh > -1$), causes an inverse $c$-shaped bisector profile (Figure \ref{fig:profile}, bottom panels), with $v_{\rm bis}$ up to 2 km$/$s. In the low-metallicity spectra, $\feh < -2$, the behaviour is inverted and the bisector profile has a $c$-shape. The bisector velocity measurements for the full asteroseismic sample are shown in Fig. \ref{fig:bisec}. The line blending explains why the bisector velocity $v_{\rm bis}$ shows a strong correlation with $\feh$. We stress, though, that we fit only the blue part of the \ha~core, so that the Co blend in the red wing has no effect on the $\wh$ measurements.

It should be noted that the \ha\ line develops strong wings in stars hotter than about $4900 - 5000$ K, depending on metallicity; this is the temperature limit where interactions with ions, electrons, and hydrogen atoms become important leading to the specific profile shape with self-broadening, van der Waals, and Stark damping wings. Thus, the simple shape, as described by the cubic exponential function may not apply, and caution is advised when fitting the Eq. 2 to the spectra of lower RGB or subgiant branch stars.
\section{Results}
\subsection{\ha\ and stellar parameters}\label{sec:massage}
\begin{figure*}
\begin{center}
\hbox{
\includegraphics[width=0.33\textwidth, angle=0]{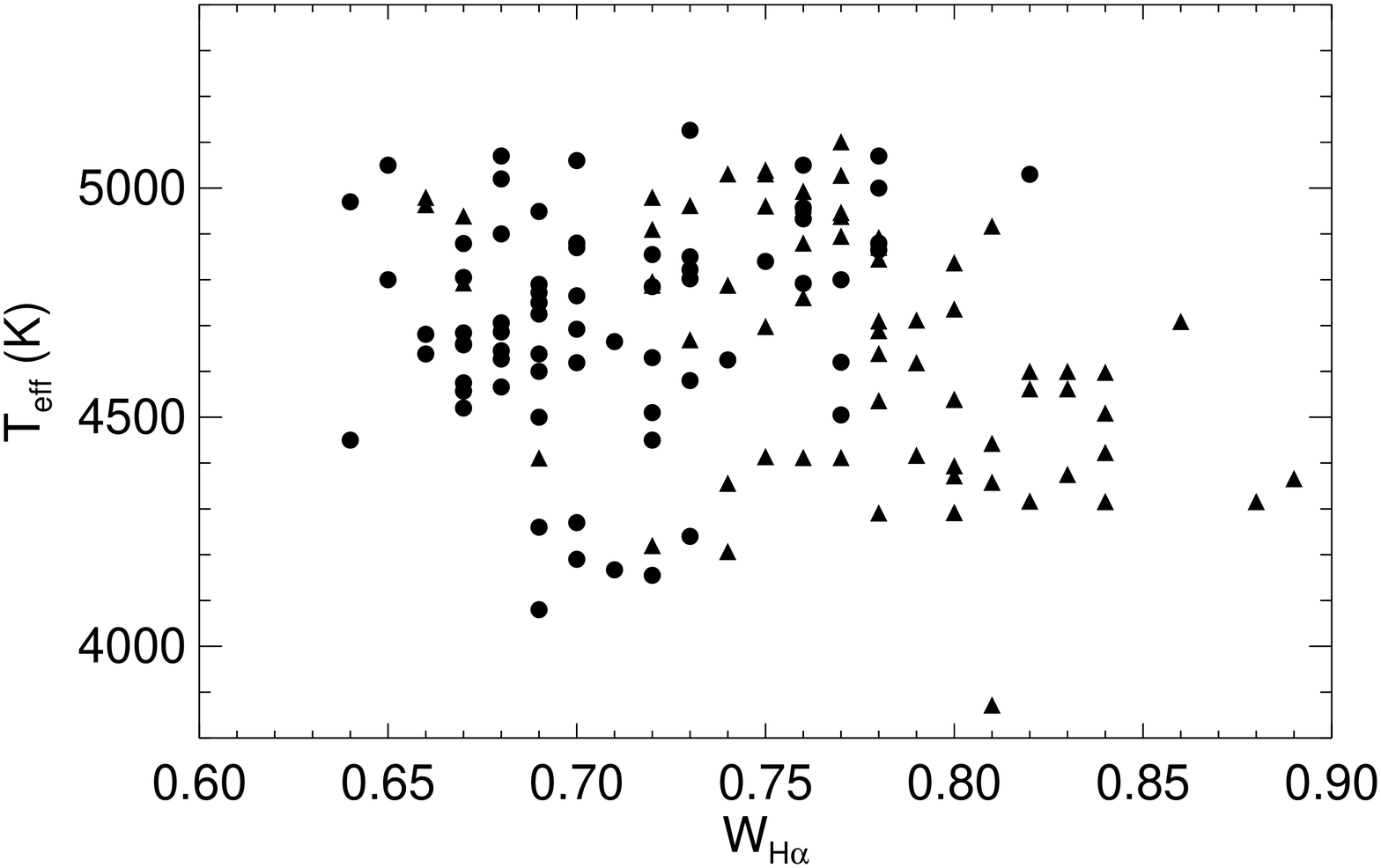} 
\includegraphics[width=0.33\textwidth, angle=0]{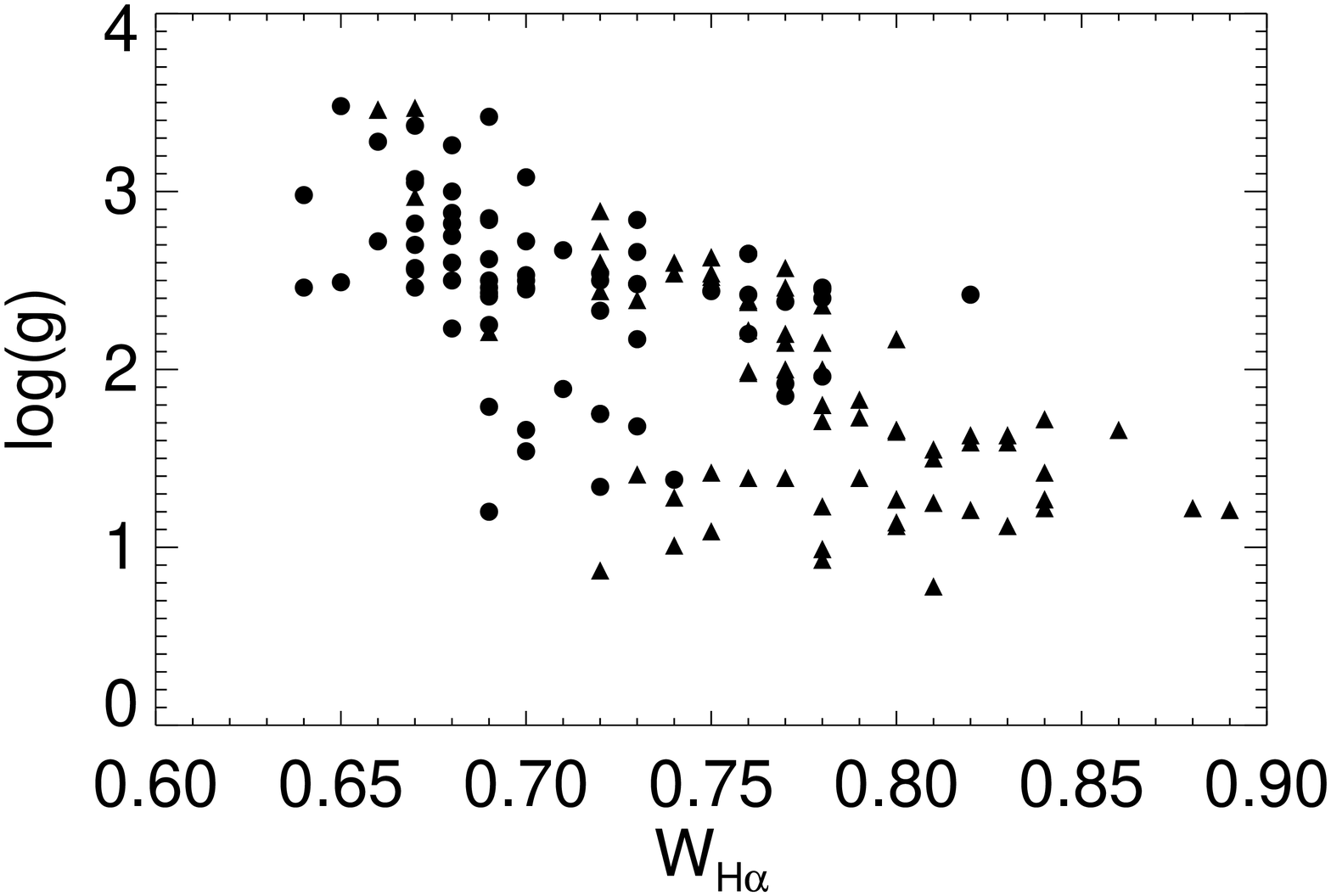}
\includegraphics[width=0.33\textwidth, angle=0]{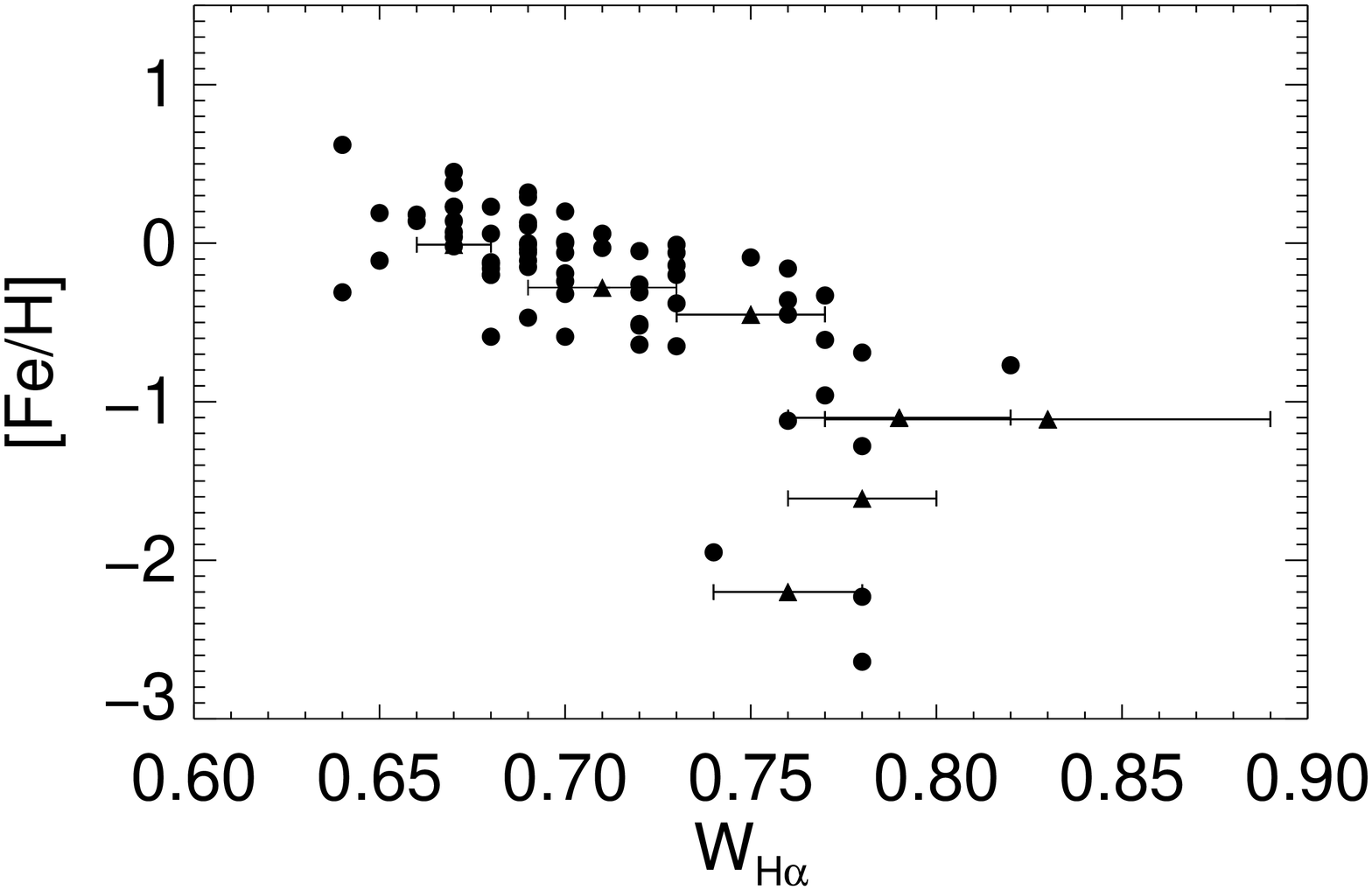}
}
\caption{The \ha~width parameter measurements in the observed stellar sample versus stellar effective temperature, surface gravity, and metallicity. The stars with asteroseismic data are shown with filled circles and the stars in the clusters with filled triangles.}
\label{fig:haw}
\end{center}
\end{figure*}

We have explored the correlation of the measured \ha~line properties with different stellar parameters. The results for our reference asteroseismic sample and for the stars in the clusters are shown in Fig. \ref{fig:haw}. In the $\feh-\wh$ plot, only the mean values of the measured cluster metallicity and $\wh$ are shown, because the observed scatter in $\feh$ (as seen in Table 6) could be caused by the measurement uncertainties\footnote{Some clusters show variations in metallicity \citep[][and references therein]{2016MNRAS.458.2122D}, however this variation is small enough to not impact our conclusions. The standard error of our metallicity measurements is $\sim 0.1$ dex, comparable to the intra-cluster metallicity variations.}.The error bars show $1 \sigma$ standard deviation of the individual $\wh$ measurements within each cluster.

The plots show that there is no clear dependence on \teff, but there are correlations between $\wh$ and $\log g$, as well as between $\wh$ and metallicity. The more metal-rich stars appear to have narrower \ha~lines (smaller $\wh$), as opposed to the more metal-poor stars, which show broader \ha~profiles and, thus, lie at large values for $\wh$. This correlation with metallicity is interesting, because it already looks quite similar to a classic age-metallicity relation (Fig. 13 in the Appendix). Also, given the $\wh$ - metallicity correlation, it is also not surprising to find the correlation of $\wh$ with asteroseismic $\log g$. Surface gravity depends on the star's mass and radius, thus this correlation may indicate there is a more fundamental underlying relationship, such as that with stellar mass.

The stars in NGC 2808 ($\feh = -1.11$) show the largest intra-cluster spread of the $\wh$ estimates that is reflected in the large uncertainty of its mean value. The most metal-poor cluster NGC 4372 ($\feh = -2.20$), too, stands out in the $\wh - \feh$ plot. The only noteworthy difference of the NGC 2808 and NGC 4372 stars with respect to the other clusters is their comparatively high luminosity (Fig. \ref{hrd}): most of the observed stars in these two clusters have $\log L/L_{\odot} \ge 2.5$, which is where one could expect the effect of asymmetries and periodic variations in the \ha~profile to become important  (Cacciari et al. 2004, Meszaros et al. 2009). However, visual inspection of the observed data did not reveal any obvious problems with the spectra. Neither do we have repeated observations of these stars to check the effect of variability. Thus, currently, we do not have a suitable explanation for the behaviour of $\wh$ in the most luminous stars in the globular clusters and set it aside as a problem to be addressed in future work.
\subsection{Mass effects in the Balmer line}\label{sec:massage}

\begin{figure}[!ht]
\includegraphics[width=0.5\textwidth]{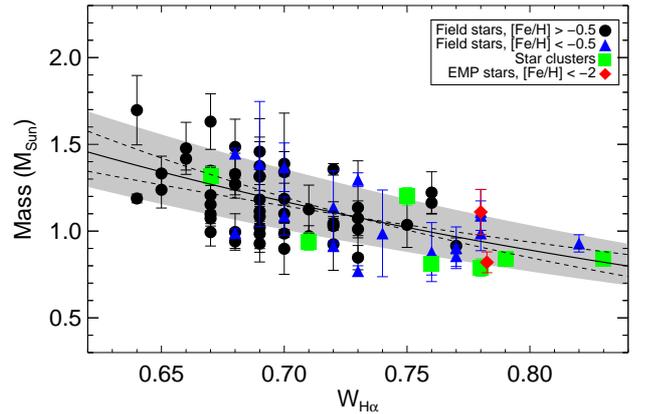}
\caption{The observed relation between the \ha~ spectral line widths versus asteroseismic masses of stars. Star  clusters from the Gaia-ESO survey are shown with green diamonds. The very metal-poor stars in the sample, $\feh < -2$, are shown with red symbols. Dashed black lines show the 1$\sigma$ uncertainty of the fit. The shaded area is the average rms between the data and the fit. See Sect. 3.}
\label{relation1}
\end{figure}

While the $\wh$ vs. $\feh$ and $\log g$ plots already suggested that $\wh$ is related to the mass and hence age of stars, we can directly probe this by plotting the dependence between $\wh$ and the independently determined asteroseismic mass (Fig. \ref{relation1}). This correlation is well-defined and it appears that all stars, also those in star clusters, follow the same $\wh$-mass trend within the errors. \ha~lines tend to be narrower in the spectra of more massive stars, but broader in the spectra of less massive stars. All we now need for a mass estimate from spectroscopy is to fit a relationship between $\wh$ and the independent mass determinations. For the sake of simplicity, we fit a linear relation between \ha~and the logarithmic mass of stars in our sample, i.e. $\log{M}= a \cdot {\wh} + b$. This is done by a $\chi^2$ fit to the bootstrapped data\footnote{We apply bootstrapping to our dataset, using sampling with replacement. The sampling process is repeated n$^2$ (about 2900) times.} in order to avoid that a few stars with small mass uncertainties dominate the results. In addition, we find an almost perfect linear relation between $a$ and $b$. Thus, $b$ can in turn be expressed as a linear function of the slope $a$, with the advantage that the uncertainty of the fit can be expressed in terms of the uncertainty in $a$, determined from the bootstrapping. The final result is:
\begin{equation}
\log{M({\rm H_\alpha})} = a \cdot \left({\wh} - 0.73\right) + 0.08; 
\end{equation}
where $a= -2.73 \pm 0.72$. This is shown in Figure~\ref{relation1} as the black solid line. Dashed black lines show the 1$\sigma$ uncertainty of the fit, as obtained from the bootstrap analysis described above. The shaded area is the average root-mean-square difference between the data and the fit.

Fig.\,\ref{relation2} (a) compares our empirical masses determined from the Eq.\,3 with the reference asteroseismic masses for the program stars. Panel (b) shows the results from the classical "stellar isochrone" approach that makes use of the stellar parameters from spectroscopy (\teff, $\logg$, $\feh$) and isochrones (Serenelli et al. 2013). The masses determined empirically from the \ha~measurements are more precise compared to the classical stellar evolution approach. The plot, however, reveals that we are slightly over-estimating masses at the low-mass end, and under-estimating masses for more massive stars. At present, it is not clear what is the origin of this systematics - the  asteroseismic estimates or some kind of a second parameter problem in \ha. More data will be needed to explore this differences. Taking into account the uncertainty in the $a$ coefficient, and by cross-validation with the reference asteroseismic masses, we conclude that the accuracy of our mass estimates is about $0.15$ M$_\odot$. 
\subsection{Stellar ages}
\begin{figure*}[!ht]
\centering
\makebox[\textwidth]{
\hbox{\includegraphics[width=0.35\textwidth, angle=0]{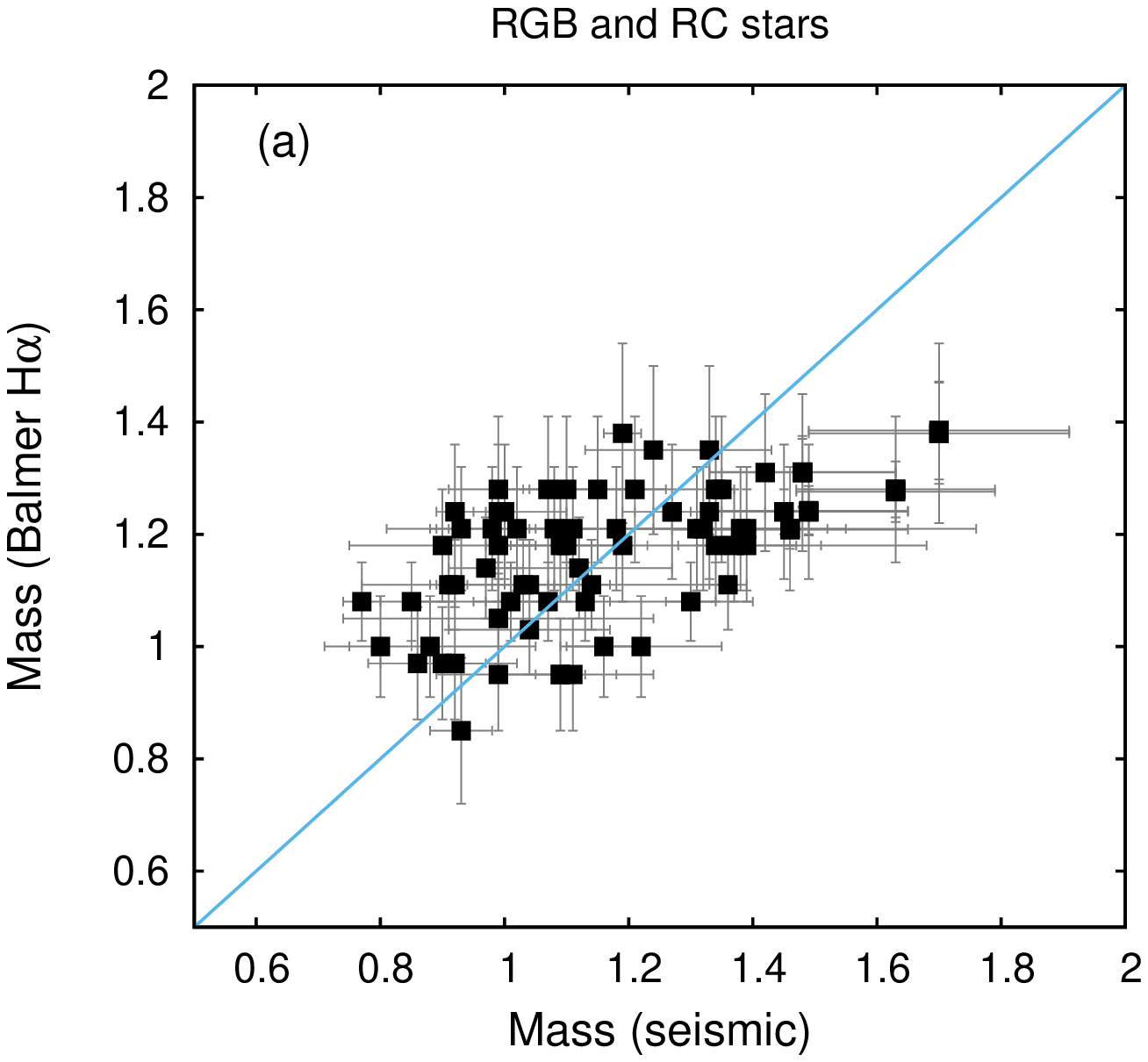}
\includegraphics[width=0.35\textwidth, angle=0]{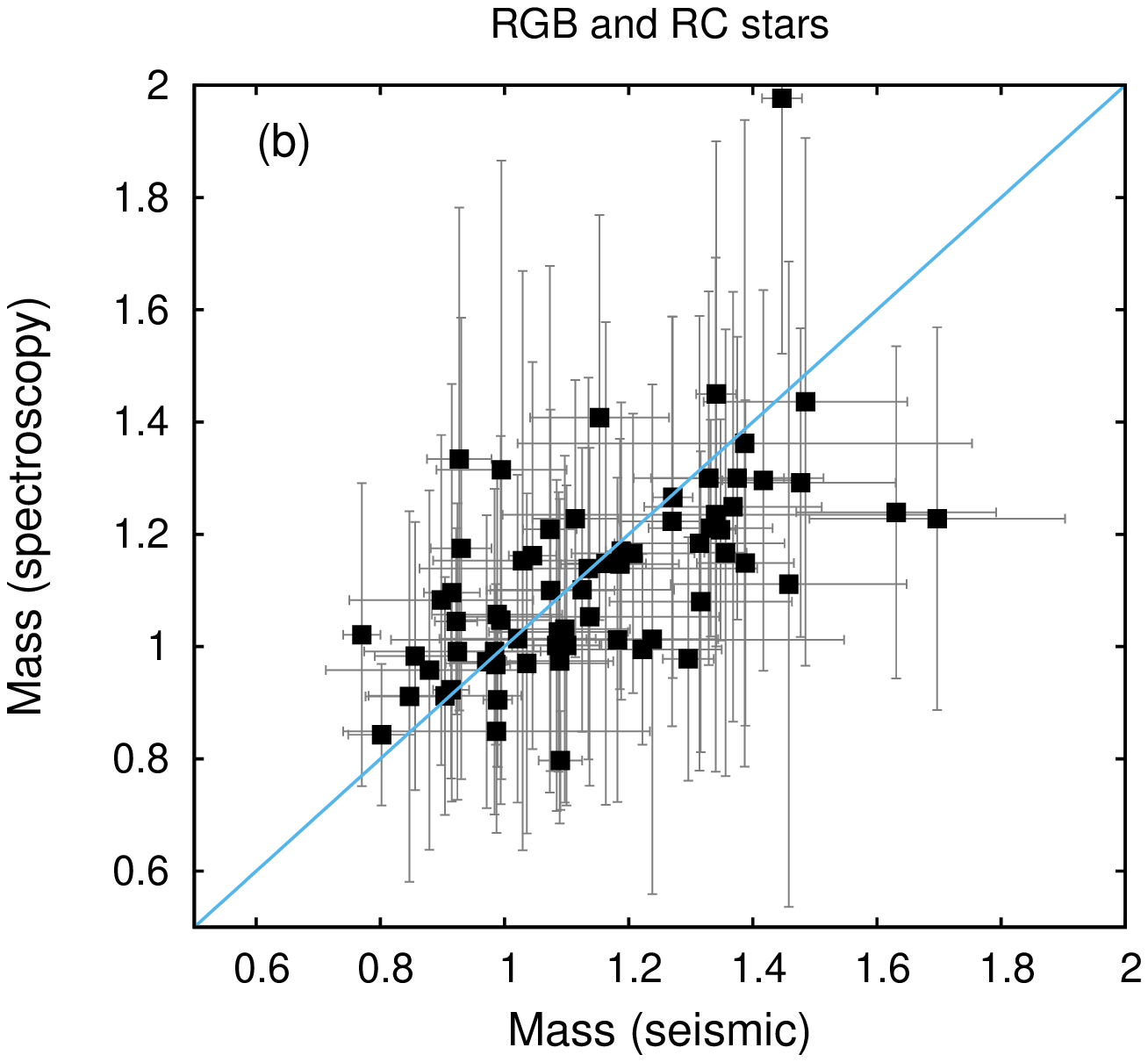}}
}
\centering
\makebox[\textwidth]{
\hbox{\includegraphics[width=0.35\textwidth, angle=0]{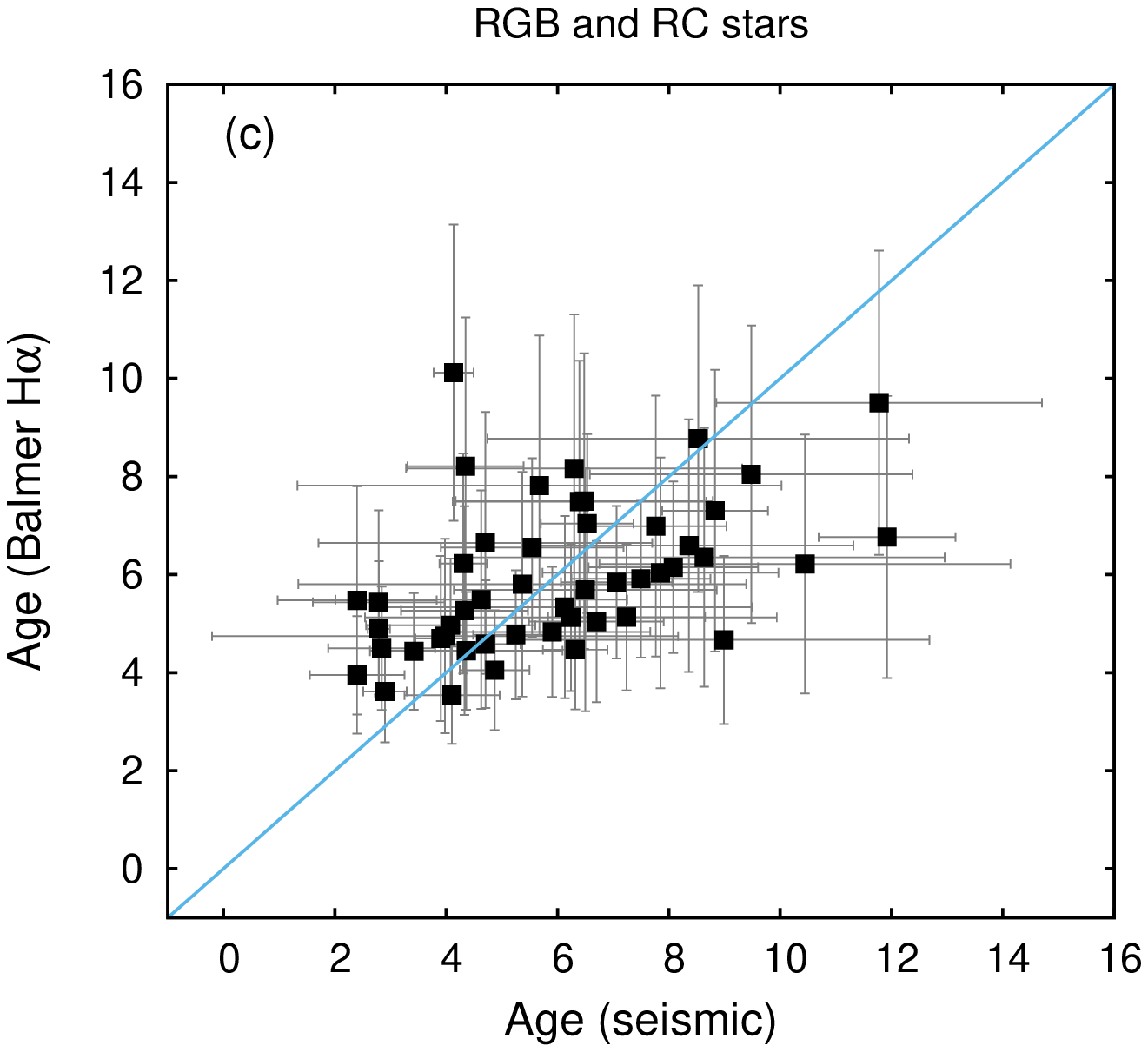}
\includegraphics[width=0.35\textwidth, angle=0]{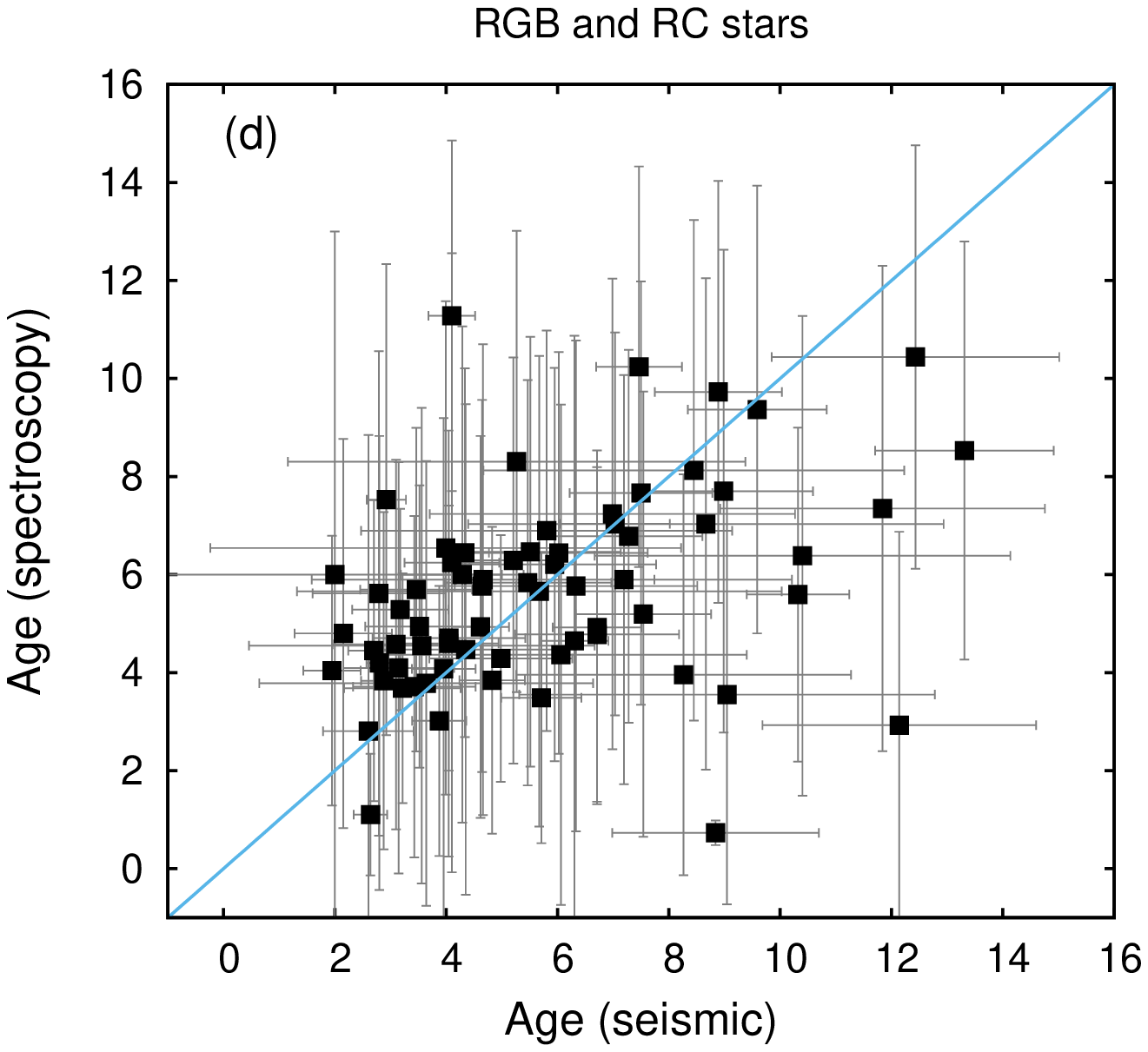}}
}
\caption{The comparison of masses (top panel) and ages (bottom panel) determined using the $\wh$ coefficient directly (a,c) and the classical method of stellar isochrones (b,d).}
\label{relation2}
\end{figure*}

In view of the relationship discussed above, another interesting point is whether the \ha~line profiles show any evidence of ageing, i.e. a star getting older. This is unlikely, even though there is observational evidence that chromospheric activity decreases with stellar age (Steiman-Cameron et al. 1985). If true, one could see some effects in the observed stellar spectra, since the degree of line core filling by the chromospheric temperature inversion would change in lockstep with age. Younger and more active stars would show brighter line cores.

However, we can use a different approach. The initial mass and chemical composition are the two fundamental quantities needed to determine the age of red giant stars. One complication is a substantial mass loss: from observations only the present-day mass can be determined, while age estimates require knowledge of the initial mass of a star. Mass loss may affect the absolute ages of very evolved ($\logg\,\lesssim$ 1.5) low mass RGB and RC stars. This difficulty, however, is universally present in any method for determining the age of red giant stars based on their mass, including asteroseismology, and addressing it is outside of the scope of the paper. In this work, we want to understand whether stellar ages derived from the \ha~width coefficient directly are consistent with the ages determined from asteroseismology, which is the best currently available method.

We use the masses determined from the \ha~line width parameter (Eq. 2) and the Eq. 3 in BeSPP, the Bayesian grid-based modelling code (see Sect. 2.3).  The resulting ages are shown in Fig. \ref{relation2} (panel c) for the stars, for which we have confidence in the reliability of the \ha~mass estimates (i.e. those with \teff $< 4900$ K, Sect. 2.5). The age uncertainties are computed as $\pm  34 \%$ around the median value\footnote{This choice is supported by running the code on the ideal set of data and quantifying the offsets assuming different types of location parameter and its dispersion. The \ha~ages are more consistent with the asteroseismic data then the ages derived using the classical method (panel d), i.e. BeSPP with masses computed from the spectroscopic estimates of \teff, $\log g$, and \feh.}. The very large errors of the ages determined with the classical method reflect the uncertainty of the input spectroscopic gravities, which are of the order $\sim 0.2$ dex.

Fig. \ref{relation2} highlights the main problem of the classical mass and age determination on the RGB. In the standard isochrone fitting method (panels b,d of Fig. \ref{relation2}), the likelihood on the RGB is nearly flat, because all isochrones line up together. To distinguish stars with the age difference of even $10$ Gyr, extreme and so far unachievable precision in \teff~ and $\log g$ ($10$ K and $0.02$ dex) is needed. In the Bayesian framework, the ages determined using uninformative data are completely dominated by the priors, especially the initial mass function (Serenelli et al. 2013, Fig. 4). For a simple stellar population, this approach may work producing an age distribution which does not look too odd, but just by chance. Moreover, this is barely applicable to any real astrophysical system, such as the Galactic disk, which is a conglomerate of stars formed in different environments in-situ or even outside the Milky Way. 
\section{Testing for biases}\label{sec:biases}
The observed correlation between the measured parameter $\wh$ and the stellar metallicities in Figure \ref{fig:haw}  raises the question if our fitting relation might be just an omitted variable bias of some age-metallicity relation, or if our mass estimates are affected by a metallicity bias, e.g. through the dependence of the \ha~line profile shape on the chemical composition of the stellar atmosphere, i.e. on metallicity. Similarly we need to check for dependencies on $\log g$ and \teff.

\begin{figure}
\centering
\includegraphics[width=0.45\textwidth, angle=0]{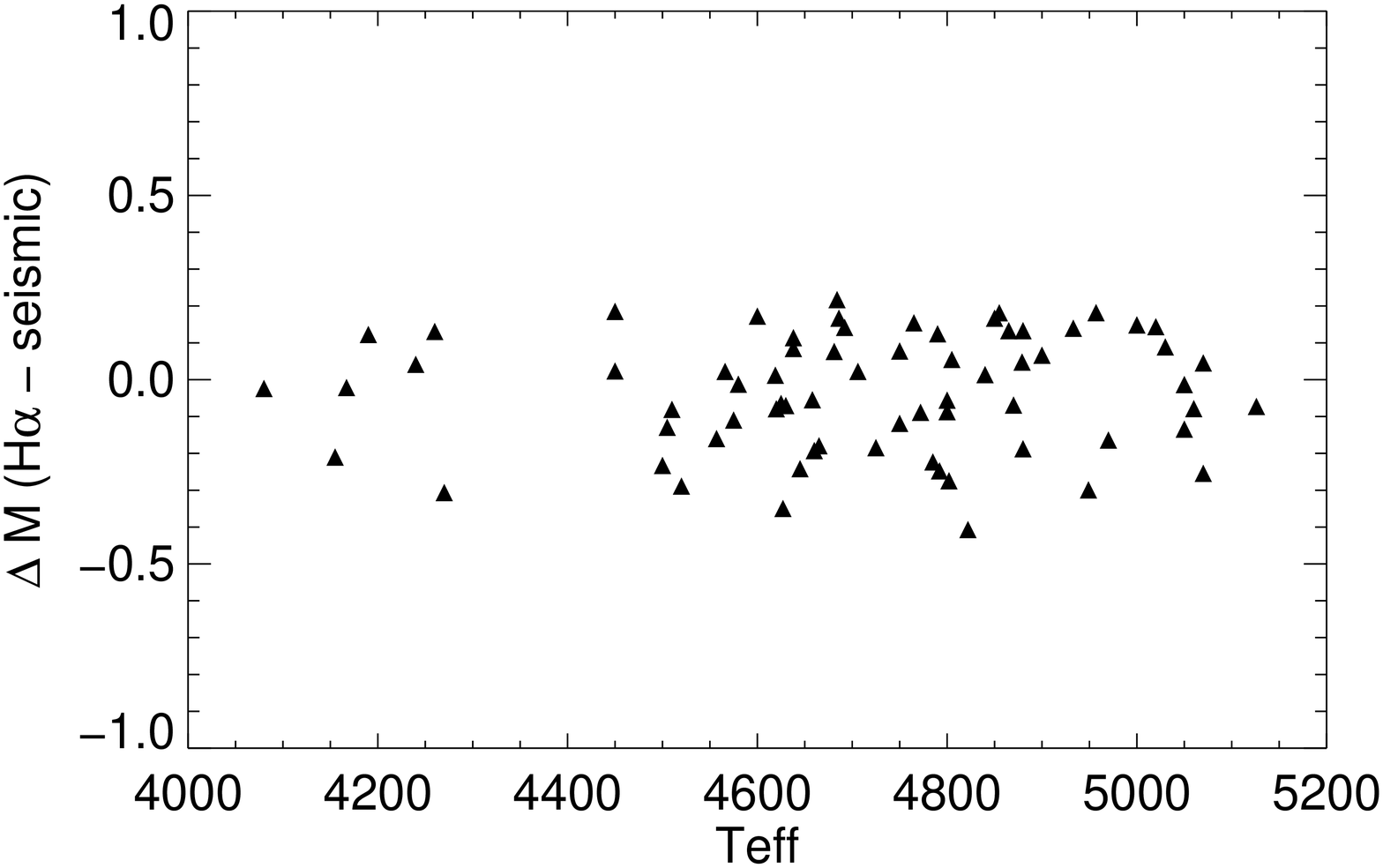}
\includegraphics[width=0.45\textwidth, angle=0]{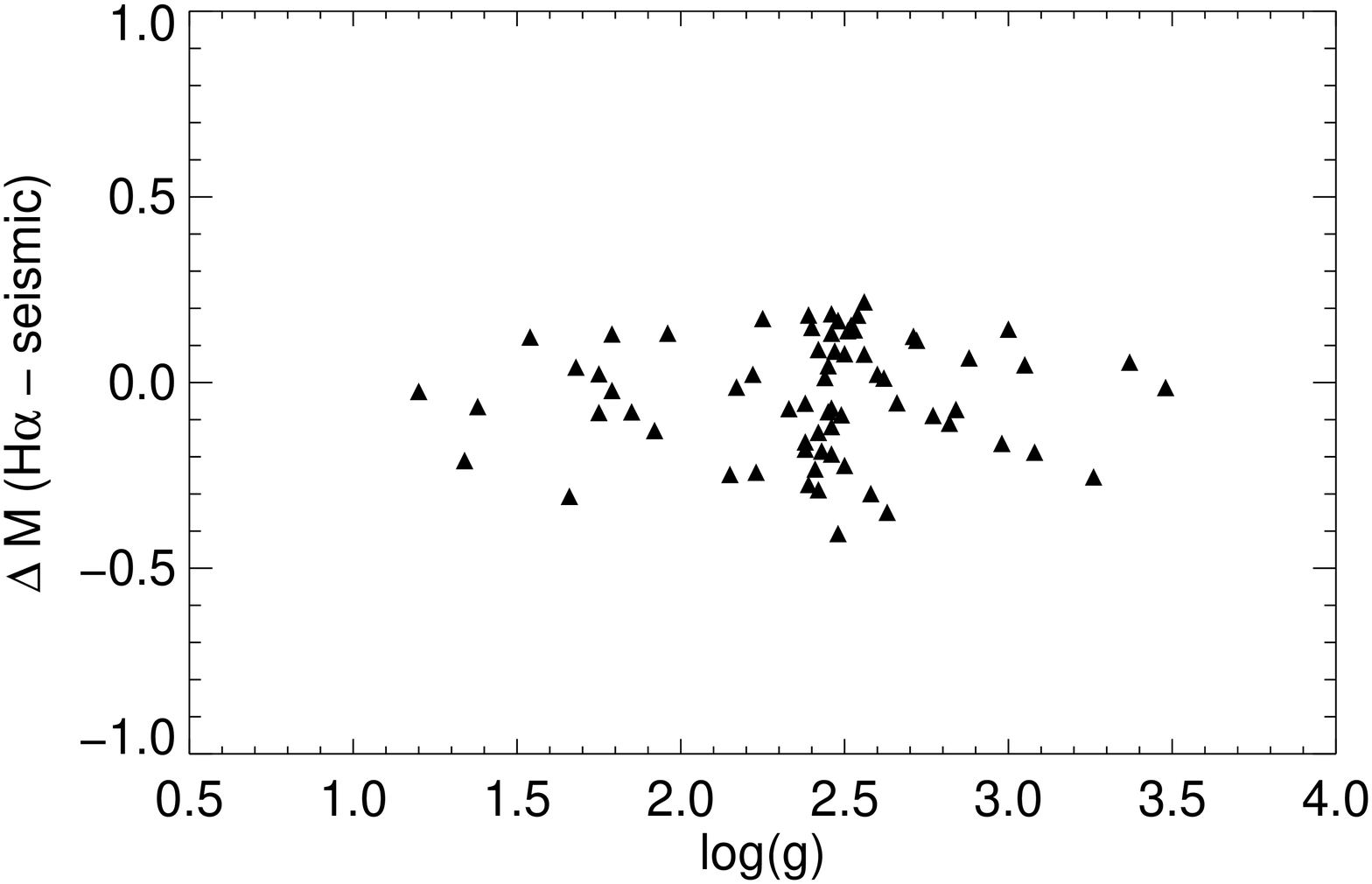}
\includegraphics[width=0.45\textwidth, angle=0]{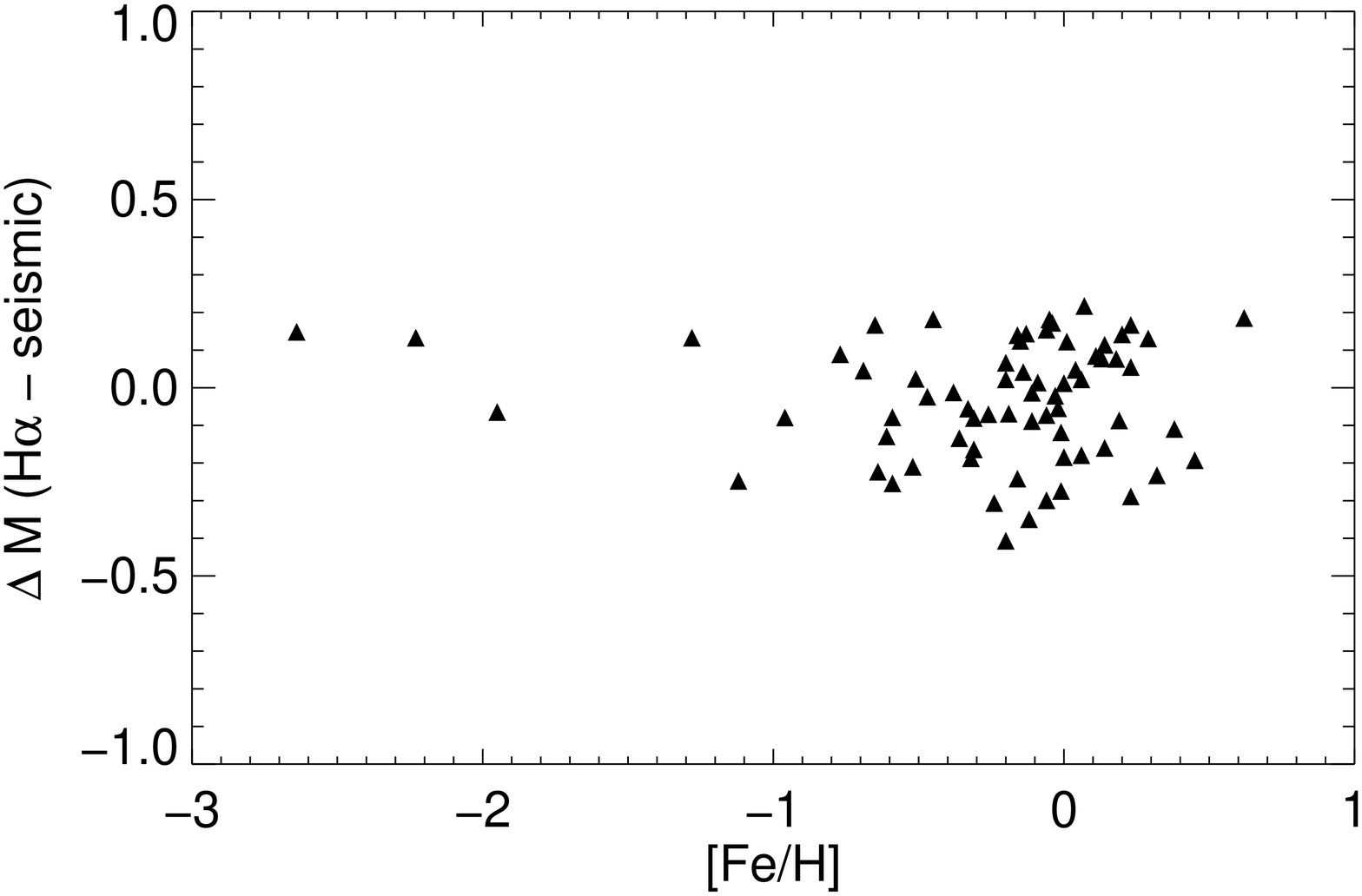}
\caption{The difference between the \ha~masses and asteroseismic masses as a function of  \teff, $\feh$, and $\log g$.}
\label{fig:correlations}
\end{figure}

We can test this quite easily: since our sample is well-populated (independently) in all dimensions (asteroseismic mass, \teff, $\feh$, $\log g$), additional dependencies would show in regression analysis  against the other variables and thus give an indication if the equation is subject to omitted variable bias.
Our relation between mass and $\wh$ was derived from the comparison with asteroseismic masses without respect to the other parameters, so we can simply test for a correlation between the residuals of the mass determination and, say, $\log g$. If $\log g$ had an impact on our $\wh$ estimates, it would show in the regression analysis and be apparent in the plot of residuals (Figure \ref{fig:correlations}). However, visual inspections as well as the sample's statistics affirm that the mass residuals between our spectroscopic method and asteroseismology show no correlation with the other stellar parameters (Figure \ref{fig:correlations}). To test this statistically, we did a linear regression with the equation
\begin{equation}
\Delta_{M} = a_i P_i + b_i + \epsilon
\end{equation}
%
 %
%
%
\begin{table*}
\caption{Left: the fit parameters in a weighted regression using the errors from asteroseismic and spectroscopic mass determinations. Right: the fit parameters from a simple linear regression.}
\label{table7}
\tabcolsep1.1mm
\begin{center}
\begin{tabular}{lrcrc}
\noalign{\smallskip}\hline\noalign{\smallskip} 
\noalign{\smallskip}\hline\noalign{\smallskip} 
parameter & $a_i$ & $b_i$ & $a'_i$ & $b'_i$ \\
\noalign{\smallskip}\hline\noalign{\smallskip}
\teff $-4500$ K & $(-1.3 \pm 1.1)10^{-4}~~$\Msun K$^{-1}$   & $(0.032 \pm 0.021)$ \Msun  & $(-1.5 \pm 1.0)10^{-4}$ & $(0.023 \pm 0.021)$ \\
$\feh$          & $ (0.077 \pm 0.036) ~~$\Msun dex$^{-1}$  & $(0.044 \pm 0.022)$ \Msun  & $(0.036 \pm 0.034)$ & $(0.017 \pm 0.021)$ \\
$\log g -2.5$   & $(-0.020 \pm 0.052) ~~$\Msun dex$^{-1}$  & $(0.020 \pm 0.019)$ \Msun  & $(0.000 \pm 0.043)$ & $(0.007 \pm 0.020)$ \\  
\noalign{\smallskip}\hline\noalign{\smallskip}
\end{tabular}
\end{center}
\end{table*}

where $\Delta_{M} = M_{H\alpha} - M_{\rm seismic}$ is the difference/residual between our masses and the asteroseismic masses, $P_i$ is the parameter in question ($\feh$, $\log g$, \teff), $a_i$ is the slope (which should be $\sim 0$ in the ideal case) and $b_i$ is the offset. The results are reported in Table \ref{table7}, where we show on the left the fit parameters in a weighted regression using the errors from asteroseismic and spectroscopic mass determinations and on the right the fit parameters from a simple linear regression. The table shows that the only trend that reaches the $2 \sigma$ significance limit is the trend with metallicity. However this trend halves when we do a simple linear regression instead of weighted least squares. It is mostly caused by mild outliers on the very metal-poor end and the most metal-rich star in the sample: any trend fully vanishes when we exclude the stars with $\feh < -2.0$ and $\feh > 0.3$. While this can well be just a lucky draw, it might point to (partly known) problems with asteroseismic scaling relations. On the other hand this test indicates that our method can be used with good confidence in the metallicity range $-2.0 < \feh < 0.3$.
\section{Discussion}\label{sec:discussion}
An important question is the applicability of our method.  Currently, the best methods for measuring the mass and age of stars use either asteroseismology or fits of turn-off or sub-giant stars to evolutionary tracks. Asteroseismology requires bright apparent magnitudes, while turn-off stars and subgiants are too faint for observations at large distances. 

Under conservative assumptions, we have shown that our method works for stars with $0.5 < \logg < 3.5$, $4000 <$ \teff ~$< 5000$~K, $-2 < \feh < 0.3$, and $\log L/ L_{\odot} \lesssim 2.5$. These stars, which comprise the RGB, HB, and AGB branches in the H-R diagram, are intrinsically bright, especially when compared to their dwarf and TO counterparts. Thus, our method may prove to be a valuable tool to measure masses and ages for spectroscopically feasible stars beyond the solar neighbourhood, extending to even dwarf galaxies and the Andromeda galaxy.

\begin{figure}
\centering
\includegraphics[width=0.3\textwidth, angle=-90]{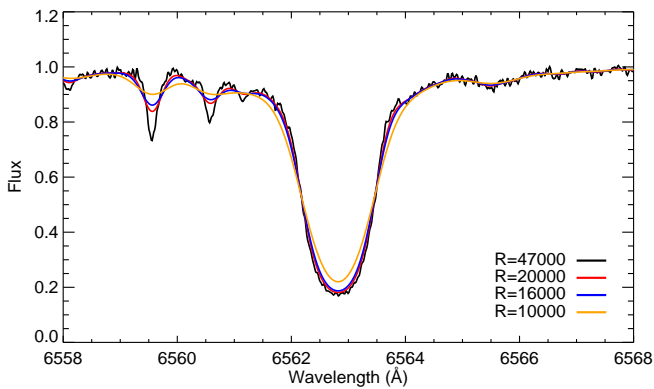}
\caption{The impact of spectral resolution on the \ha~line profile. The observed UVES spectrum of the star $19251846+0016550$ was taken from the Gaia-ESO archive.}
\label{fig:resolution1}
\end{figure}

\begin{figure}
\centering
\includegraphics[width=0.4\textwidth, angle=0]{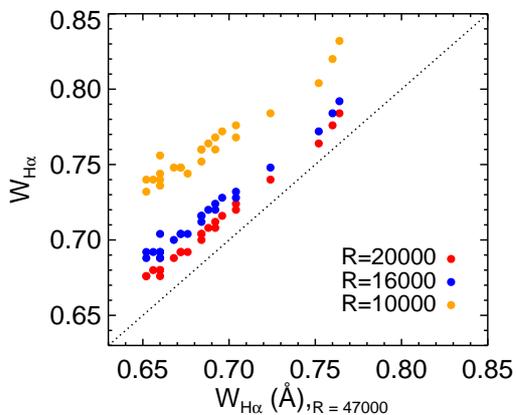}
\caption{The comparison of the \ha~width parameter measurements for different values of spectral resolving power.}
\label{fig:resolution2}
\end{figure}

To examine the effect of spectral resolution of the observed data, the $\wh$ measurements for the Gaia-ESO stellar sample are shown in Fig. \ref{fig:resolution2}. The original UVES spectra were degraded from $R = 47\,000$ to several values of resolving power, representative of other instrument facilities (Fig. \ref{fig:resolution1}), such as the medium-resolution Giraffe spectrograph at the VLT ($R=16\,000$) and the high-resolution mode of the WEAVE and 4MOST facilities ($R=20\,000$). No changes were made to the fitting procedure or the \ha~linemask. Interestingly, degrading the data to about half ($R=20\,000$) or one third ($R=16\,000$) of the original resolution does not have a significant effect on the \ha~measurements. The offset from the high-resolution data is of the order $0.02$ respectively $0.03$ \AA~that can be taken into account by shifting the zero-point of the $\wh$-mass relationship (Eq. 3). For resolving powers $R = 10\,000$ and lower, the differences between the original resolution and degraded measurements are large enough to introduce systematic errors in the mass estimates. However, the relationship is still there, thus it may be possible to calibrate the Eq.~3 on low-resolution spectra.

\begin{figure}
\centering
\includegraphics[width=0.3\textwidth, angle=-90]{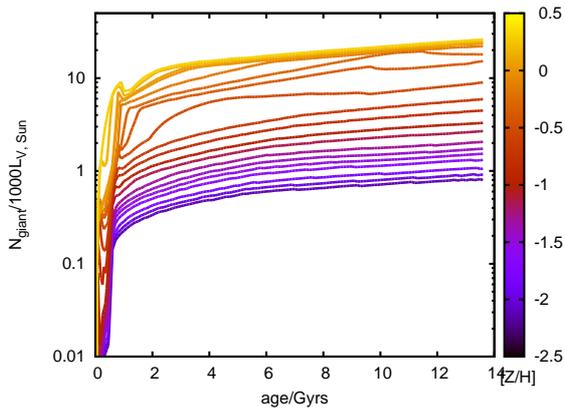}
\caption{The number of red giant stars per $1000 L_{\odot}$ in the Johnson-Cousins $V$ band, depending on age and metallicity (colour coding).}
\label{fig:popsyn}
\end{figure}

In terms of galaxy formation, an interesting question is what is the fraction of observable stars in a typical L$^*$ type galaxy that could be potentially studied with this technique.  Fig. \ref{fig:popsyn} shows the number of stars in our \teff-$\logg$ parameter space per $1000 L_{\odot}$ in the Johnson-Cousins $V$ band, as a function of age (x-axis) and metallicity (colour coding). The plot was created using the population synthesis code underlying Sch{\"o}nrich \& Bergemann (2014), and it represents a typical Galactic disc, calculated with the updated version \citep{2016arXiv160502338S} of the \citet{2009MNRAS.396..203S} model.
The plot shows that all stellar populations with ages above $1$ Gyr are well-sampled at any metallicity, from the metal-rich $\feh +0.5$ to the most metal-poor $\feh -2.5$. The selection is up to an order of magnitude more efficient for metal-rich populations, because the cooler effective temperatures move larger parts of the giant branches into our \teff - $\logg$ parameter region. The gradient mostly reflects the radial metallicity gradient, with more stars being selected in the more metal-rich inner disc regions. This bias must be taken into account when comparing to galaxy models or when an unbiased dataset is constructed. Nevertheless, the selection probability is sufficiently large for all ages and metallicities, allowing to sample any population with age $\tau > 1$ Gyr in units of $\geq 10^5$ solar masses.

To test for omitted variable biases in our empirical formula, we tested the residuals between masses derived from our method and the astroseismic values for trends with the stellar parameters \teff, $\log g$, $\feh$. This test confirmed that no significant trends could be found. I.e. our method does not have any significant bias with any stellar parameter, arguing against omitted variable biases and supporting the validity of the empirical formula.

To conclude, we have shown that our method samples well all stellar populations with ages above $1$ Gyr and still delivers sufficient sample sizes for moderate mass dwarf galaxies. Targeting bright giants, the method allows to obtain simultaneous age and chemical abundance (by applying the usual spectroscopic method of model atmospheres to the observed stellar spectrum) information far deeper than would be possible with asteroseismology of red giants, extending the possible survey volume to remote regions of the Milky Way and even to neighbouring galaxies like Andromeda or the Magellanic Clouds with present instrumentation on telescopes, like the VLT, Keck, or LBT. For example, with UVES/VLT or with the HIRES instrument at Keck \citep{vogt2011}, optical high- or medium-resolution spectra can be acquired for stars with V magnitude $\sim 19$; in contrast the majority of red giants with asteroseismic ages from CoRoT or Kepler missions are brighter than $14$ mag (Mosser et al. 2010, Batalha et al. 2010, Hekker et al. 2011, Huber et al. 2014). In the future, our method opens a novel possibility to measure directly \textit{and} consistently ages and chemical abundances of individual stars in more distant galaxies in the Local Group, e.g. with 4MOST, WEAVE, and E-ELT (Zerbi et al. 2014), which will reach stars with $V \sim 21$.

\section{Summary and conclusions}
We have analysed hydrogen (\ha) line profiles of $69$ red giant stars with high-resolution optical spectra obtained with different instruments. A part of the stellar spectra were acquired within the Gaia-ESO spectroscopic survey (Gilmore et al. 2012, Randich et al. 2013). For all these stars, asteroseismic data from CoRoT and Kepler observations are available, allowing us to compute their mass and age. We also include $73$ stars in $7$ open and globular clusters; their masses are derived from the CMD fitting. The sample covers a wide range range in metallicity, from metal-rich $+0.5$ to very metal-poor, $\feh < -2.5$, stars.

We find that the observed \ha~profile presents a systematic dependence on fundamental stellar parameters. The profile of the unblended blue wing can be represented by a simple cubic-exponential function, which provides an excellent fit to the observed line profiles. The only parameters in the fit are the steepness, or width, of the \ha\ line and the minimum flux in the line core (Eq.~1). Both values appear to correlate with surface gravity and metallicity that suggests a more fundamental underlying relationship with the mass of a star. Indeed, there is a well-defined correlation between the steepness of \ha\ and stellar mass that can be described by a simple linear model (Eq.~2). This allows determination of masses with the accuracy 10-15\% independent of stellar atmosphere and interior modelling. Although masses can be determined quite precisely with our method, it should be kept in mind that the estimates are slightly offset from the asteroseismic scale, such that at the low-mass end stellar masses are on average too high by $\sim 0.1$ M$_{\odot}$ and slightly lower at the high-mass end. This discrepancy could be caused by the asteroseismic mass estimates or by the second parameter problem of the \ha~line. Still, the empirical masses can be used to determine ages of stars. The uncertainties of ages derived from the \ha\ line profile measurements are respectively $20-30$\%, better than the classical stellar isochrone methods.

We cannot yet identify the physical mechanism underlying the relationship, because there are no stellar model atmospheres built from first principles that could be used to reproduce the observed \ha~lines in red giants and explore their sensitivity to stellar mass. The standard 1D LTE models predict too weak \ha~profiles compared to the observations. The more advanced 1D NLTE models with dynamics and winds are parametrised and do not explain the formation of the \ha~lines, because they are designed to match the observed \ha~profile shapes. 3D MHD chromospheric models have been computed for the Sun, but are not yet available for quantitative spectroscopy of stars with very different properties, like the RGB and RC stars. Having such models for a few red giants would be essential to understand the slope of the \ha-mass relation. Most likely, this would require global convection modelling and the inclusion of chromosphere. One other possibility is that stellar model atmospheres lack information about the stellar interior structure which, nevertheless, has an impact on the stellar atmosphere properties as recent work \citep{2014ApJS..215...19P, 2014MNRAS.445.3685C} suggests. One interesting problem for the follow-up work would be to test the effect of variability in the \ha~line on the $\wh$ mass diagnostic. There is observational evidence (e.g. Cacciari et al. 2004, Meszaros et al. 2009 and references therein) that metal-poor stars ($\feh \lesssim -1.0$) more luminous than $\log (L/L_{\odot}) \sim 2.5$ show time-dependent emission in the \ha~wings, as well as asymmetries in the line core. There is currently no evidence for such processes in more metal-rich stars. Most red giants in our asteroseismic sample are fainter than this luminosity threshold and do not show emission or core shifts in the \ha~line. With a large sample of stars for which spectra taken at different epochs are available it would be possible to test the effect of variability and possibly extend the method to very luminous red giants.

Our empirical results have interesting implications for spectroscopic observations of distant evolved stars: extremely metal-poor first stars, stars in the 'dark' part of the Galaxy - the halo, in distant star clusters, Magellanic Clouds and in other galaxies of the Local Group, which will be routinely observable with new facilities like 4MOST and E-ELT. The instruments will reach targets as faint as V $\sim 21$ mag, that is $5$ times fainter compared to what is currently possible with asteroseismic methods, e.g. using Kepler or CoRoT mission data, and even with future missions like TESS. The observed optical spectra, in particular the conspicuous \ha\ line, may directly provide mass and age determinations for these stars, eliminating the need for model-dependent fitting methods based on stellar evolution and population synthesis models. We have also shown that our method, based on the spectral line \ha, samples well all stellar populations with ages above $1$ Gyr and still delivers sufficient sample sizes for moderate mass dwarf galaxies. Thus, the applications of our method are numerous, and extend our ability to measure the mass and age of stars to a much larger volume.
\begin{acknowledgements}
We thank Luca Casagrande for the IRFM temperatures of the program stars and Achim Weiss for the data for the very metal-poor stars. Based on data products from observations made with ESO Telescopes at the La Silla Paranal Observatory under programme ID 188.B-3002. These data products have been processed by the Cambridge Astronomy Survey Unit (CASU) at the Institute of Astronomy, University of Cambridge, and by the FLAMES/UVES reduction team at INAF/Osservatorio Astrofisico di Arcetri. These data have been obtained from the Gaia-ESO Survey Data Archive, prepared and hosted by the Wide Field Astronomy Unit, Institute for Astronomy, University of Edinburgh, which is funded by the UK Science and Technology Facilities Council. This work was partly supported by the European Union FP7 programme through ERC grant number 320360 and by the Leverhulme Trust through grant RPG-2012-541, and grants 2014SGR-1458 (Generalitat de Catalunya), ESP2014-56003-R and ESP2015-66134-R. We acknowledge the support from INAF and Ministero dell' Istruzione, dell' Universit\`a' e della Ricerca (MIUR) in the form of the grant "Premiale VLT 2012". The results presented here benefit from discussions held during the Gaia-ESO workshops and conferences supported by the ESF (European Science Foundation) through the GREAT Research Network Programme. The research leading to the presented results has received funding from the European Research Council under the European Community's Seventh Framework Programme (FP7/2007-2013) / ERC grant agreement no 338251 (StellarAges).
\end{acknowledgements}
\section{Appendix material}
\newpage
\begin{figure}
\centering
\includegraphics[width=0.35\textwidth, angle=-90]{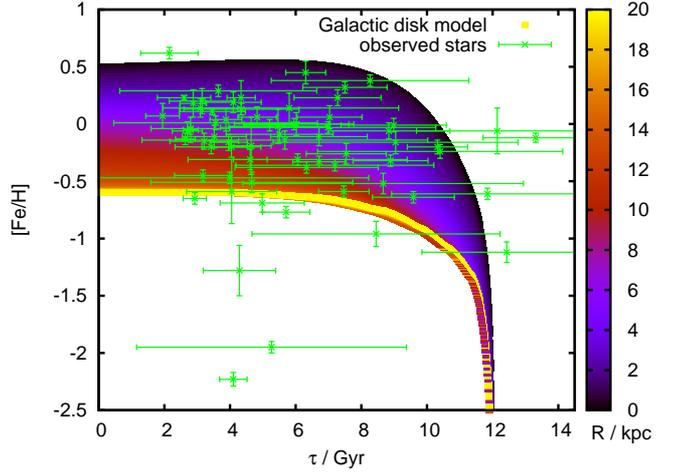}
\caption{The age-metallicity relation (AMR) for our observed stellar sample in the Milky Way disk, computed using the asteroseismic masses and spectroscopic metallicities. The AMR for different Galactocentric radii (R in kpc) from a chemical evolution model of the Galactic disk \citep{2009MNRAS.396..203S} is represented by the colour map.}
\label{fig:amr}
\end{figure}

\newpage
%
%
%
\begin{table*}
\begin{minipage}[t][100mm]{\textwidth}
\centering
\caption{Observational data for the Gaia-ESO stars.}
\label{table1}
\tabcolsep1.1mm
\begin{center}
  \begin{tabular}{rr rccr}
 \noalign{\smallskip}\hline\noalign{\smallskip} 
 \noalign{\smallskip}\hline\noalign{\smallskip} 
 CoRoT ID & Gaia-ESO ID & SNR & RAJ2000 & DECJ2000 & RV  \\
 \noalign{\smallskip}\hline\noalign{\smallskip}
          &             &     & deg   &  deg   & km$/$s  \\ 
      100922474 & 19251846+0016550 &  160 &  +291.3269167   &   +0.2819444  & 145 \\
      100974118 & 19253501+0022086 &   46 &  +291.3958750   &   +0.3690556  &  66 \\
      100864569 & 19250002+0026244 &   52 &  +291.2500833   &   +0.4401111  &  21 \\
      100856697 & 19245756+0052282 &   66 &  +291.2398333   &   +0.8745000  &   2 \\
      100853452 & 19245652+0031116 &   71 &  +291.2355000   &   +0.5198889  &  13 \\
      100597609 & 19232660+0127026 &   62 &  +290.8608333   &   +1.4507222  &  13 \\
      100821572 & 19244648+0119504 &  122 &  +291.1936667   &   +1.3306667  & $-$38 \\
      100733870 & 19241853+0053232 &   54 &  +291.0772083   &   +0.8897778  &  11 \\
      101023768 & 19255284+0012484 &  222 &  +291.4701667   &   +0.2134444  &  51 \\
      100813799 & 19244402+0121257 &  134 &  +291.1834167   &   +1.3571389  &  59 \\
      100826123 & 19244789+0127475 &   43 &  +291.1995417   &   +1.4631944  &  57 \\
      100888944 & 19250775+0014218 &   17 &  +291.2822917   &   +0.2393889  & $-$22 \\
      101080756 & 19261922+0023210 &   35 &  +291.5800833   &   +0.3891667  &   4 \\
      100761750 & 19242747+0045070 &  151 &  +291.1144583   &   +0.7519444  &$-$160 \\
      100610961 & 19233129+0141224 &   58 &  +290.8803750   &   +1.6895556  &  85 \\
      100784327 & 19243467+0050077 &  119 &  +291.1444583   &   +0.8354722  &  14 \\
      100500736 & 19225173+0122202 &   31 &  +290.7155417   &   +1.3722778  &  13 \\
      101100065 & 19262648+0029588 &   77 &  +291.6103333   &   +0.4996667  &  22 \\
      101193334 & 19270157+0035230 &   44 &  +291.7565417   &   +0.5897222  &  18 \\
      100596299 & 19232616+0145326 &   21 &  +290.8590000   &   +1.7590556  &  50 \\
      101594554 & 19294723+0007019 &   39 &  +292.4467917   &   +0.1171944  &  40 \\  
\noalign{\smallskip}\hline\noalign{\smallskip}
\end{tabular}
\end{center}
\end{minipage}
\end{table*}

%
%
\begin{table*}
\begin{minipage}[t][180mm]{\textwidth}
\centering
\caption{Stellar parameters and \ha~ measurements for the Gaia-ESO field stars. Surface gravities are determined using the asteroseismic data.}
\label{table2}
\tabcolsep1.1mm
\begin{center}
\begin{tabular}{rrcc r ccccr ccrcc}
\noalign{\smallskip}\hline\noalign{\smallskip} 
\noalign{\smallskip}\hline\noalign{\smallskip} 
 CoRoT ID & Gaia-ESO ID &  $\wh$  & $f_0$ & age & error & mass & error & \teff & $\sigma$ & $\logg$ & $\sigma$ & $\feh$ & $\sigma$ & $\log \frac{L}{L_{\rm Sun}}$ \\
\noalign{\smallskip}\hline\noalign{\smallskip}
          &             &  \AA\   & \AA\  & Gyr &       & M$_{\odot}$ &    & K &  & dex &  & dex &  & dex \\
%
      100922474 &  19251846+0016550 &  0.76 &  0.83 &  3.17 &  0.87 &  1.22 &  0.13 & 4957 &  23 &  2.39 &  0.01 & -0.45 &  0.05 &  1.73 \\
      100974118 &  19253501+0022086 &  0.76 &  0.83 &  3.96 &  0.57 &  1.16 &  0.06 & 4933 &  58 &  2.51 &  0.01 & -0.16 &  0.07 &  1.51 \\
      100864569 &  19250002+0026244 &  0.69 &  0.83 &  6.71 &  1.47 &  1.11 &  0.07 & 4772 &  52 &  2.77 &  0.01 & -0.11 &  0.08 &  1.32 \\
      100856697 &  19245756+0052282 &  0.66 &  0.81 &  2.88 &  0.41 &  1.42 &  0.07 & 4681 & 134 &  2.56 &  0.01 &  0.18 &  0.12 &  1.64 \\
      100853452 &  19245652+0031116 &  0.71 &  0.82 &  7.03 &  0.99 &  0.97 &  0.06 & 4665 &  68 &  2.38 &  0.01 &  0.06 &  0.10 &  1.64 \\
      100597609 &  19232660+0127026 &  0.69 &  0.82 &  3.46 &  1.01 &  1.32 &  0.15 & 4638 &  62 &  2.47 &  0.02 &  0.11 &  0.03 &  1.74 \\
      100821572 &  19244648+0119504 &  0.68 &  0.81 & 13.31 &  1.60 &  0.92 &  0.03 & 4627 &  75 &  2.63 &  0.01 & -0.12 &  0.04 &  1.39 \\
      100733870 &  19241853+0053232 &  0.67 &  0.80 &  5.80 &  3.33 &  1.10 &  0.10 & 4557 & 105 &  2.38 &  0.01 &  0.14 &  0.13 &  1.58 \\
      101023768 &  19255284+0012484 &  0.70 &  0.82 &  2.70 &  0.46 &  1.39 &  0.08 & 4765 &  57 &  2.52 &  0.01 & -0.06 &  0.08 &  1.71 \\
      100813799 &  19244402+0121257 &  0.66 &  0.81 &  3.10 &  1.14 &  1.48 &  0.15 & 4638 &  79 &  2.72 &  0.01 &  0.14 &  0.08 &  1.65 \\
      100826123 &  19244789+0127475 &  0.68 &  0.81 &  2.60 &  0.81 &  1.49 &  0.16 & 4686 &  36 &  2.48 &  0.02 &  0.23 &  0.02 &  1.68 \\
      100888944 &  19250775+0014218 &  0.69 &  0.82 & 12.14 &  2.46 &  0.93 &  0.05 & 4949 & 181 &  2.58 &  0.01 & -0.06 &  0.20 &  1.42 \\
      101080756 &  19261922+0023210 &  0.67 &  0.81 &  5.47 &  1.50 &  1.21 &  0.10 & 4658 &  43 &  2.66 &  0.01 & -0.02 &  0.15 &  1.41 \\
      100761750 &  19242747+0045070 &  0.76 &  0.82 & 12.43 &  2.58 &  0.80 &  0.05 & 4792 &  58 &  2.15 &  0.02 & -1.12 &  0.09 &  1.92 \\
      100610961 &  19233129+0141224 &  0.69 &  0.83 &  3.21 &  1.04 &  1.38 &  0.14 & 4790 &  36 &  2.71 &  0.02 & -0.15 &  0.07 &  1.54 \\
      100784327 &  19243467+0050077 &  0.71 &  0.82 &  6.99 &  3.28 &  1.12 &  0.15 & 4167 &  87 &  1.79 &  0.02 & -0.03 &  0.05 &  2.16 \\
      100500736 &  19225173+0122202 &  0.73 &  0.82 &  8.98 &  1.61 &  0.85 &  0.07 & 4802 &  49 &  2.39 &  0.01 & -0.01 &  0.06 &  1.64 \\
      101100065 &  19262648+0029588 &  0.67 &  0.81 &  7.28 &  1.32 &  0.99 &  0.08 & 4520 &  45 &  2.42 &  0.01 &  0.23 &  0.07 &  1.59 \\
      101193334 &  19270157+0035230 &  0.70 &  0.83 &  6.02 &  1.60 &  1.19 &  0.09 & 4619 &  46 &  2.62 &  0.02 &  0.00 &  0.10 &  1.47 \\
      100596299 &  19232616+0145326 &  0.67 &  0.83 &  1.95 &  0.51 &  1.63 &  0.16 & 4684 &  42 &  2.56 &  0.02 &  0.07 &  0.11 &  1.76 \\
      101594554 &  19294723+0007019 &  0.68 &  0.83 &  4.04 &  0.90 &  1.27 &  0.08 & 4566 &  63 &  2.22 &  0.01 & -0.20 &  0.09 &  1.87 \\

\noalign{\smallskip}\hline\noalign{\smallskip}
\end{tabular}
\end{center}
\end{minipage}
\end{table*}

%
\begin{table}
\begin{minipage}[t][100mm]{\textwidth}
\centering
\caption{Observational data for the Gaia-ESO stars in clusters.}
\label{table3}
\tabcolsep1.1mm
\begin{center}
\begin{tabular}{rr rccr}
\noalign{\smallskip}\hline\noalign{\smallskip} 
\noalign{\smallskip}\hline\noalign{\smallskip} 
 Cluster & Gaia-ESO ID & SNR & RAJ2000 & DECJ2000 & RV   \\
 \noalign{\smallskip}\hline\noalign{\smallskip}
         &            &     & deg & deg & (km$/$s) \\
M67     &    08510838+1147121  &  163 &  +132.7849167  &   +11.7866944   &  33  \\
M67     &    08513045+1148582  &  209 &  +132.8768750  &   +11.8161667   &  41  \\
M67     &    08513577+1153347  &  191 &  +132.8990417  &   +11.8929722   &  34  \\
M67     &    08514507+1147459  &  228 &  +132.9377917  &   +11.7960833   &  33  \\
NGC1851 &    05133868-4007395  &   77 &   +78.4111667  &   $-$40.1276389 & 319  \\
NGC1851 &    05134382-4001154  &   64 &   +78.4325833  &   $-$40.0209444 & 316  \\
NGC1851 &    05134740-4004098  &   50 &   +78.4475000  &   $-$40.0693889 & 320  \\
NGC1851 &    05135599-4004536  &   52 &   +78.4832917  &   $-$40.0815556 & 318  \\
NGC1851 &    05135634-4003448  &  117 &   +78.4847500  &   $-$40.0624444 & 325  \\
NGC1851 &    05135918-4002496  &  154 &   +78.4965833  &   $-$40.0471111 & 310  \\
NGC1851 &    05135946-4005226  &   64 &   +78.4977500  &   $-$40.0896111 & 315  \\
NGC1851 &    05135977-4002009  &   82 &   +78.4990417  &   $-$40.0335833 & 312  \\
NGC1851 &    05140069-4003242  &   90 &   +78.5028750  &   $-$40.0567222 & 320  \\
NGC1851 &    05140180-4002525  &   61 &   +78.5075000  &   $-$40.0479167 & 311  \\
NGC1851 &    05140376-4001458  &  136 &   +78.5156667  &   $-$40.0293889 & 315  \\
NGC1851 &    05141054-4003192  &   80 &   +78.5439167  &   $-$40.0553333 & 315  \\
NGC1851 &    05141074-4004189  &   95 &   +78.5447500  &   $-$40.0719167 & 328  \\
NGC1851 &    05141171-3959545  &   51 &   +78.5487917  &   $-$39.9984722 & 313  \\
NGC1851 &    05141447-4001109  &   92 &   +78.5602917  &   $-$40.0196944 & 321  \\
NGC1851 &    05141566-4000059  &   49 &   +78.5652500  &   $-$40.0016389 & 314  \\
NGC1851 &    05141576-4003299  &  194 &   +78.5656667  &   $-$40.0583056 & 319  \\
NGC1851 &    05141615-4001502  &   71 &   +78.5672917  &   $-$40.0306111 & 313  \\
NGC1851 &    05141638-4003542  &   65 &   +78.5682500  &   $-$40.0650556 & 317  \\
NGC1851 &    05141957-4004055  &   95 &   +78.5815417  &   $-$40.0681944 & 327  \\
NGC1851 &    05141979-4006446  &   44 &   +78.5824583  &   $-$40.1123889 & 320  \\
NGC1851 &    05142070-4004195  &   43 &   +78.5862500  &   $-$40.0720833 & 321  \\
NGC1851 &    05142480-4002227  &   63 &   +78.6033333  &   $-$40.0396389 & 317  \\
NGC1851 &    05142875-4003159  &   63 &   +78.6197917  &   $-$40.0544167 & 324  \\
NGC1851 &    05142892-4004454  &   78 &   +78.6205000  &   $-$40.0792778 & 324  \\
NGC1851 &    05135946-4005226  &   62 &   +78.4977500  &   $-$40.0896111 & 314  \\
NGC1851 &    05140180-4002525  &   60 &   +78.5075000  &   $-$40.0479167 & 310  \\
NGC1851 &    05141054-4003192  &   78 &   +78.5439167  &   $-$40.0553333 & 315  \\
NGC1851 &    05142875-4003159  &   61 &   +78.6197917  &   $-$40.0544167 & 324  \\
NGC2243 &    06292300-3117299  &   67 &   +97.3457917  &   $-$31.2916111 &  58  \\
NGC2243 &    06292939-3115459  &   75 &   +97.3724583  &   $-$31.2627500 &  59  \\
NGC2243 &    06294149-3114360  &   69 &   +97.4228750  &   $-$31.2433333 &  59  \\
NGC2243 &    06294582-3115381  &   82 &   +97.4409167  &   $-$31.2605833 &  59  \\
NGC2243 &    06290541-3117025  &   60 &   +97.2725417  &   $-$31.2840278 &  59  \\
NGC2243 &    06290934-3110325  &  150 &   +97.2889167  &   $-$31.1756944 &  60  \\
NGC2243 &    06291101-3120394  &  140 &   +97.2958750  &   $-$31.3442778 &  60  \\
NGC2243 &    06292300-3117299  &  196 &   +97.3457917  &   $-$31.2916111 &  59  \\
NGC2243 &    06292939-3115459  &  165 &   +97.3724583  &   $-$31.2627500 &  58  \\
NGC2243 &    06293009-3116587  &  131 &   +97.3753750  &   $-$31.2829722 &  59  \\
NGC2243 &    06293240-3117294  &  121 &   +97.3850000  &   $-$31.2915000 &  58  \\
NGC2243 &    06293518-3117239  &   87 &   +97.3965833  &   $-$31.2899722 &  60  \\
NGC2243 &    06294149-3114360  &  191 &   +97.4228750  &   $-$31.2433333 &  59  \\
NGC2243 &    06294582-3115381  &  207 &   +97.4409167  &   $-$31.2605833 &  60  \\
NGC2808 &    09110169-6451360  &   75 &  +137.7570417  &   $-$64.8600000 & 118  \\
NGC2808 &    09112752-6451312  &   55 &  +137.8646667  &   $-$64.8586667 & 103  \\
NGC2808 &    09115120-6448375  &   91 &  +137.9633333  &   $-$64.8104167 & 110  \\
NGC2808 &    09121405-6448429  &   89 &  +138.0585417  &   $-$64.8119167 & 109  \\
NGC2808 &    09122114-6447139  &   71 &  +138.0880833  &   $-$64.7871944 & 105  \\
NGC2808 &    09123097-6456085  &   59 &  +138.1290417  &   $-$64.9356944 & 102  \\
NGC2808 &    09123170-6449222  &   72 &  +138.1320833  &   $-$64.8228333 &  96  \\
NGC2808 &    09123679-6451451  &   54 &  +138.1532917  &   $-$64.8625278 & 112  \\
NGC2808 &    09123986-6455430  &   88 &  +138.1660833  &   $-$64.9286111 &  99  \\
NGC2808 &    09124112-6446258  &   80 &  +138.1713333  &   $-$64.7738333 & 109  \\
NGC2808 &    09124587-6453014  &   75 &  +138.1911250  &   $-$64.8837222 &  90  \\
NGC2808 &    09125432-6445045  &  146 &  +138.2263333  &   $-$64.7512500 &  95  \\
NGC2808 &    09115120-6448375  &   68 &  +137.9633333  &   $-$64.8104167 & 110  \\
NGC2808 &    09120415-6450224  &   75 &  +138.0172917  &   $-$64.8395556 & 102  \\
NGC2808 &    09123986-6455430  &   68 &  +138.1660833  &   $-$64.9286111 &  98  \\
NGC6752 &    19103866-5954507  &  123 &  +287.6610833  &   $-$59.9140833 & $-$27  \\
NGC6752 &    19104208-6005293  &  133 &  +287.6753333  &   $-$60.0914722 & $-$22  \\
NGC6752 &    19114113-5959266  &   96 &  +287.9213750  &   $-$59.9907222 & $-$32  \\
NGC4372 &    12250660-7239224  &   92 &      186.5506  &   $-$72.67316   &  66  \\
NGC4372 &    12253419-7235252  &  126 &      186.5506  &   $-$72.67316   &  70  \\
NGC4372 &    12253882-7245095  &   77 &      186.5506  &   $-$72.67316   &  75  \\
NGC4372 &    12264293-7241576  &   83 &      186.5506  &   $-$72.67316   &  69  \\
NGC4372 &    12264875-7239413  &   80 &      186.5506  &   $-$72.67316   &  75  \\
NGC5927 &    15272429-5037134  &   28 &      232.0144  &     -50.70945   & $-$106 \\
NGC5927 &    15275926-5039023  &   28 &  +231.9969167  &  $-$50.6506389  & $-$103 \\
\noalign{\smallskip}\hline\noalign{\smallskip}
\end{tabular}
\end{center}
\end{minipage}
\end{table}

%
%
\begin{table*}
\begin{minipage}[t][180mm]{\textwidth}
\centering
\caption{Stellar parameters and \ha~measurements for the Gaia-ESO cluster stars. See Sect. 2.5 for more details.}
\label{table4}
\tabcolsep1.1mm
\begin{center}
\begin{tabular}{rrcc r ccccr cc}
\noalign{\smallskip}\hline\noalign{\smallskip} 
 Cluster & Gaia-ESO ID &  $\wh$  & $f_0$ & \teff & $\sigma$ & $\logg$ & $\sigma$ & $\feh$ & $\sigma$  \\
 \noalign{\smallskip}\hline\noalign{\smallskip}
          &             &  \AA\   & \AA\  & K &  & dex &  & dex &   \\ 
M67      &  08510838+1147121  &   0.66 &  0.81 &   4980 &   37 &   3.46 &   0.16 &  -0.05 &   0.07 \\
M67      &  08513045+1148582  &   0.67 &  0.81 &   4939 &   50 &   3.47 &   0.08 &   0.05 &   0.09 \\
M67      &  08513577+1153347  &   0.66 &  0.82 &   4964 &   25 &   3.46 &   0.10 &  -0.03 &   0.05 \\
M67      &  08514507+1147459  &   0.67 &  0.82 &   4793 &   44 &   2.97 &   0.18 &   0.01 &   0.11 \\
NGC1851  &  05133868-4007395  &   0.83 &  0.83 &   4375 &   24 &   1.12 &   0.36 &  -1.23 &   0.10 \\
NGC1851  &  05134382-4001154  &   0.78 &  0.81 &   4871 &   55 &   2.00 &   0.05 &  -1.18 &   0.02 \\
NGC1851  &  05134740-4004098  &   0.78 &  0.83 &   4892 &   51 &   2.36 &   0.01 &  -1.00 &   0.03 \\
NGC1851  &  05135599-4004536  &   0.80 &  0.83 &   4837 &   24 &   2.17 &   0.14 &  -0.98 &   0.04 \\
NGC1851  &  05135634-4003448  &   0.78 &  0.86 &   4291 &   34 &   0.99 &   0.38 &  -1.16 &   0.09 \\
NGC1851  &  05135918-4002496  &   0.81 &  0.81 &   4917 &   35 &   1.55 &   0.04 &  -1.19 &   0.02 \\
NGC1851  &  05135946-4005226  &   0.80 &  0.83 &   4539 &  111 &   1.66 &   0.16 &  -1.00 &   0.12 \\
NGC1851  &  05135977-4002009  &   0.84 &  0.83 &   4509 &    6 &   1.42 &   0.13 &  -1.09 &   0.03 \\
NGC1851  &  05140069-4003242  &   0.78 &  0.83 &   4536 &    5 &   1.23 &   0.17 &  -1.20 &   0.06 \\
NGC1851  &  05140180-4002525  &   0.83 &  0.84 &   4562 &   90 &   1.63 &   0.21 &  -1.04 &   0.13 \\
NGC1851  &  05140376-4001458  &   0.79 &  0.83 &   4619 &   35 &   1.83 &   0.07 &  -0.98 &   0.04 \\
NGC1851  &  05141054-4003192  &   0.76 &  0.84 &   4412 &   47 &   1.39 &   0.08 &  -1.09 &   0.08 \\
NGC1851  &  05141074-4004189  &   0.80 &  0.82 &   4372 &   19 &   1.14 &   0.26 &  -1.27 &   0.11 \\
NGC1851  &  05141171-3959545  &   0.77 &  0.82 &   4938 &    9 &   2.20 &   0.31 &  -1.17 &   0.07 \\
NGC1851  &  05141447-4001109  &   0.82 &  0.85 &   4317 &    6 &   1.21 &   0.57 &  -1.10 &   0.13 \\
NGC1851  &  05141566-4000059  &   0.77 &  0.82 &   4947 &   28 &   2.15 &   0.03 &  -1.06 &   0.00 \\
NGC1851  &  05141576-4003299  &   0.89 &  0.86 &   4366 &   36 &   1.21 &   0.33 &  -1.18 &   0.07 \\
NGC1851  &  05141615-4001502  &   0.76 &  0.83 &   4880 &   37 &   2.22 &   0.09 &  -1.08 &   0.06 \\
NGC1851  &  05141638-4003542  &   0.76 &  0.82 &   4761 &   50 &   1.99 &   0.15 &  -1.02 &   0.07 \\
NGC1851  &  05141957-4004055  &   0.80 &  0.82 &   4394 &    9 &   1.12 &   0.22 &  -1.22 &   0.11 \\
NGC1851  &  05141979-4006446  &   0.76 &  0.83 &   4993 &   48 &   2.38 &   0.19 &  -1.04 &   0.08 \\
NGC1851  &  05142070-4004195  &   0.76 &  0.83 &   4949 &   98 &   2.38 &   0.15 &  -1.10 &   0.10 \\
NGC1851  &  05142480-4002227  &   0.79 &  0.81 &   4712 &  160 &   1.73 &   0.49 &  -1.20 &   0.15 \\
NGC1851  &  05142875-4003159  &   0.83 &  0.83 &   4600 &    6 &   1.59 &   0.12 &  -1.06 &   0.07 \\
NGC1851  &  05142892-4004454  &   0.81 &  0.85 &   4443 &   17 &   1.50 &   0.07 &  -0.97 &   0.02 \\
NGC1851  &  05135946-4005226  &   0.80 &  0.83 &   4539 &  111 &   1.66 &   0.16 &  -1.00 &   0.12 \\
NGC1851  &  05140180-4002525  &   0.82 &  0.85 &   4562 &   90 &   1.63 &   0.21 &  -1.04 &   0.13 \\
NGC1851  &  05141054-4003192  &   0.77 &  0.84 &   4412 &   47 &   1.39 &   0.08 &  -1.09 &   0.08 \\
NGC1851  &  05142875-4003159  &   0.82 &  0.83 &   4600 &    6 &   1.59 &   0.12 &  -1.06 &   0.07 \\
NGC2243  &  06292300-3117299  &   0.75 &  0.83 &   5039 &   43 &   2.63 &   0.11 &  -0.48 &   0.02 \\
NGC2243  &  06292939-3115459  &   0.74 &  0.83 &   5031 &   28 &   2.54 &   0.16 &  -0.42 &   0.02 \\
NGC2243  &  06294149-3114360  &   0.72 &  0.83 &   4788 &   48 &   2.60 &   0.26 &  -0.44 &   0.04 \\
NGC2243  &  06294582-3115381  &   0.73 &  0.83 &   4962 &   55 &   2.39 &   0.14 &  -0.49 &   0.06 \\
NGC2243  &  06290541-3117025  &   0.75 &  0.84 &   4961 &   24 &   2.52 &   0.11 &  -0.43 &   0.04 \\
NGC2243  &  06290934-3110325  &   0.72 &  0.83 &   4910 &   36 &   2.72 &   0.10 &  -0.41 &   0.04 \\
NGC2243  &  06291101-3120394  &   0.77 &  0.82 &   4895 &   30 &   2.46 &   0.18 &  -0.43 &   0.05 \\
NGC2243  &  06292300-3117299  &   0.75 &  0.82 &   5039 &   43 &   2.63 &   0.11 &  -0.48 &   0.02 \\
NGC2243  &  06292939-3115459  &   0.75 &  0.83 &   5031 &   28 &   2.54 &   0.16 &  -0.42 &   0.02 \\
NGC2243  &  06293009-3116587  &   0.78 &  0.83 &   4689 &   49 &   2.15 &   0.39 &  -0.48 &   0.07 \\
NGC2243  &  06293240-3117294  &   0.77 &  0.83 &   5028 &   47 &   2.57 &   0.13 &  -0.43 &   0.03 \\
NGC2243  &  06293518-3117239  &   0.72 &  0.82 &   4980 &   35 &   2.89 &   0.08 &  -0.42 &   0.07 \\
NGC2243  &  06294149-3114360  &   0.74 &  0.82 &   4788 &   48 &   2.60 &   0.26 &  -0.44 &   0.04 \\
NGC2243  &  06294582-3115381  &   0.76 &  0.82 &   4962 &   55 &   2.39 &   0.14 &  -0.49 &   0.06 \\
NGC2808  &  09110169-6451360  &   0.74 &  0.83 &   4356 &   59 &   1.28 &   0.26 &  -1.11 &   0.11 \\
NGC2808  &  09112752-6451312  &   0.86 &  0.83 &   4709 &   19 &   1.66 &   0.24 &  -0.97 &   0.05 \\
NGC2808  &  09115120-6448375  &   0.80 &  0.83 &   4292 &   29 &   1.27 &   0.07 &  -1.14 &   0.05 \\
NGC2808  &  09121405-6448429  &   0.84 &  0.86 &   4423 &   41 &   1.27 &   0.25 &  -1.14 &   0.03 \\
NGC2808  &  09122114-6447139  &   0.94 &  0.81 &   4520 &    8 &   1.49 &   0.10 &  -1.19 &   0.02 \\
NGC2808  &  09123097-6456085  &   0.84 &  0.82 &   4598 &    8 &   1.72 &   0.04 &  -1.08 &   0.02 \\
NGC2808  &  09123170-6449222  &   0.91 &  0.85 &   4384 &   34 &   1.27 &   0.30 &  -1.12 &   0.12 \\
NGC2808  &  09123679-6451451  &   0.78 &  0.83 &   4710 &   40 &   1.71 &   0.15 &  -1.02 &   0.02 \\
NGC2808  &  09123986-6455430  &   0.88 &  0.85 &   4316 &   62 &   1.22 &   0.30 &  -1.15 &   0.10 \\
NGC2808  &  09124112-6446258  &   0.81 &  0.86 &   4358 &   52 &   1.25 &   0.19 &  -1.13 &   0.08 \\
NGC2808  &  09124587-6453014  &   0.79 &  0.82 &   4417 &   38 &   1.39 &   0.09 &  -1.21 &   0.21 \\
NGC2808  &  09125432-6445045  &   0.81 &  0.90 &   3872 &   36 &   0.78 &   0.25 &  -1.00 &   0.08 \\
NGC2808  &  09115120-6448375  &   0.80 &  0.84 &   4292 &   29 &   1.27 &   0.07 &  -1.14 &   0.05 \\
NGC2808  &  09120415-6450224  &   0.74 &  0.87 &   4207 &   62 &   1.01 &   0.22 &  -1.09 &   0.10 \\
NGC2808  &  09123986-6455430  &   0.84 &  0.83 &   4316 &   62 &   1.22 &   0.30 &  -1.15 &   0.10 \\
NGC6752  &  19103866-5954507  &   0.78 &  0.81 &   4845 &   62 &   1.80 &   0.10 &  -1.62 &   0.02 \\
NGC6752  &  19104208-6005293  &   0.80 &  0.81 &   4736 &   59 &   1.65 &   0.26 &  -1.62 &   0.03 \\
NGC6752  &  19114113-5959266  &   0.76 &  0.81 &   4966 &   21 &   1.98 &   0.11 &  -1.58 &   0.02 \\
NGC4372  &  12250660-7239224  &   0.73 &  0.81 &   4669 &   41 &   1.41 &   0.43 &  -2.20 &   0.11 \\
NGC4372  &  12253419-7235252  &   0.75 &  0.90 &   4414 &  100 &   1.09 &   0.50 &  -2.19 &   0.20 \\
NGC4372  &  12253882-7245095  &   0.77 &  0.82 &   5101 &  442 &   2.00 &   0.83 &  -1.91 &   0.42 \\
NGC4372  &  12264293-7241576  &   0.75 &  0.80 &   4698 &   21 &   1.42 &   0.15 &  -2.29 &   0.02 \\
NGC4372  &  12264875-7239413  &   0.78 &  0.83 &   4639 &   23 &   0.93 &   0.21 &  -2.43 &   0.05 \\
NGC5927  &  15272429-5037134  &   0.72 &  0.84 &   4796 &   83 &   2.44 &   0.25 &  -0.35 &   0.09 \\
NGC5927  &  15275926-5039023  &   0.69 &  0.84 &   4411 &   98 &   2.21 &   0.12 &  -0.21 &   0.11 \\
\noalign{\smallskip}\hline\noalign{\smallskip}
\end{tabular}
\end{center}
\end{minipage}
\end{table*}

%
%
%
\begin{table*}
\begin{minipage}[t][180mm]{\textwidth}
\centering
\caption{Stellar parameters and \ha~measurements for the Kepler stars from Thygesen et al. (2012). The stars marked with a star symbol are targets of Silva Aguirre et al. (2016, in preparation). Surface gravities are determined using the asteroseismic data.}
\label{table6}
\tabcolsep1.1mm
\begin{center}
\begin{tabular}{rrcc r ccccr cc cc}
\noalign{\smallskip}\hline\noalign{\smallskip} 
\noalign{\smallskip}\hline\noalign{\smallskip} 
Kepler ID &  $\wh$  & $f_0$ & age & error & mass & error & \teff & $\sigma$ & $\log g$ & $\feh$ & $\sigma$ & & type \\
\noalign{\smallskip}\hline\noalign{\smallskip}
          &  \AA\  & \AA\  & Gyr &       & M$_{\odot}$ &     & K &  & dex & dex &  &  &  \\
   2425631 &   0.69 &  0.83 &  2.79 &  1.19 &  1.46 &  0.19 & 4600 &  46 &  2.25 & $-$0.04 &  0.05 &  1.96  &      \\
   2714397 &   0.70 &  0.82 &  4.05 &  1.37 &  1.09 &  0.11 & 5060 &  36 &  2.45 & $-$0.59 &  0.28 &  1.79  &      \\
   3429205 &   0.65 &  0.81 &  3.52 &  0.98 &  1.33 &  0.10 & 5050 &  37 &  3.48 & $-$0.11 &  0.10 &  0.85  &      \\
   3430868 &   0.73 &  0.83 &  8.84 &  1.86 &  1.01 &  0.06 & 5126 &  36 &  2.84 & $-$0.06 &  0.07 &  1.39  &      \\
   3744043 &   0.64 &  0.81 &  4.62 &  0.51 &  1.19 &  0.03 & 4970 &  47 &  2.98 & $-$0.31 &  0.09 &  1.27  &  RGB \\
   3748691 &   0.69 &  0.83 &  4.35 &  1.75 &  1.31 &  0.14 & 4750 &  36 &  2.50 &    0.13 &  0.05 &  1.72  &  RGB \\
   3955590 &   0.68 &  0.82 &  9.05 &  3.73 &  1.00 &  0.10 & 4645 &  36 &  2.23 & $-$0.16 &  0.10 &  1.83  &      \\
   4072740 &   0.67 &  0.81 &  4.34 &  0.43 &  1.35 &  0.02 & 4805 &  53 &  3.37 &    0.23 &  0.15 &  0.88  &      \\
   4177025 &   0.70 &  0.83 & 10.40 &  3.73 &  0.90 &  0.15 & 4270 &  70 &  1.66 & $-$0.24 &  0.06 &  2.21  &      \\
   4262505 &   0.68 &  0.83 &  3.43 &  1.10 &  1.33 &  0.12 & 4900 &  86 &  2.88 & $-$0.20 &  0.06 &  1.40  &  RGB \\
   4283484 &   0.82 &  0.83 &  5.71 &  0.72 &  0.93 &  0.05 & 5030 &  36 &  2.42 & $-$0.77 &  0.05 &  1.75  &      \\
   4480358 &   0.77 &  0.84 &  8.45 &  3.78 &  0.90 &  0.12 & 4620 &  36 &  1.85 & $-$0.96 &  0.11 &  2.16  &      \\
   4659706 &   0.64 &  0.80 &  2.15 &  0.87 &  1.70 &  0.21 & 4450 &  61 &  2.46 &    0.62 &  0.05 &  1.75  &      \\
  4671239$^{*}$ &   0.78 &  0.79 &   NaN &   NaN &  1.11 &  0.13 & 5000 & 100 &  2.40 & $-$2.64 &  0.22 &   NaN  &   	 \\        
   5113061 &   0.70 &  0.82 &  3.56 &  3.10 &  1.34 &  0.34 & 4190 &  36 &  1.54 &    0.01 &  0.06 &  2.47  &      \\
   5113910 &   0.72 &  0.84 &  6.06 &  3.34 &  1.03 &  0.14 & 4510 &  55 &  1.75 & $-$0.31 &  0.05 &  2.27  &      \\
   5284127 &   0.67 &  0.81 &  6.31 &  0.61 &  1.07 &  0.04 & 4660 &  36 &  2.46 &    0.45 &  0.10 &  1.63  &   RC \\
   5612549 &   0.77 &  0.84 &  6.71 &  0.80 &  0.92 &  0.05 & 4800 &  77 &  2.38 & $-$0.33 &  0.05 &  1.70  &   RC \\
   5701829 &   0.70 &  0.82 &  8.89 &  1.14 &  0.99 &  0.02 & 4880 &  59 &  3.08 & $-$0.32 &  0.05 &  1.06  &      \\
   5779724 &   0.73 &  0.83 &  5.67 &  4.35 &  1.13 &  0.27 & 4240 &  36 &  1.68 & $-$0.14 &  0.08 &  2.28  &      \\
   5859492 &   0.65 &  0.81 &  4.10 &  0.85 &  1.24 &  0.11 & 4800 &  36 &  2.49 &    0.19 &  0.09 &  1.72  &      \\
   5866965 &   0.72 &  0.85 &  8.66 &  4.27 &  0.92 &  0.15 & 4155 &  75 &  1.34 & $-$0.52 &  0.09 &  2.49  &      \\
   6125893 &   0.69 &  0.82 &  3.64 &  3.00 &  1.39 &  0.37 & 4260 &  41 &  1.79 &    0.29 &  0.05 &  2.26  &      \\
   6547007 &   0.72 &  0.82 &  9.59 &  1.25 &  0.91 &  0.03 & 4785 &  36 &  2.50 & $-$0.64 &  0.05 &  1.57  &      \\
   6579998 &   0.78 &  0.83 &  4.98 &  1.28 &  0.99 &  0.10 & 5070 &  36 &  2.45 & $-$0.69 &  0.08 &  1.76  &   RC \\
   6680734 &   0.73 &  0.82 &  6.33 &  2.18 &  1.07 &  0.10 & 4580 &  76 &  2.17 & $-$0.38 &  0.05 &  1.90  &      \\
   6690139 &   0.68 &  0.82 &  2.64 &  0.30 &  1.45 &  0.03 & 5020 &  42 &  3.00 & $-$0.13 &  0.05 &  1.35  &  RGB \\
   6696436 &   0.72 &  0.83 &  7.54 &  1.22 &  1.04 &  0.04 & 4630 &  74 &  2.33 & $-$0.26 &  0.09 &  1.74  &      \\
   6837256 &   0.73 &  0.83 &  2.92 &  0.35 &  1.30 &  0.04 & 4850 &  48 &  2.48 & $-$0.65 &  0.05 &  1.77  &      \\
   7006979 &   0.70 &  0.82 &  4.64 &  0.61 &  1.10 &  0.05 & 4870 &  96 &  2.46 & $-$0.19 &  0.25 &  1.72  &   RC \\
   7340724 &   0.67 &  0.81 &  3.88 &  0.49 &  1.34 &  0.03 & 4879 & 112 &  3.05 &    0.04 &  0.10 &  1.22  &  RGB \\
  7693833$^{*}$ &   0.78 &  0.79 &  4.10 &  0.42 &  1.09 &  0.04 & 4880 &  49 &  2.46 & $-$2.23 &  0.06 &  1.72  &  RGB \\
   7812552 &   0.68 &  0.82 &  7.47 &  0.77 &  0.99 &  0.02 & 5070 &  78 &  3.26 & $-$0.59 &  0.05 &  0.95  &      \\
   8017159 &   0.74 &  0.82 &  5.26 &  4.11 &  0.99 &  0.25 & 4625 &  36 &  1.38 & $-$1.95 &  0.05 &  2.67  &      \\
   8210100 &   0.70 &  0.82 &  3.15 &  1.00 &  1.37 &  0.14 & 4692 &  36 &  2.53 &    0.20 &  0.11 &  1.68  &   RC \\
   8211551 &   0.73 &  0.82 & 10.32 &  0.92 &  0.77 &  0.03 & 4822 &  36 &  2.48 & $-$0.20 &  0.06 &  1.53  &      \\
   8476245 &   0.78 &  0.81 &  4.29 &  1.10 &  1.09 &  0.09 & 4865 & 159 &  1.96 & $-$1.28 &  0.22 &  2.22  &      \\
   8873797 &   0.69 &  0.81 &  7.50 &  1.28 &  0.98 &  0.07 & 4500 &  40 &  2.41 &    0.32 &  0.05 &  1.59  &   RC \\
   9288026 &   0.76 &  0.83 &  7.19 &  3.02 &  0.88 &  0.17 & 5050 &  36 &  2.42 & $-$0.36 &  0.05 &  1.73  &      \\
   9474021 &   0.69 &  0.85 &  3.99 &  4.23 &  1.18 &  0.37 & 4080 &  36 &  1.20 & $-$0.47 &  0.05 &  2.71  &      \\
  10186608 &   0.69 &  0.83 &  5.95 &  1.82 &  1.02 &  0.13 & 4725 &  49 &  2.43 &    0.00 &  0.05 &  1.67  &   RC \\
  10323222 &   0.68 &  0.82 &  4.83 &  0.59 &  1.27 &  0.03 & 4706 &  75 &  2.60 &    0.06 &  0.09 &  1.59  &  RGB \\
  10403036 &   0.77 &  0.83 & 11.84 &  2.92 &  0.86 &  0.06 & 4505 &  87 &  1.92 & $-$0.61 &  0.05 &  2.02  &      \\
  10404994 &   0.72 &  0.83 &  2.80 &  0.23 &  1.36 &  0.03 & 4855 &  42 &  2.54 & $-$0.05 &  0.11 &  1.73  &   RC \\
  11045542 &   0.72 &  0.84 &  4.66 &  3.08 &  1.14 &  0.21 & 4450 &  36 &  1.75 & $-$0.51 &  0.09 &  2.29  &      \\
  11342694 &   0.67 &  0.80 &  8.27 &  3.01 &  1.15 &  0.11 & 4575 &  36 &  2.82 &    0.38 &  0.05 &  1.27  &      \\
  11444313 &   0.69 &  0.82 &  5.21 &  0.79 &  1.08 &  0.07 & 4750 &  36 &  2.46 & $-$0.01 &  0.05 &  1.67  &   RC \\
  11657684 &   0.75 &  0.83 &  5.51 &  1.63 &  1.04 &  0.13 & 4840 &  36 &  2.44 & $-$0.09 &  0.07 &  1.71  &   RC \\
\noalign{\smallskip}\hline\noalign{\smallskip}
\end{tabular}
\end{center}
\end{minipage}
\end{table*}


\begin{thebibliography}{99}
\bibitem[Alfv{\'e}n(1942)]{1942Natur.150..405A} Alfv{\'e}n, H.\ 1942, \nat, 150, 405
\bibitem[Babcock(1961)]{1961ApJ...133..572B} Babcock, H.~W.\ 1961, \apj, 133, 572
\bibitem[Bagnulo et al.(2003)]{2003Msngr.114...10B} Bagnulo, S., Jehin, E., Ledoux, C., et al.\ 2003, The Messenger, 114, 10
\bibitem[Baker(2008)]{baker}Baker 2008, Natl. Inst. Stand. Technol. Technical Note 1612, 8 pp
\bibitem[Barklem et al.(2000)]{2000A&A...355L...5B} Barklem, P.~S., Piskunov, N., \& O'Mara, B.~J.\ 2000, \aap, 355, L5
\bibitem[Batalha et al.(2010)]{batalha2010} Batalha, N.~M., Borucki, W.~J., Koch, D.~G., et al.\ 2010, \apjl, 713, L109 
\bibitem[Belkacem et al.(2011)]{2011A&A...530A.142B} Belkacem, K., Goupil, M.~J., Dupret, M.~A., et al.\ 2011, \aap, 530, A142
\bibitem[Bergemann et al.(2012)]{2012MNRAS.427...27B} Bergemann, M., Lind, K., Collet, R., Magic, Z., \& Asplund, M.\ 2012, \mnras, 427, 27
\bibitem[Bergemann et al.(2014)]{2014A&A...565A..89B} Bergemann, M., Ruchti, G.~R., Serenelli, A., et al.\ 2014, \aap, 565, A89
\bibitem[Bruntt et al.(2012)]{2012MNRAS.423..122B} Bruntt, H., Basu, S., Smalley, B., et al.\ 2012, \mnras, 423, 122
\bibitem[Cacciari et al.(2004)]{2004A&A...413..343C} Cacciari, C., Bragaglia, A., Rossetti, E., et al.\ 2004, \aap, 413, 343
\bibitem[Carlsson \& Stein(2002)]{2002ApJ...572..626C} Carlsson, M., \& Stein, R.~F.\ 2002, \apj, 572, 626
\bibitem[Carlsson et al.(2007)]{2007PASJ...59S.663C} Carlsson, M., Hansteen, V.~H., de Pontieu, B., et al.\ 2007, \pasj, 59, 663
\bibitem[Casagrande et al.(2014)]{2014MNRAS.439.2060C} Casagrande, L., Portinari, L., Glass, I.~S., et al.\ 2014, \mnras, 439, 2060 
\bibitem[Chaplin et al.(2014)]{2014ApJS..210....1C} Chaplin, W.~J., Basu, S., Huber, D., et al.\ 2014, \apjs, 210, 1
\bibitem[Christensen-Dalsgaard et al.(2014)]{2014MNRAS.445.3685C} Christensen-Dalsgaard, J., Silva Aguirre, V., Elsworth, Y., \& Hekker, S.\ 2014, \mnras, 445, 3685 
\bibitem[Coelho et al.(2015)]{2015MNRAS.451.3011C} Coelho, H.~R., Chaplin, W.~J., Basu, S., et al.\ 2015, \mnras, 451, 3011 
\bibitem[Collet et al.(2011)]{2011A&A...528A..32C} Collet, R., Hayek, W., Asplund, M., et al.\ 2011, \aap, 528, A32
\bibitem[D'Antona et al.(2016)]{2016MNRAS.458.2122D} D'Antona, F., Vesperini, E., D'Ercole, A., et al.\ 2016, \mnras, 458, 2122
\bibitem[De Angeli et al.(2005)]{2005AJ....130..116D} De Angeli, F., Piotto, G., Cassisi, S., et al.\ 2005, \aj, 130, 116
\bibitem[Dupree et al.(1984)]{1984ApJ...281L..37D} Dupree, A.~K., Hartmann, L., \& Avrett, E.~H.\ 1984, \apjl, 281, L37
\bibitem[Dupree et al.(2009)]{2009AJ....138.1485D} Dupree, A.~K., Smith, G.~H., \& Strader, J.\ 2009, \aj, 138, 1485
\bibitem[Dupree et al.(2016)]{dupree16} Dupree, A.~K., Avrett, E.~H., \& Kurucz, R.~L.\ 2016, \apjl, 821, L7 
\bibitem[Epstein et al.(2014)]{2014ApJ...785L..28E} Epstein, C.~R., Elsworth, Y.~P., Johnson, J.~A., et al.\ 2014, \apjl, 785, L28
\bibitem[Fontenla et al.(2009)]{2009ApJ...707..482F} Fontenla, J.~M., Curdt, W., Haberreiter, M., Harder, J., \& Tian, H.\ 2009, \apj, 707, 482
\bibitem[Garc{\'{\i}}a et al.(2014)]{2014A&A...572A..34G} Garc{\'{\i}}a, R.~A., Ceillier, T., Salabert, D., et al.\ 2014, \aap, 572, A34 
\bibitem[Ghezzi \& Johnson(2015)]{2015ApJ...812...96G} Ghezzi, L., \& Johnson, J.~A.\ 2015, \apj, 812, 96
\bibitem[Gilmore et al.(2012)]{2012Msngr.147...25G} Gilmore, G., Randich, S., Asplund, M., et al.\ 2012, The Messenger, 147, 25
\bibitem[Gudiksen et al.(2011)]{2011A&A...531A.154G} Gudiksen, B.~V., Carlsson, M., Hansteen, V.~H., et al.\ 2011, \aap, 531, A154
\bibitem[Gustafsson et al.(2008)]{2008A&A...486..951G} Gustafsson, B., Edvardsson, B., Eriksson, K., et al.\ 2008, \aap, 486, 951
\bibitem[Hansteen(2004)]{2004IAUS..223..385H} Hansteen, V.~H.\ 2004, Multi-Wavelength Investigations of Solar Activity, 223, 385
\bibitem[Hansteen et al.(2015)]{2015ApJ...811..106H} Hansteen, V., Guerreiro, N., De Pontieu, B., \& Carlsson, M.\ 2015, \apj, 811, 106
\bibitem[Hartmann \& MacGregor(1980)]{1980ApJ...242..260H} Hartmann, L., \& MacGregor, K.~B.\ 1980, \apj, 242, 260
\bibitem[Hekker et al.(2011)]{hekker2011} Hekker, S., Gilliland, R.~L., Elsworth, Y., et al.\ 2011, \mnras, 414, 2594
\bibitem[Huber et al.(2014)]{huber2014} Huber, D., Silva Aguirre,  V., Matthews, J.~M., et al.\ 2014, \apjs, 211, 2
\bibitem[J{\o}rgensen \& Lindegren(2005)]{2005A&A...436..127J} J{\o}rgensen, B.~R., \& Lindegren, L.\ 2005, \aap, 436, 127
\bibitem[Kurucz(1979)]{1979ApJS...40....1K} Kurucz, R.~L.\ 1979, \apjs, 40, 1
\bibitem[Leenaarts(2010)]{2010MmSAI..81..576L} Leenaarts, J.\ 2010, \memsai, 81, 576
\bibitem[Leenaarts et al.(2012)]{2012ApJ...749..136L} Leenaarts, J., Carlsson, M., \& Rouppe van der Voort, L.\ 2012, \apj, 749, 136
\bibitem[Leenaarts et al.(2015)]{2015ApJ...802..136L} Leenaarts, J., Carlsson, M., \& Rouppe van der Voort, L.\ 2015, \apj, 802, 136 
\bibitem[Magic et al.(2013)a]{2013A&A...557A..26M} Magic, Z., Collet, R., Asplund, M., et al.\ 2013, \aap, 557, A26
\bibitem[Magic et al.(2013)b]{2013A&A...560A...8M} Magic, Z., Collet, R., Hayek, W., \& Asplund, M.\ 2013, \aap, 560, A8
\bibitem[Martig et al.(2016)]{2016MNRAS.456.3655M} Martig, M., Fouesneau, M., Rix, H.-W., et al.\ 2016, \mnras, 456, 3655
\bibitem[Mashonkina et al.(2008)]{mashonkina2008} Mashonkina, L., Zhao, G., Gehren, T., et al.\ 2008, \aap, 478, 529
\bibitem[Masseron \& Gilmore(2015)]{2015MNRAS.453.1855M} Masseron, T., \& Gilmore, G.\ 2015, \mnras, 453, 1855
\bibitem[Meibom et al.(2015)]{2015Natur.517..589M} Meibom, S., Barnes, S.~A., Platais, I., et al.\ 2015, \nat, 517, 589
\bibitem[Meszaros et al.(2009)]{meszaros09} Meszaros, S., Avrett, E., \& Dupree, A.~K.\ 2009, \aj, 138, 615
\bibitem[Miglio et al.(2012)]{2012MNRAS.419.2077M} Miglio, A., Brogaard, K., Stello, D., et al.\ 2012, \mnras, 419, 2077
\bibitem[Mosser et al.(2010)]{2010A&A...517A..22M} Mosser, B., Belkacem, K., Goupil, M.-J., et al.\ 2010, \aap, 517, A22
\bibitem[Ness et al.(2015)]{2015arXiv151108204N} Ness, M., Hogg, D.~W., Rix, H., et al.\ 2015, arXiv:1511.08204
\bibitem[Pinsonneault et al.(2014)]{2014ApJS..215...19P} Pinsonneault, M.~H., Elsworth, Y., Epstein, C., et al.\ 2014, \apjs, 215, 19
\bibitem[Pont \& Eyer(2004)]{2004MNRAS.351..487P} Pont, F., \& Eyer, L.\ 2004, \mnras, 351, 487 
\bibitem[Przybilla \& Butler(2004)]{przybilla04} Przybilla, N., \& Butler, K.\ 2004, \apjl, 610, L61
\bibitem[Randich et al.(2013)]{2013Msngr.154...47R} Randich, S., Gilmore, G., \& Gaia-ESO Consortium 2013, The Messenger, 154, 47
\bibitem[Rutten(2008)]{2008ASPC..397...54R} Rutten, R.~J.\ 2008, First Results From Hinode, 397, 54
\bibitem[Sch{\"o}nrich \& Binney(2009)]{2009MNRAS.396..203S} Sch{\"o}nrich, R., \& Binney, J.\ 2009, \mnras, 396, 203 
\bibitem[Rutten \& Uitenbroek(2012)]{2012A&A...540A..86R} Rutten, R.~J., \& Uitenbroek, H.\ 2012, \aap, 540, A86
\bibitem[Sacco et al.(2014)]{2014A&A...565A.113S} Sacco, G.~G., Morbidelli, L., Franciosini, E., et al.\ 2014, \aap, 565, A113 
\bibitem[Salaris et al.(2004)]{2004A&A...414..163S} Salaris, M., Weiss, A., \& Percival, S.~M.\ 2004, \aap, 414, 163 
\bibitem[Sch{\"o}nrich \& Binney(2009)]{2009MNRAS.396..203S} Sch{\"o}nrich, R., \& Binney, J.\ 2009, \mnras, 396, 203
\bibitem[Sch{\"o}nrich \& Bergemann(2014)]{2014MNRAS.443..698S} Sch{\"o}nrich, R., \& Bergemann, M.\ 2014, \mnras, 443, 698 
\bibitem[Sch{\"o}nrich \& McMillan(2016)]{2016arXiv160502338S} Sch{\"o}nrich, R., \& McMillan, P.\ 2016, arXiv:1605.02338
\bibitem[Serenelli et al.(2013)]{2013MNRAS.429.3645S} Serenelli, A.~M., Bergemann, M., Ruchti, G., \& Casagrande, L.\ 2013, \mnras, 429, 3645
\bibitem[Smiljanic et al.(2014)]{2014A&A...570A.122S} Smiljanic, R., Korn, A.~J., Bergemann, M., et al.\ 2014, \aap, 570, A122
\bibitem[Soderblom(2010)]{2010ARA&A..48..581S} Soderblom, D.~R.\ 2010, \araa, 48, 581 
\bibitem[Soderblom(2015)]{2015Natur.517..557S} Soderblom, D.~R.\ 2015, \nat, 517, 557
\bibitem[Steiman-Cameron et al.(1985)]{1985ApJ...291L..51S} Steiman-Cameron, T.~Y., Johnson, H.~R., \& Honeycutt, R.~K.\ 1985, \apjl, 291, L51 
\bibitem[Weiss \& Schlattl(2008)]{2008Ap&SS.316...99W} Weiss, A., \& Schlattl, H.\ 2008, \apss, 316, 99 
\bibitem[Thygesen et al.(2012)]{2012A&A...543A.160T} Thygesen, A.~O., Frandsen, S., Bruntt, H., et al.\ 2012, \aap, 543, A160
\bibitem[Thygesen et al.(2012)]{2012yCat..35430160T} Thygesen, A.~O., Frandsen, S., et al.\ 2012, VizieR Online Data Catalog, 354
\bibitem[VandenBerg et al.(2013)]{2013ApJ...775..134V} VandenBerg, D.~A., Brogaard, K., Leaman, R., \& Casagrande, L.\ 2013, \apj, 775, 134
\bibitem[Vogt et al.(2011)]{vogt2011} Vogt, S. S., Steve L. Allen, Bruce C. Bigelow, L. Bresee, B. Brown, T. Cantrall, Albert Conrad, M. Couture, C. Delaney, Harland W. Epps, Darrie Hilyard, David F. Hilyard, E. Horn, Neal Jern, D. Kanto, Michael J. Keane, Robert I. Kibrick, James W. Lewis, Jack Osborne, G. H. Pardeilhan, T. Pfister, T. Ricketts, Lloyd B. Robinson, Richard J. Stover, D. Tucker, J. Ward, Ming Zhi Wei. "HIRES: the high-resolution echelle spectrometer on the Keck 10-m Telescope," pp. 362-375, 1994, S.P.I.E., 2198, 362
\bibitem[White et al.(2015)]{2015EPJWC.10106068W} White, T.~R., Silva Aguirre, V., Boyajian, T., et al.\ 2015, European Physical Journal Web of  Conferences, 101, 06068
\bibitem[Zerbi et al.(2014)]{2014SPIE.9147E..23Z} Zerbi, F.~M., Bouchy, F.,  Fynbo, J., et al.\ 2014, \procspie, 9147, 914723

\end{thebibliography}
\end{document}